\documentclass[fleqn,usenatbib]{mnras}

\usepackage[T1]{fontenc}
\usepackage{ae,aecompl}

\usepackage{graphicx}	
\usepackage{amsmath}	
\usepackage{amssymb}	
\usepackage{newtxtext,newtxmath}

\usepackage{times}
\PassOptionsToPackage{hyphens}{url}
\usepackage{color}
\definecolor{Brown}{rgb}{0.647,0.165,0.165}
\definecolor{NavyBlue}{rgb}{0.0,0,0.5}
\definecolor{Burgundy}{rgb}{0.5,0.0,0.125}
\hypersetup{
    breaklinks=true,
    colorlinks=true,
    citecolor={Burgundy},
    linkcolor={NavyBlue},
    urlcolor={Brown}
}

\def\apj{Astrophys.\ J.}

\def\mnras{Mon.\ Not.\ R.\ Astron.\ Soc.}
\def\aap{Astron.\ Astrophys.}
\def\apjl{Astrophys.\ J.\ Lett.}

\def\prl{Phys.\ Rev.\ Lett.}

\def\apjs{ApJS}

\def\pasa{PASA}

\def\nat{Nature}

\def\araa{Ann.\ Rev.\ Astron.\ Astrophys.}

\newcommand{\lb}{\ell_b} 

\newcommand{\cs}{c_\mathrm{s}}           
\newcommand{\Mach}{\mathcal{M}}      
\newcommand{\MachA}{\mathcal{M}_{\text A}}      
\newcommand{\Pm}{\text{Pm}}
\renewcommand{\Re}{\text{Re}} 
\newcommand{\Rm}{\text{Rm}} 

\renewcommand{\vec}[1]{\mathbf{#1}}	
\newcommand{\dd}{\mathrm{d}}        

\newcommand{\cm}{\,{\rm cm}}    
\newcommand{\m}{\,{\rm m}}      
\newcommand{\pc}{\,{\rm pc}}     
\newcommand{\kpc}{\,{\rm kpc}}  

\newcommand{\mkG}{\mu{\rm G}} 
\newcommand{\muG}{\,\mu{\rm G}} 

\newcommand{\Jy}{\,{\rm Jy}}    
\newcommand{\mJy}{\,{\rm mJy}}          

\newcommand{\rad}{\,{\rm rad}} 

\newcommand{\brms}{\,b_{\rm rms}}

\newcommand{\urms}{\,u_{\rm rms}}

\renewcommand{\ne}{n_{\rm e}}

\newcommand{\RM}{\text{RM}}

\newcommand{\DM}{\text{DM}}

\newcommand{\obsb}{1.232 \, \RM/\DM}
\newcommand{\obsbabs}{1.232 \, |\RM|/\DM}
\newcommand{\bpar}{\langle b_\parallel \rangle}
\newcommand{\sk}{\mathcal{S}}
\newcommand{\ku}{\mathcal{K}}
\newcommand{\CO}{{^{12} \mathrm {CO}}}
\newcommand{\NHI}{{\rm N_{HI}}}
\newcommand{\IHa}{{\rm I_{H\alpha}}}

\newcommand\Eq[1]{Eq.\,\ref{#1}}
\newcommand\Fig[1]{Fig.~\ref{#1}}
\newcommand\Sec[1]{Sec.~\ref{#1}}
\newcommand\Tab[1]{Table~\ref{#1}}
\newcommand\App[1]{Appendix~\ref{#1}}

\graphicspath{{./}{figs/}}

\newcommand\rev[1]{#1}
\newcommand\reva[1]{#1}
\newcommand\revb[1]{#1}

\title[Galactic magnetic fields and $n_{\rm e}$--$B$ correlation]{Magnetic fields in the Milky Way from pulsar observations: effect of the correlation between thermal electrons and magnetic fields}

\author[Seta and Federrath]{
Amit Seta 
\thanks{E-mail: amit.seta@anu.edu.au}
and Christoph Federrath
\\
Research School of Astronomy and Astrophysics, 
Australian National University, Canberra, ACT 2611, Australia\\
}

\date{Accepted XXX. Received YYY; in original form ZZZ}

\pubyear{2020}

\begin{document}
\label{firstpage}
\pagerange{\pageref{firstpage}--\pageref{lastpage}}
\maketitle

\begin{abstract}
Pulsars can act as an excellent probe of the Milky Way magnetic field. The average strength of the Galactic magnetic field component parallel to the line of sight can be estimated as $\langle B_\parallel \rangle = 1.232 \, \text{RM}/\text{DM}$, where $\text{RM}$ and $\text{DM}$ are the rotation and dispersion measure of the pulsar. However, this assumes that the thermal electron density and magnetic field of the interstellar medium are uncorrelated. Using numerical simulations and observations, we test the validity of this assumption. Based on magnetohydrodynamical simulations of driven turbulence, we show that the correlation between the thermal electron density and the small-scale magnetic field increases with increasing Mach number of the turbulence. We find that the assumption of uncorrelated thermal electron density and magnetic fields is valid only for subsonic and transsonic flows, but for supersonic turbulence, the field strength can be severely overestimated by using $1.232 \, \text{RM}/\text{DM}$. We then correlate existing pulsar observations from the Australia Telescope National Facility with regions of enhanced thermal electron density and magnetic fields probed by ${^{12} \mathrm {CO}}$ data of molecular clouds, magnetic fields from the Zeeman splitting of the 21~cm line, neutral hydrogen column density, and H$\alpha$ observations. Using these observational data, we show that the thermal electron density and magnetic fields are largely uncorrelated over kpc scales. Thus, we conclude that the relation $\langle B_\parallel \rangle  = 1.232 \, \text{RM}/\text{DM}$ provides a good estimate of the magnetic field on Galactic scales, but might break down on sub--kpc scales.
\end{abstract}




\section{Introduction} \label{sec:intro}
Magnetic fields are an important component of the Milky Way (or in general spiral galaxies) because they provide additional pressure support to gas against gravity \citep{BoularesC1990}, heat the gas via reconnection \citep{Raymond1992}, alter the morphology of the gas and gas flows \citep{ShettyO2016}, reduce the star-formation efficiency \citep{MestelS1956, KrumholzF2019}, control the propagation of relativistically charged particles \citep{Cesarsky1980, ShukurovEA2017}, and suppress galactic outflows \citep{EvirgenEA17}. The Milky Way magnetic fields can be observationally probed using optical polarisation, polarised synchrotron radiation, Zeeman effect, polarised emission from dust and molecules, and Faraday rotation measure of extragalactic sources, pulsars, and Fast Radio Bursts \citep[see Chapter~3 in][]{KleinFletcher2015}. 

Observationally, galactic magnetic fields are divided into large- and small-scale components (or equivalently mean and random fluctuating components). The large-scale magnetic fields, which are coherent over kpc scales, are primarily studied using the linearly polarized synchrotron intensity and mean Faraday rotation measure. On the other hand, the small-scale random fields have a coherence length of the order of $100 \pc$ and are usually probed using depolarisation of synchrotron radiation and rotation measure fluctuations. Other observational probes include starlight polarisation \citep{FosalbaEA2002}, Zeeman splitting of spectral lines \citep{CrutcherK2019}, and polarised emission from dust \citep{PlanckLSF2016}. The observed large- and small-scale magnetic field strengths in the Milky Way are typically (around the Solar neighborhood) of the order of $1~\text{--}~2 \muG$ and $3~\text{--}~5 \muG$, respectively \citep{Haverkorn2015, Beck2016}. The total field strength increases close to the centre of the Galaxy \citep[$\sim 50~\text{--}~100 \muG$ near the Galactic center,][]{CrockerEA2010} and in the denser regions of the interstellar medium (ISM) \citep[$\gtrsim 5 \muG$,][]{CrutcherEA2010}. The observed magnetic fields are even weaker in the halo of the Milky Way \citep{MaoEA2010,Haverkorn2015}. 

Besides the strength, it is also important to study the structure of the magnetic field. In the Galactic disc, the large-scale field roughly follows the spiral arms and most studies propose the presence of an axisymmetric component with a reversal in the Solar neighborhood \citep[other modes and more than one reversal are also possible; see table~1 in][]{Haverkorn2015}. In the halo of the Milky Way, the estimated average large-scale magnetic field strength is approximately $4 \muG$ and the magnetic field scale height is around $2 \kpc$ \citep{SobeyEA2019}, but the large-scale field has a significant vertical component only towards the southern Galactic pole \citep{MaoEA2010}, and thus the halo magnetic field has an even more complicated structure than the disc. Physically, some of the properties of the large-scale fields in the disc and halo can be explained by the mean-field dynamo theory \citep{ShukurovEA2019}. The small-scale magnetic field is expected to be spatially intermittent with the field concentrated in filaments and sheets and its strength and structure can be understood using the small-scale dynamo theory \citep{SetaEA2020}. Besides the small-scale dynamo, the small-scale magnetic field can also be generated by the tangling of the large-scale field \citep[Sec.
~4.1 in][]{SetaF2020} and compression due to shocks \citep[Appendix~A in][]{SetaEA2018}. The presence of filamentary magnetic fields is also favored by various recent observational results \citep{HaverkornKdB2004, Gaenslar2011,ZaroubiEA2015,PillaiEA2015,Planck2016filg,Planck2016fil,FederrathEA2016,KalberlaEA2017,KierdorfEA2020}.

Most observational techniques provide magnetic field information only along the perpendicular or parallel component to the line of sight, and most likely requires additional information for extracting the magnetic field properties from observables (such as information about relativistic electron density is required for synchrotron observations and about the thermal electron density for rotation measure studies). In the absence of such information, assumptions are made to infer magnetic fields from observables. Such assumptions can have limitations and biases (see \cite{SetaB2019} for discussion on equipartition assumption required for synchrotron observations and \cite{BeckEA2003} for possible effects of assumptions about thermal electron density required for Faraday rotation measurements). 

Pulsars can be excellent probes of the Milky Way magnetic field. The pulsed signals from the pulsars are delayed to lower frequencies due to thermal electrons in the ISM along the path between the pulsar and the observer \citep{HewishEA1968}. The observed dispersion in the pulse is quantified in terms of the Dispersion Measure $\DM$ and is expressed in terms of thermal electron density $\ne$ as
\begin{align} \label{eq:dm}
\frac{\DM}{\pc \cm^{-3}} = \int_{L/\pc} \frac{\ne}{\cm^{-3}} \, \dd \left(\frac{l}{\pc}\right),
\end{align}
where $L$ is the path length (or equivalently the distance to the pulsar). Besides the dispersion, the plane of linearly polarised pulsar emission is also rotated due to $\ne$ and the Galactic magnetic field component along the line of sight $B_{\parallel}$, and this observed rotation is quantified in terms of the Rotation Measure $\RM$, which is expressed as\footnote{\rev{\Eq{eq:rm}, in general, is for the Faraday depth, but for pulsars, since the source and Faraday rotating regions are well-separated, the Faraday depth is equal to $\RM$.}} 
\begin{align} \label{eq:rm}
\frac{\RM}{\rad \m^{-2}} = 0.812 \int_{L/\pc} \frac{\ne}{\cm^{-3}} \,  \frac{B_{\parallel}}{\muG} \, \dd \left(\frac{l}{\pc}\right).
\end{align}
The $\RM$ and $\DM$ (both observables) can be used to determine the average Galactic magnetic field along the path length as
\begin{align} \label{eq:obsb}
\frac{\langle B_{\parallel} \rangle}{\muG}  = \frac{\int_L \ne \, B_{\parallel} \, \dd l}{\int_L \ne \, \dd l} = 1.232 \, \frac{\RM / \rad \m^{-2}}{\DM / \cm^{-3} \phantom{-}}. 
\end{align}
This technique has been used to measure Galactic magnetic fields from pulsar observations, especially to study the properties of the large-scale magnetic fields. \citep{Smith1968,Manchester1972,Manchester1974,LyneS1989,RandK1989,HanMQ1999,IndraniD1999,MitraEA2003,HanEA2006,HanEA2018,SobeyEA2019}.

However, determining magnetic fields using \Eq{eq:obsb} assumes that the thermal electron density and the magnetic field are uncorrelated, which need not be the case throughout the ISM. We might expect that in regions with high compressibility (quantified by the Mach number $\Mach = \urms/\cs$, where $\urms$ is the RMS turbulent velocity and $\cs$ is the sound speed), the thermal electron density and the magnetic field will be correlated (especially in regions where the magnetic field is enhanced due to shock compression). Using simulations and various observations, we aim to explore the effect of such a correlation on the estimated Milky Way magnetic field.

\cite{BeckEA2003} studied the bias in magnetic field estimates introduced by a correlation between thermal electrons and magnetic fields using analytical models and \cite{WuEA2009,WuKR2015} studied it using numerical simulations of subsonic and transsonic ($\Mach \le 2$) magnetohydrodynamic (MHD) turbulence. Their simulations show that the correlation does not have much effect on the magnetic field estimate, the $\RM$ distribution is Gaussian, and that the standard deviation of the $\RM$ distribution can be related to the strength of the initial uniform mean field. However, in the strong mean-field regime of theirs, the small-scale magnetic field is largely generated by the tangling of the mean field, which most likely will not be correlated with the thermal electron density. Moreover, on increasing the initial field strength, the root-mean square (RMS) strength of small-scale magnetic fields decreases because the turbulence finds it harder to tangle the mean field \citep[see][for further details]{Federrath2016, BeattieFS2020}. This is most likely the reason for the anti-correlation between the standard deviation of $\RM$ and the mean field in their work \citep[see Fig.~5 and Eq.~3 in][]{WuEA2009}. In addition to that, their simulations are ideal and thus it might be difficult to control the viscous and dissipative effects, which shape the small-scale properties of the velocity and magnetic field.

Here we study the effect of the correlation between the thermal electron density and the magnetic field using non-ideal MHD simulations of driven subsonic, transsonic, and supersonic ($\Mach \le 10$) turbulence. \rev{This is motivated by the fact that the analysis of the linear polarisation maps of regions in the Milky Way suggests subsonic and transsonic turbulence \citep{Gaenslar2005,BurkhartLG2012,SunEA2014}, whereas the ratio of the turbulent to thermal energies in external spiral galaxies, for example IC342 \citep{Beck2015}, suggests supersonic turbulence \citep[see Sec.~4.2 in][for further details]{Beck2016}. Also, a recent high-resolution numerical simulation of driven turbulence shows subsonic turbulence on smaller scales and supersonic turbulence on larger scales, with the transition happening at the sonic scale \citep{FederrathEA2020}. Thus, depending on the length scales, the nature of turbulence might differ. However, one also needs to take into account the various phase transitions and associated temperature variations in the ISM.}

The methods are described in \Sec{sec:nummet} and results from the numerical simulations are presented and discussed in \Sec{sec:simres}. We seed the simulations with a very weak random field with zero mean. The seed field evolves due to the driven turbulence and eventually becomes dynamically important. \rev{In our numerical simulations, we only have small-scale random magnetic fields, which do not significantly contribute to the mean $\RM$ (which is primarily a consequence of the large-scale magnetic field), but contribute significantly to the higher-order statistical moments (for example, standard deviation and kurtosis) of the $\RM$ distribution \reva{\citep[also see][for a discussion on the standard deviation of rotation measure fluctuations under different physical conditions]{OnEA2019}}. Thus, throughout this study, we compute and discuss only the statistical properties of the $\RM$, $\DM$, and $\obsb$ distributions, but do not consider the actual values of $\RM$ or $\DM$.} In \Sec{sec:obs}, we explore the correlation and its effect on the magnetic field estimated using pulsar observations from the Australia Telescope National Facility (ATNF) pulsar catalog \citep{ManchesterEA2005} \footnote{The data used here is from version~1.63 of the updated catalog, which is available at \href{http://www.atnf.csiro.au/research/pulsar/psrcat}{http://www.atnf.csiro.au/research/pulsar/psrcat}.}, $\CO$ observations of molecular clouds \citep{MivilleME2017}, Zeeman effect \citep{HeilesT2004}, neutral hydrogen column density observations of the Milky Way \citep{HI4PI2016}, and ionised hydrogen intensity (H$\alpha$) observations of the Milky Way \citep{Finkbeiner2003}. Finally, we discuss a few additional aspects of our study in \Sec{sec:dis} (where we also use the data for $\RM$ variations along the pulse profile from \citet{IlieJW2019} and the $\RM$ and $\DM$ of Fast Radio Bursts (FRBs) from the FRB catalog; \citet{PetroffEA2016}\footnote{The data used in this paper is from the updated FRB catalog available at \href{http://www.frbcat.org}{http://www.frbcat.org}.}) and then conclude in \Sec{sec:con}. \rev{In \App{app:m51}, we also discuss the properties of the $\RM$ distribution in an external galaxy (M51), using the data from \cite{KierdorfEA2020}.}

\section{Numerical methods} \label{sec:nummet}
To quantify the correlation between the thermal electron density and the magnetic field, and to study its effect on the magnetic field estimate via \Eq{eq:obsb}, we numerically solve the equations of non-ideal compressible MHD (\Eq{eq:ce} -- \Eq{eq:div}) in three-dimensions on a uniform triply-periodic grid using an updated version of the FLASH code, version~4 \citep{FryxellEA2000, DubeyEA2008}. We assume that the gas is isothermal and the thermal electron density $\ne$ is a fraction of the thermal gas density. The correlation between thermal electrons and magnetic fields is largely controlled by varying the Mach number $\Mach$. For simplicity, we assume that both the viscosity $\nu$ and resistivity $\eta$ are constant in space and time. We solve the following equations,
\begin{align}
	&\frac{\partial \rho}{\partial t} + \nabla \cdot (\rho \vec{u}) = 0,  \label{eq:ce} \\
	&\frac{\partial (\rho \vec{u})}{\partial t} + \nabla \cdot \left(\rho \vec{u} \otimes \vec{u} - \frac{1}{4 \pi} \vec{b} \otimes \vec{b}\right) + 
	\nabla \left(\cs^2 \rho + \frac{\vec{b}^2}{8 \pi}\right) =  \nonumber \\ 
	& \hspace{0.675\columnwidth} \nabla \cdot (2 \nu \rho \vec{S}) + \rho \vec{F}, \label{eq:ns} \\
	&\frac{\partial \vec{b}}{\partial t} = \nabla \times (\vec{u} \times \vec{b}) + \eta \nabla^2 \vec{b},  \label{eq:ie} \\
	&\nabla \cdot \vec{b} = 0 \label{eq:div},
\end{align}
where $\rho$ is the gas density, $\vec{u}$ is the turbulent velocity, $\vec{b}$ is the small-scale random field (with mean zero), $\cs$ is the sound speed, $S_{ij} = (1/2) \, (u_{i,j} + u_{j,i} - (2/3) \, \delta_{ij} \, \nabla \cdot \vec{u})$ is the rate of strain tensor, and $\vec{F}$ is the prescribed acceleration field to drive turbulence.

We use the HLL3R (3-wave approximate) Riemann solver to solve the equation above \citep{BouchutKW2007,BouchutKW2010,WaaganFK2011}, and drive turbulence in a numerical box of side length $L$ with $512^3$ points via $\vec{F}$ using the Ornstein-Uhlenbeck process. The flow is driven solenoidally on larger scales $1 \le kL/2 \pi \le 3$, where $k$ represents the Fourier wavenumber, with a parabolic function for the power, which peaks at $kL/2\pi=2$ and reaches zero at $kL/2\pi=1$ and $kL/2\pi=3$. Thus, the effective driving scale of turbulence is $L/2$. We select the auto-correlation time of $\vec{F}$ to be the eddy turnover time $t_0$, defined with respect to the driving scale of turbulence, i.e., $t_0 = L/2\urms$, where $\urms$ is the RMS turbulent velocity. This method of driving turbulence is described in detail in \citet{FederrathEA2010}.

The dissipative effects of the kinematic viscosity and magnetic resistivity are characterised by the fluid ($\Re$) and magnetic ($\Rm$) Reynolds numbers. These are defined with respect to the RMS velocity $\urms$ and the driving scale of the turbulence $L/2$ as $\Re = \urms L / 2 \nu$ and $\Rm = \urms L / 2 \eta$. For all our runs, we choose $\Re=\Rm=2000$. Thus, the magnetic Prandtl number $\Pm=\Rm/\Re$ is unity. The compressibility of the medium is quantified in terms of the turbulent Mach number $\Mach=\urms/\cs$, which we set to $\Mach=0.1$ for subsonic, $2$ for transsonic, and $5$ and $10$ for supersonic simulations, by varying $\urms$ (the temperature and thus $\cs$ are fixed between the different simulations).

At time $t=0$, we prescribe a constant density $\rho~(t=0) \equiv \rho_0$, zero velocity, and a very weak small-scale random seed magnetic field with zero mean. The seed field follows a power-law spectrum with slope $3/2$ and the strength $\brms~(t=0)$ is chosen such that the Alfv\'enic Mach number $\MachA = (4 \pi \rho_0)^{1/2} \urms/\brms~(t=0) = 3.5 \times 10^{5}$. The properties of the seed field do not affect the final statistically steady state magnetic field \citep{SetaF2020}. The initial plasma beta $\beta~(t=0) = \cs^2 \rho_0/(\brms^2/8 \pi)=2.5 \times 10^{13}$. All these parameters are the same for all runs and we only vary the Mach number of the turbulent flow from $0.1$ to $10$. We run each simulation till the magnetic field reaches saturation, which in our setup for the given set of parameters takes approximately $80\,t_0$.

\section{Simulation results} \label{sec:simres}

\subsection{Correlation between $\ne$ and $b$ on varying $\Mach$} \label{sec:simres:corr}

\begin{figure*} 
\includegraphics[width=\columnwidth]{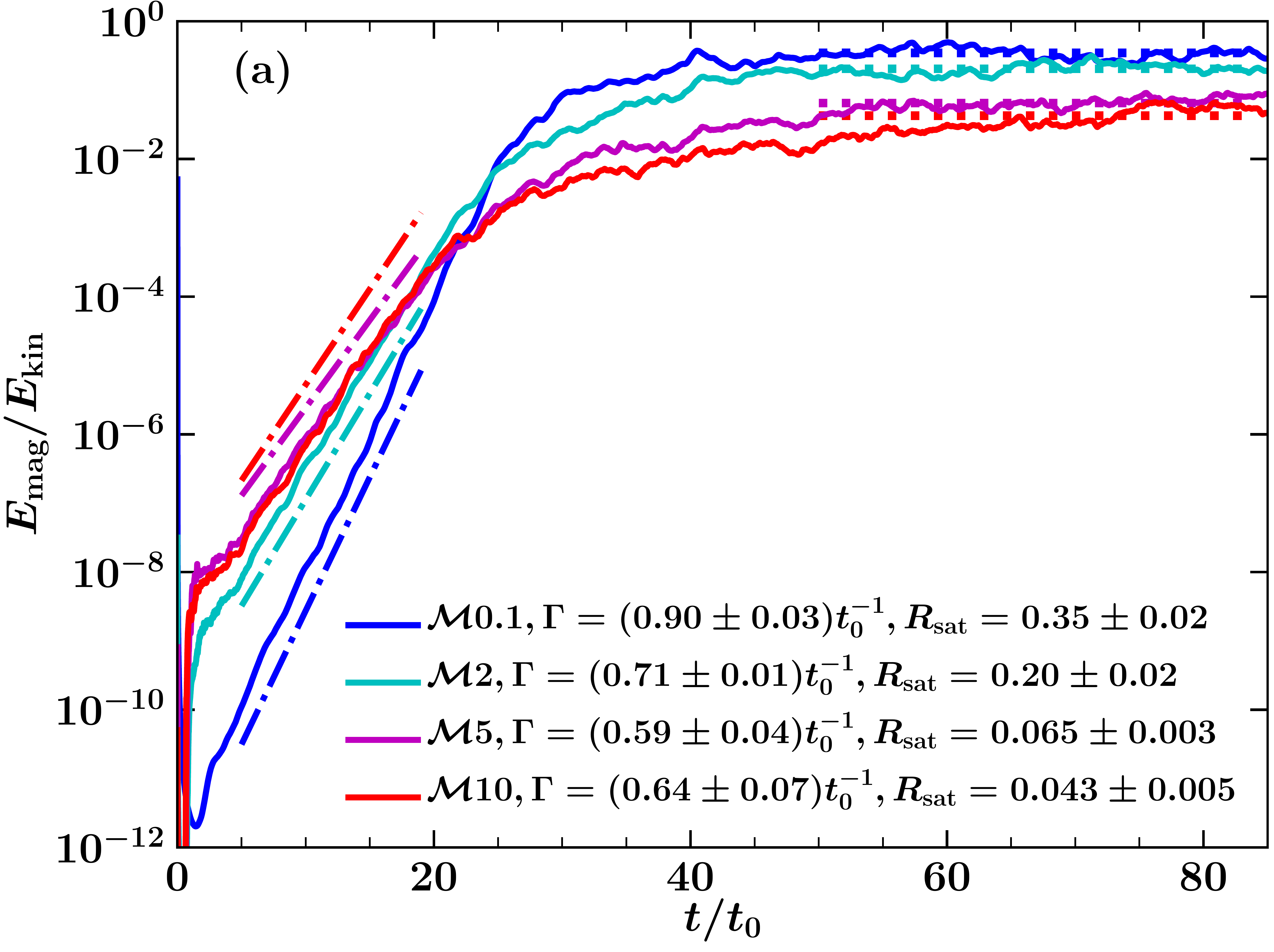} \hspace{0.5cm}
\includegraphics[width=\columnwidth]{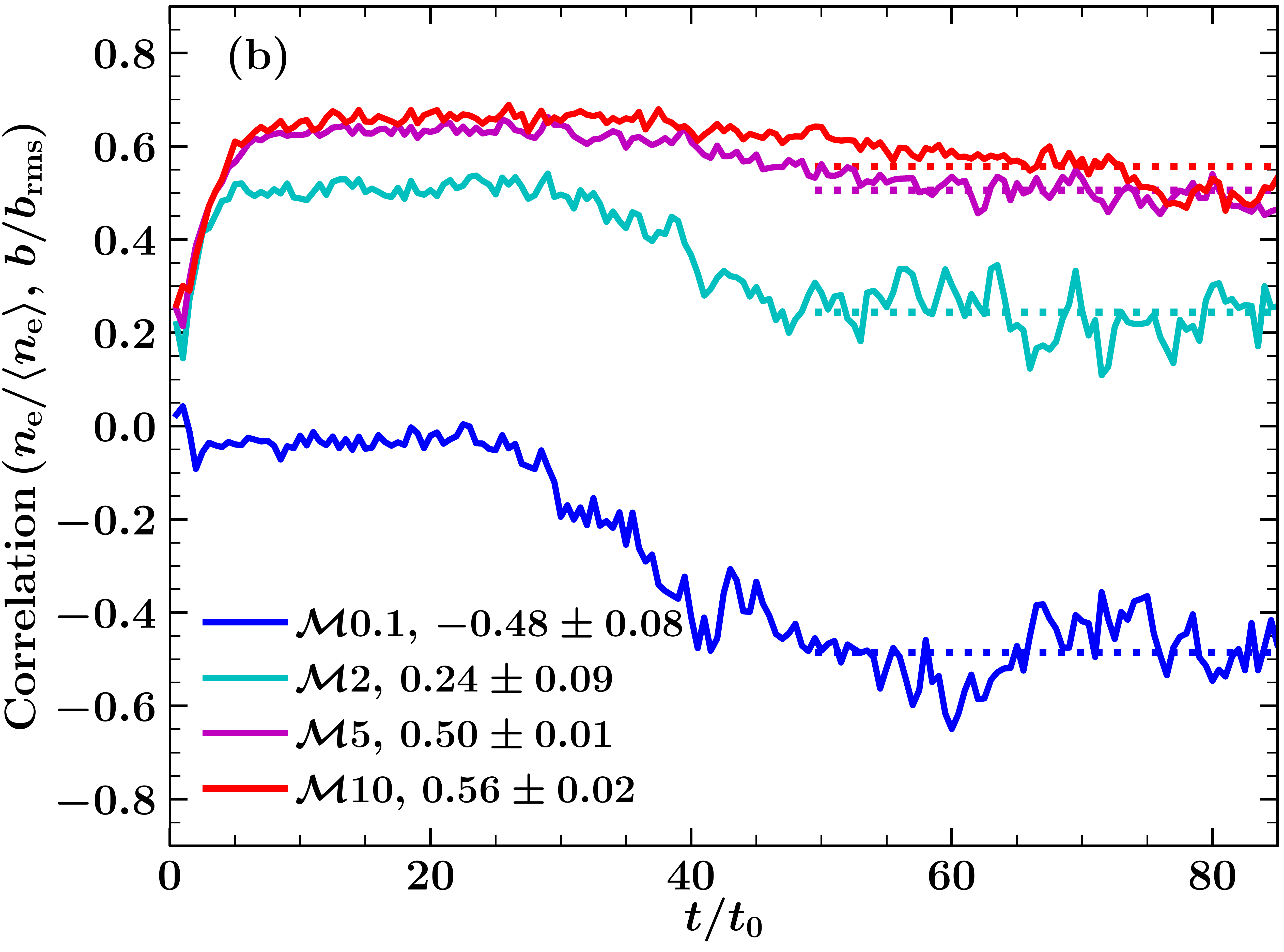}
\caption{(a) Time evolution (in units of the eddy turnover time, $t_0$) of the ratio of the magnetic to kinetic energy, $E_{\rm mag}/E_{\rm kin}$, for $\Mach=0.1$ (blue), $\Mach=2$ (cyan), $\Mach=5$ (magenta), and $\Mach=10$ (red). After the initial transient phase, for each case, the magnetic field grows exponentially (dashed, dotted line of the same color for each case) with a growth rate $\Gamma \, [t_0^{-1}]$. The growth rate decreases as the Mach number increases. Eventually, the magnetic field saturates and the saturation level ($R_{\rm sat}$, dotted line for each color) also decreases with the Mach number. (b) Time evolution of the correlation coefficient between $\ne$ and $b$, calculated using \Eq{eq:corr} for each Mach number. The correlation coefficient is negative for $\Mach0.1$ run and positive for $\Mach2, \Mach5,$ and $\Mach10$ runs. As the field becomes dynamically important (slow down of exponential increases in panel~a), the correlation also decreases as the magnetic pressure, which is now significant, pushes the gas locally. The correlation coefficient also reaches a statistically steady value (dotted line for each simulation model) as the magnetic field saturates. The correlation coefficient increases as the Mach number increases, showing the enhancement in the correlation between $\ne$ and $b$ with increasing Mach number.
}
\label{fig:ts}
\end{figure*}

\begin{table*}
	\caption{Main simulation parameters and outcomes. All simulations are performed in a numerical domain of size $L$ with $512^3$ points. In all cases, the flow is driven on large scales (the effective forcing scale is $L/2$) with $\Re=\Rm=2000$ and only the Mach number $\Mach$ of the turbulent flow is varied, with the simulation name reflecting the Mach number. The columns in the table are follows: 1.~simulations name, 2.~Mach number, $\Mach$, 3.
~growth rate of the magnetic field, $\Gamma\,[t_0^{-1}]$ (\Fig{fig:ts}~(a)), 4.~saturated level of the ratio of the magnetic to kinetic energy,  $R_{\rm sat} = E_{\rm mag}/E_{\rm kin}$ (\Fig{fig:ts}~(a)), 5.~plasma beta in the saturated state, $\beta_{\rm sat} = \langle \cs^2 \rho /(\brms^2/8 \pi) \rangle$, 6. correlation length of the random magnetic field in the saturated state $\ell_b/L$ calculated using \Eq{eq:lb}, and 7.~saturated correlation coefficient, ${\rm Correlation} \, (\ne/\langle \ne \rangle, \, b/\brms)_{\rm sat}$ (\Fig{fig:ts}~(b)). The reported error for each quantity is the maximum of the fitting and systematic errors.}
	\label{tab:corr}
	\begin{tabular}{lcccccc} 
		\hline
		\hline
		Simulation Name & $\Mach$ & $\Gamma \, [t_0^{-1}]$ & $R_{\rm sat}$ & $\beta_{\rm sat}$ & $\ell_b/L$ & ${\rm Correlation} \, (\ne/\langle \ne \rangle, \, b/\brms)_{\rm sat}$ \\
		\hline
		$\Mach 0.1$ & $0.1$ & $0.90 \pm 0.03$ & $0.35\phantom{0} \pm 0.02\phantom{0}$ & 
		$(5.84 \pm 0.81) \times 10^{2\phantom{+}}$ 
		& $0.186 \pm 0.010$ & $-0.48 \pm 0.08$ \\
		$\Mach 2$ & $2$ & $0.71 \pm 0.01$ & $0.20\phantom{0} \pm 0.02\phantom{0}$ & 
		$(2.08 \pm 0.21)\times 10^{0\phantom{+}}$ 
		& $0.129 \pm 0.006$ & $\phantom{0}0.24 \pm 0.09$ \\
		$\Mach 5$ & $5$ & $0.59 \pm 0.04$ & $0.065 \pm 0.003$ & 
		$(1.10 \pm 0.16) \times 10^{0\phantom{+}}$ 
		& $0.120 \pm 0.009$ & $\phantom{0}0.50 \pm 0.01$ \\
		$\Mach 10$ & $10$ & $0.64 \pm 0.07$ & $0.043 \pm 0.005$ & 
		$(4.41 \pm 1.45) \times 10^{-1}$ 
		& $0.122 \pm 0.009$ & $\phantom{0}0.56 \pm 0.02$ \\
		\hline
		\hline
   \end{tabular}
\end{table*}
\Fig{fig:ts}~(a) shows the time evolution of the ratio of the magnetic to kinetic energy $E_{\rm mag}/E_{\rm kin}$ as a function of time for different Mach numbers. After an inital adjustment phase, the magnetic field amplifies exponentially (dash-dotted lines) and finally saturates (dotted lines) due to the back reaction by the Lorentz force on the flow \citep[see][for details of the saturation mechanism]{SetaEA2020}. The growth rate $\Gamma\,[t_0^{-1}]$ and saturation level $R_{\rm sat} = E_{\rm mag}/E_{\rm kin}$ are given in the figure legend and are listed in \Tab{tab:corr}. As the Mach number increases, the growth rate and the saturation level both decrease, because the small-scale dynamo efficiency decreases with increasing compressibility \citep{HaugenBM2004,FederrathEA2011,FederrathEA2014,Federrath2016, AfonsoMV2019}. \rev{We also calculate the correlation length of the small-scale random magnetic field $\ell_b/L$ from the magnetic power spectra $M_k$ as}
\begin{align} \label{eq:lb}
\ell_b/L = \frac{\int_0^{\infty}  k^{-1} M_k \, \dd k}{\int_0^{\infty} M_k \, \dd k}.
\end{align}
\rev{The correlation length for all our runs is listed in \Tab{tab:corr}. We have also listed the saturated value of the plasma beta, $\beta_{\rm sat} = \langle \cs^2 \rho /(\brms^2/8 \pi) \rangle$ (where $\langle \rangle$ denote the average over the entire domain), in \Tab{tab:corr}.}

The spatial correlation between the thermal electron density $\ne$ and the magnetic field $b$ can be quantified using their correlation coefficient,
\begin{align} \label{eq:corr}
{\rm Correlation} \, (\ne/\langle \ne \rangle, \, b/\brms) =&  \\ & \hspace{-2cm}\nonumber \frac{\nonumber \left\langle{\ne/\langle \ne \rangle \times b/\brms} \right\rangle - \langle {\ne/\langle \ne \rangle} \rangle \times \langle{b/\brms}\rangle}{\sigma_{\ne/\langle \ne \rangle} \times \sigma_{b/\brms}}, 
\end{align}
where the angle brackets $\langle \rangle$ and $\sigma$ denote the mean and standard deviation of the quantity over the entire domain, respectively. The correlation coefficient can vary between $1$ (perfect correlation) to $-1$ (perfect anti-correlation), with $0$ indicating no correlation. 
\Fig{fig:ts}~(b) shows the time evolution of this correlation coefficient. During the exponential growth phase (c.f., panel a), the correlation is negative for the low Mach number run $\Mach0.1$ and positive for high Mach number runs ($\Mach2, \Mach5,$ and $\Mach10$). This is because, for higher Mach number runs, the magnetic field is locally amplified due to compression where the density is high\footnote{\citet{GreteOB2020} find a negative correlation because they study $\Mach \lesssim 0.6$.}. In the saturation regime, the correlation decreases for all Mach numbers. This is because as the magnetic field becomes dynamically important, the magnetic pressure resists the compression, and therefore reduces the the correlation between $\ne$ and $b$.

\begin{figure*}
\includegraphics[width=\columnwidth]{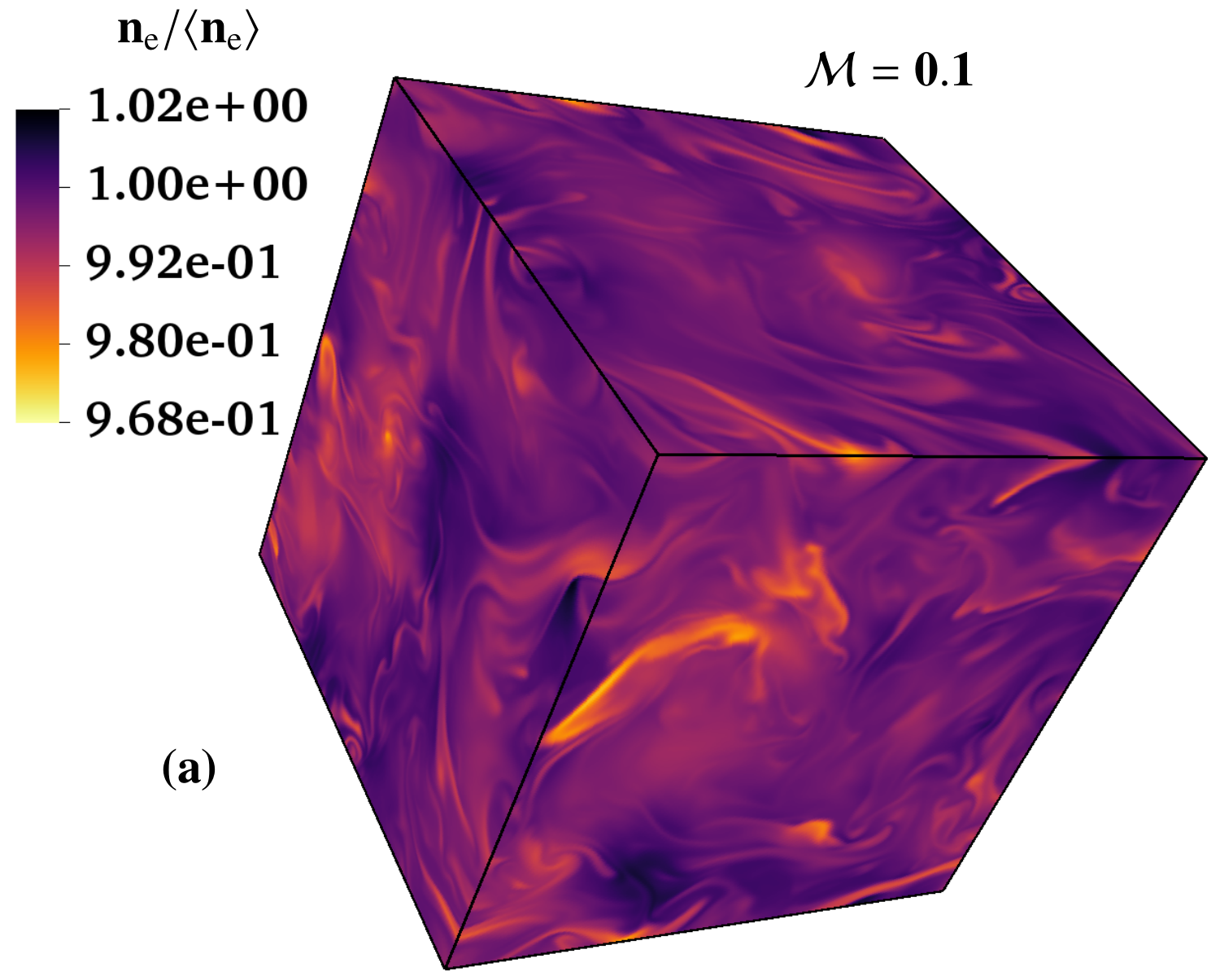}
\includegraphics[width=\columnwidth]{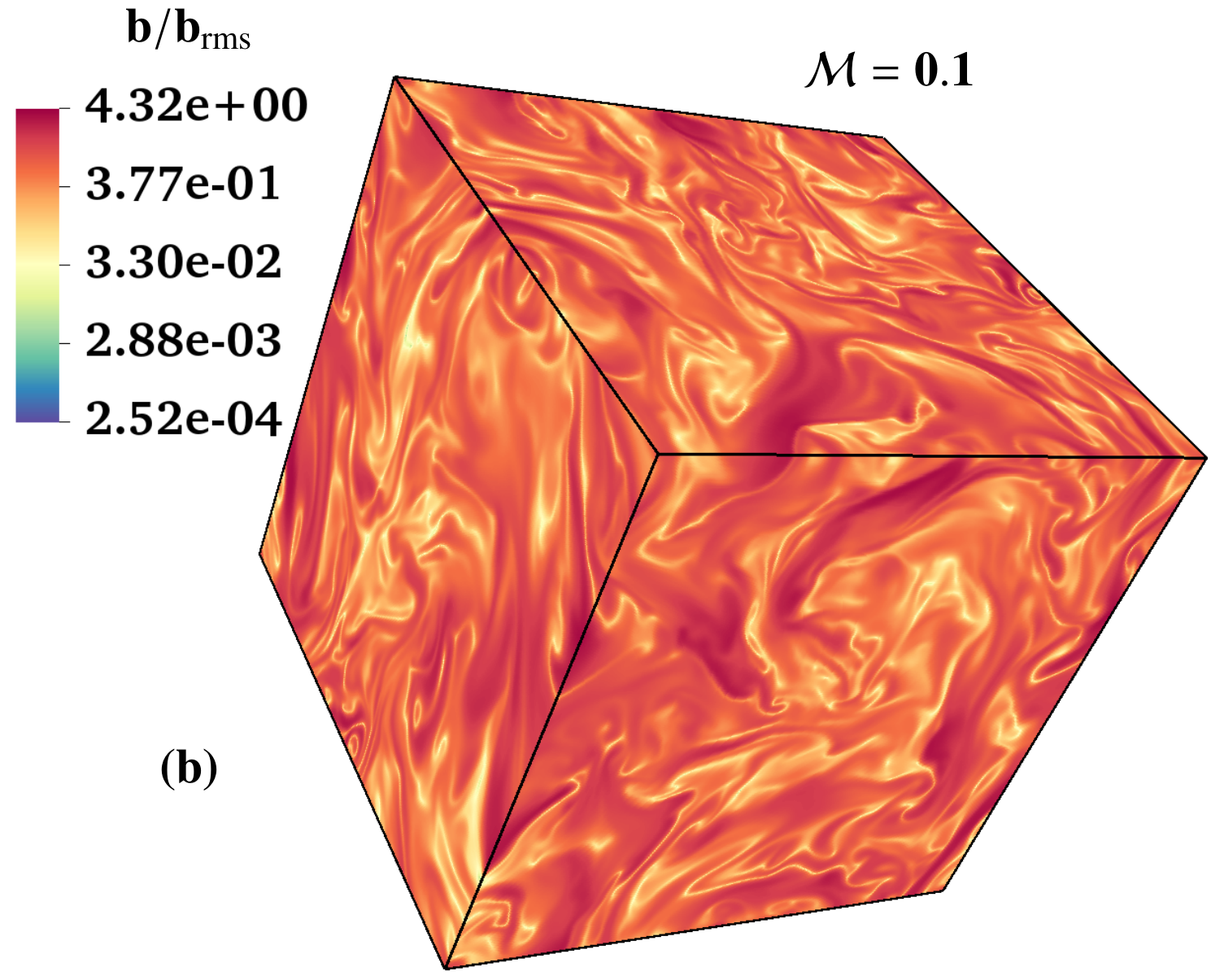}
\includegraphics[width=\columnwidth]{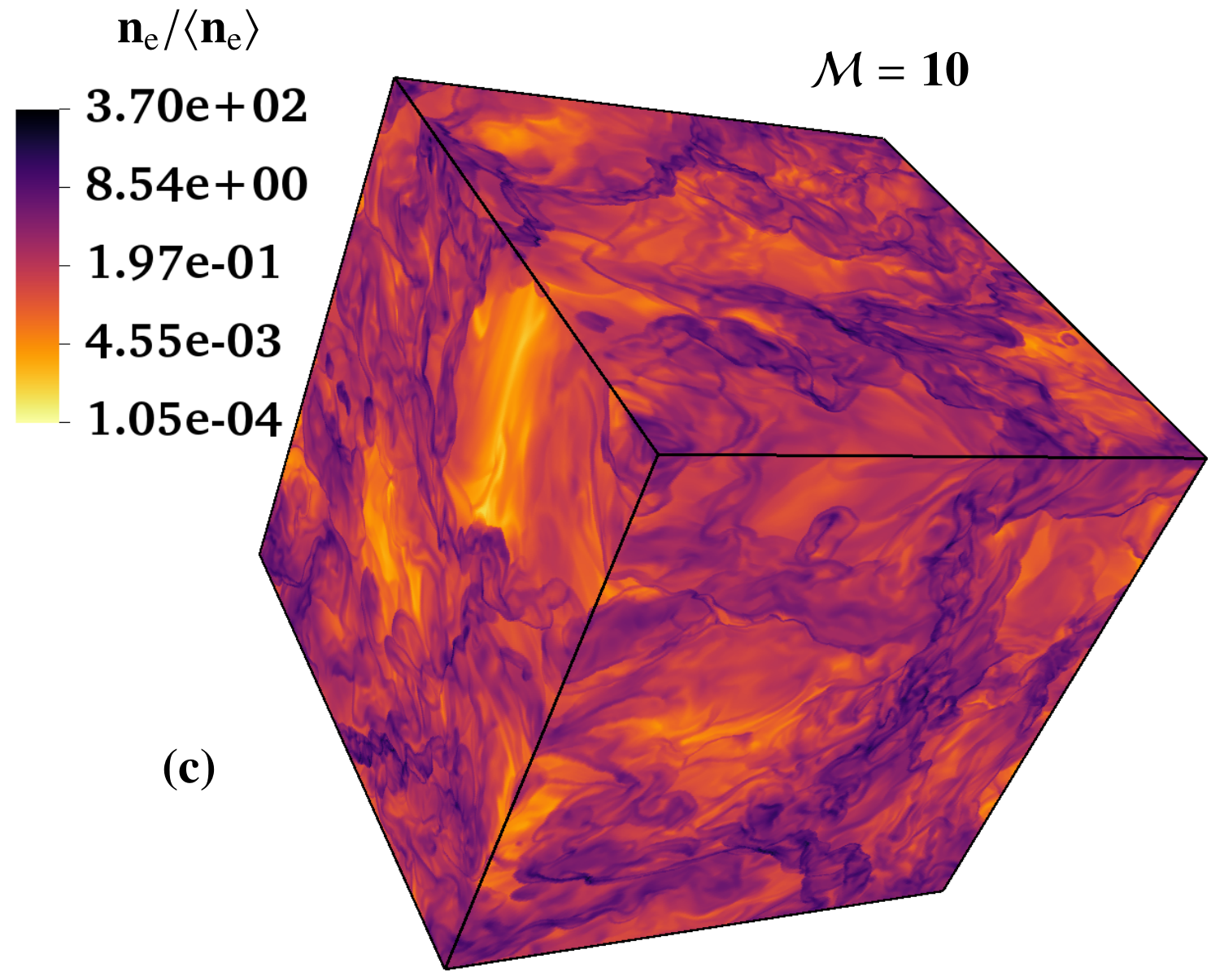}
\includegraphics[width=\columnwidth]{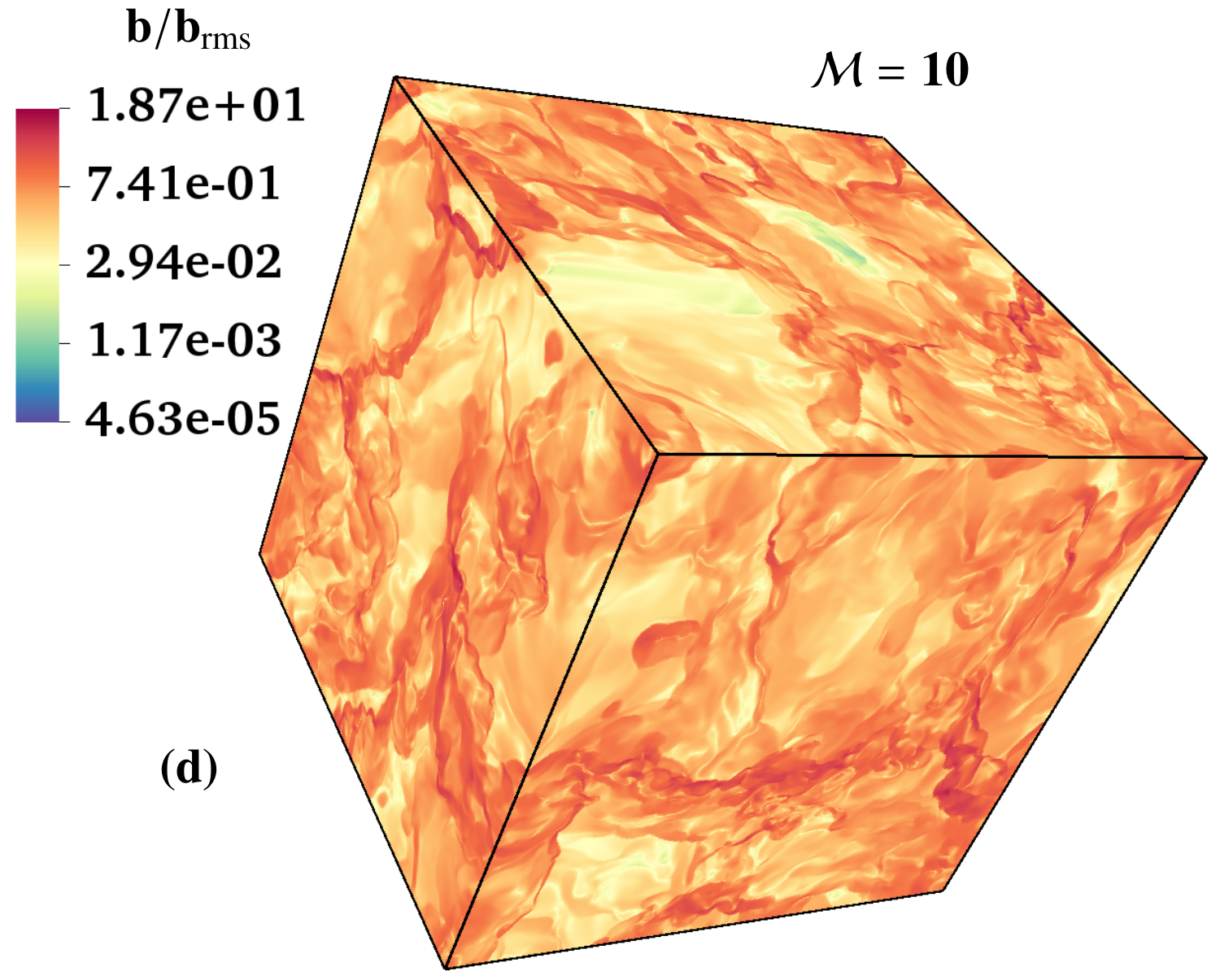}
\caption{Three-dimensional structure of $\ne/\langle \ne \rangle$ and $b/\brms$ in the saturated state (c.f., \Fig{fig:ts}a) for $\Mach=0.1$ (panels a and b) and $\Mach=10$ (panels c and d), respectively. For $\Mach=0.1$, $\ne$ does not vary much due to low compressibility at low $\Mach$, but $b$ varies over four orders of magnitude. We see that $\ne$ and $b$ are spatially anti-correlated, i.e., the regions with relatively high densities have low magnetic fields. For $\Mach=10$, both $\ne$ and $b$ vary over about six orders of magnitude and their spatial correlation is high, visually confirming the quantitative analysis of the correlation coefficient shown in \Fig{fig:ts}b.
}
\label{fig:struc3d}
\end{figure*}

In \Fig{fig:struc3d}, we show the three-dimensional structure of $\ne/\langle\ne\rangle$ and $b/\brms$, in the statistically saturated phase, for the $\Mach=0.1$ and $\Mach=10$ simulations, respectively. For $\Mach = 0.1$ (panels a and b), $\ne$ does not vary significantly and is anti-correlated with $b$, which is concentrated in filaments and sheets. Thus, the magnetic field is spatially intermittent and follows a non-Gaussian distribution \cite[see Sec.~3 in][]{SetaEA2020}. On the other hand, for the $\Mach=10$ simulation, both $\ne$ and $b$ vary over six orders of magnitude and are strongly correlated. The presence of strong shocks shows the concentration of $\ne$ and $b$ in thinner filamentary and sheet-like structures, at roughly the same spatial location.

Now that we have quantified the correlation between $\ne$ and $b$ as a function of the turbulent Mach number, we can study the effect of such a correlation on the rotation measure ($\RM$), the dispersion measure ($\DM$), and the estimated magnetic field from them, i.e., $b_\parallel=\obsb$, based on \Eq{eq:obsb}.

\subsection{Effect of the $\ne$--$b$ correlation on the relation $\bpar=\obsb$} \label{sec:simcorr:effect}
We use $\ne$ and $b$ from the saturated phase of the simulations to compute $\RM$ and $\DM$ using \Eq{eq:dm} and \Eq{eq:rm}, respectively. In order to apply these equations and to provide RM and DM in typical observational units, we scale all of our dimensionless simulations to physical units. We picked $\brms=5\muG$, $\langle \ne \rangle=0.02\cm^{-3}$ \citep{BerkhuijsenM2008,GaenslerEA2008,YaoEA2017}, and $L=600 \pc$. However, one could scale them to any other physical values. The important point is that we apply the same scaling to all of the simulations, retaining comparability between the different simulations (different $\Mach$). From $\RM$ and $\DM$ we then calculate the estimated magnetic field $\bpar=\obsb$ (\Eq{eq:obsb}) and compare it with the true average magnetic field $\bpar$ computed directly from each simulation. If the correlation between $\ne$ and $b$ does not affect the magnetic field estimate, we expect $\bpar = \obsb$. If $\bpar\neq\obsb$, we can quantify the effect of the correlation between $\ne$ and $b$. We repeat this comparison of $\bpar$ and $\obsb$ for different times in the simulations ($70 \le t/t_0 \le 80$) and compute the time average and standard deviation of all quantities ($\RM$, $\DM$, $\obsb$, and $\bpar$) over this time interval.

\begin{figure*}
\includegraphics[width=\columnwidth]{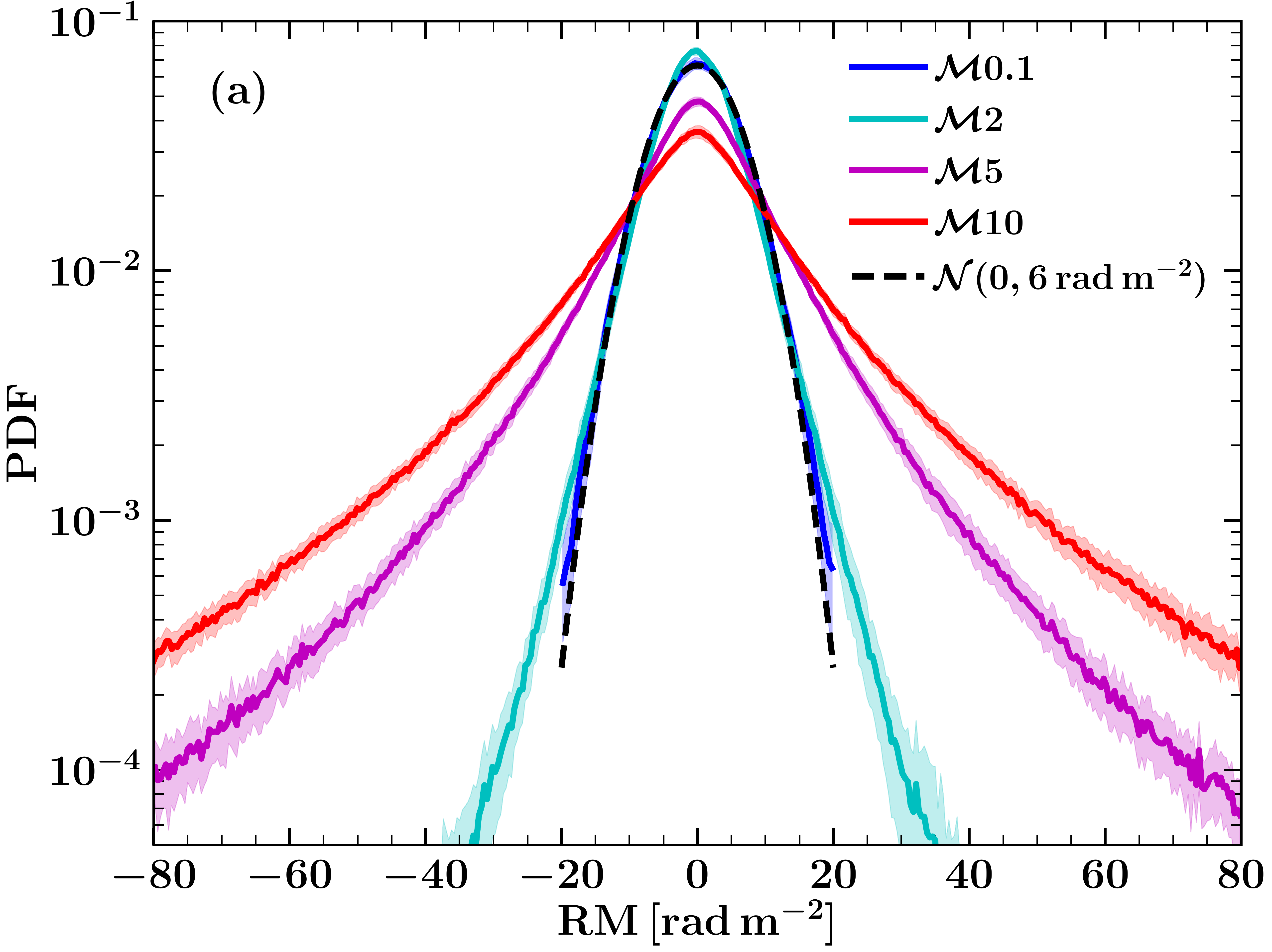} \hspace{0.5cm}
\includegraphics[width=\columnwidth]{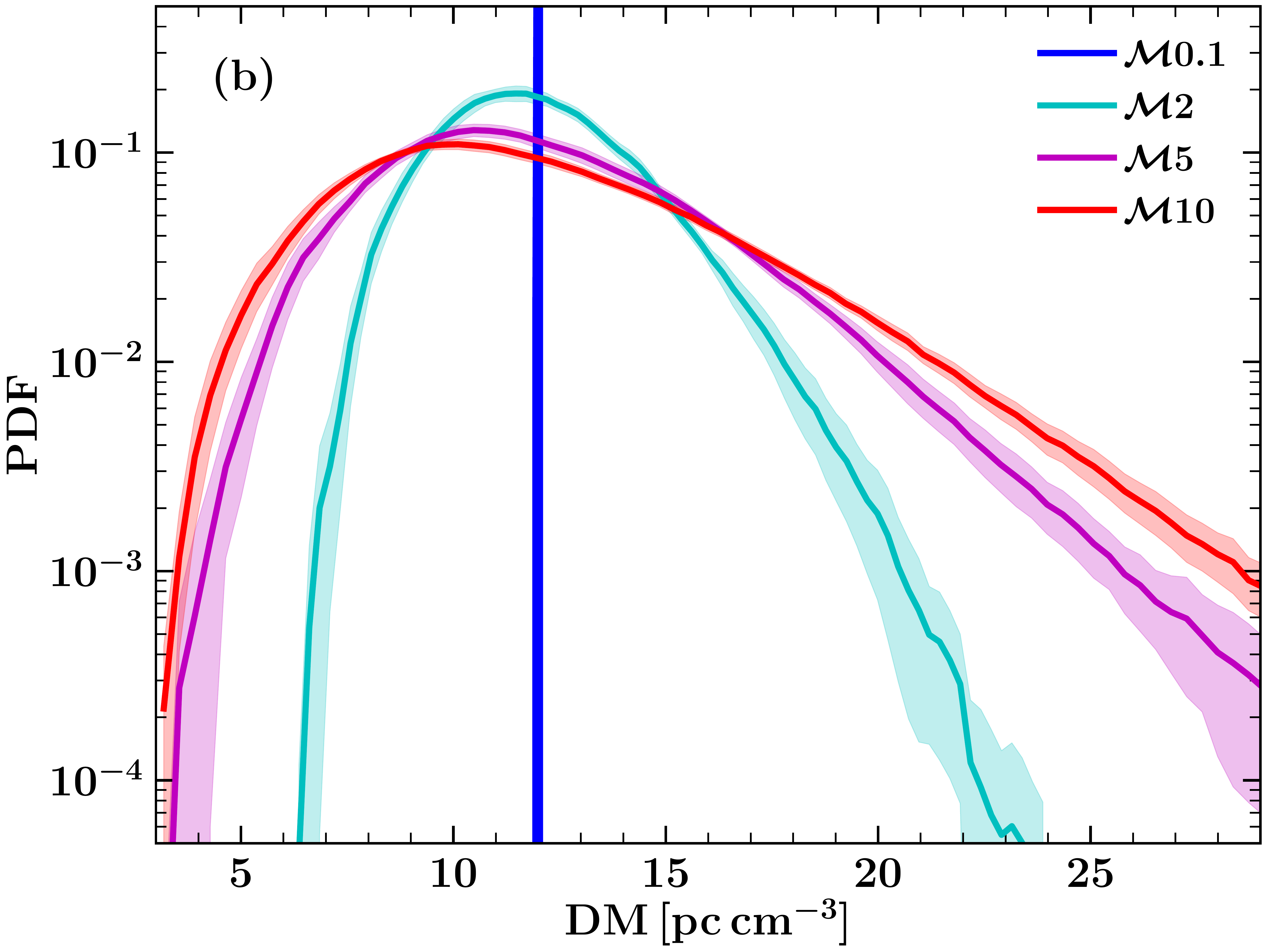}
\caption{(a) The probability density function (PDF) of the rotation measure $\RM \, [\rad \m^{-2}]$ for different Mach numbers runs: $\Mach0.1$ (blue), $\Mach2$ (cyan), $\Mach5$ (magenta), and $\Mach10$ (red). In each case, the solid line shows the mean and the shaded regions (of the same color) shows the one-sigma variations when averaged over ten eddy turnover times in the range $70 \le t/t_0 \le 80$ (\Fig{fig:ts}~(a)). The $\RM$ PDFs for all Mach numbers have a mean very close to zero since there is no large-scale field. For $\Mach = 0.1$ (subsonic case), the PDF is very close to a Gaussian distribution (dashed black line) and the $\RM$ can be completely characterised using the mean ($\sim 0$) and the standard deviation. However, as the Mach number increases the distribution becomes non-Gaussian and develops long heavy tails at higher values of $\RM$. This also increases the standard deviation of the distribution and the distribution can no longer be completely described by the mean and standard deviation for transsonic and supersonic cases ($\Mach=2, 5,$ and $10$). The computed standard deviation also increases with $\Mach$ (see the column~2 in \Tab{tab:stats}). (b) Same as (a) but for the dispersion measure $\DM$. For $\Mach=0.1$, $\DM$ is practically constant at $12 \pc \cm^{-3}$ since $\ne$ hardly varies (see \Fig{fig:struc3d}~(a)). As the Mach number increases the distribution widens. The mean of the $\DM$ distribution remains approximately equal to $12 \pc \cm^{-3}$ (see the column~4 \Tab{tab:stats}) for all Mach numbers because $\langle \ne \rangle = 0.02 \cm^{-3}$ and $L = 600 \pc$ are constant for all runs.}
\label{fig:rmdm}
\end{figure*}

\begin{figure*}
\includegraphics[width=2\columnwidth]{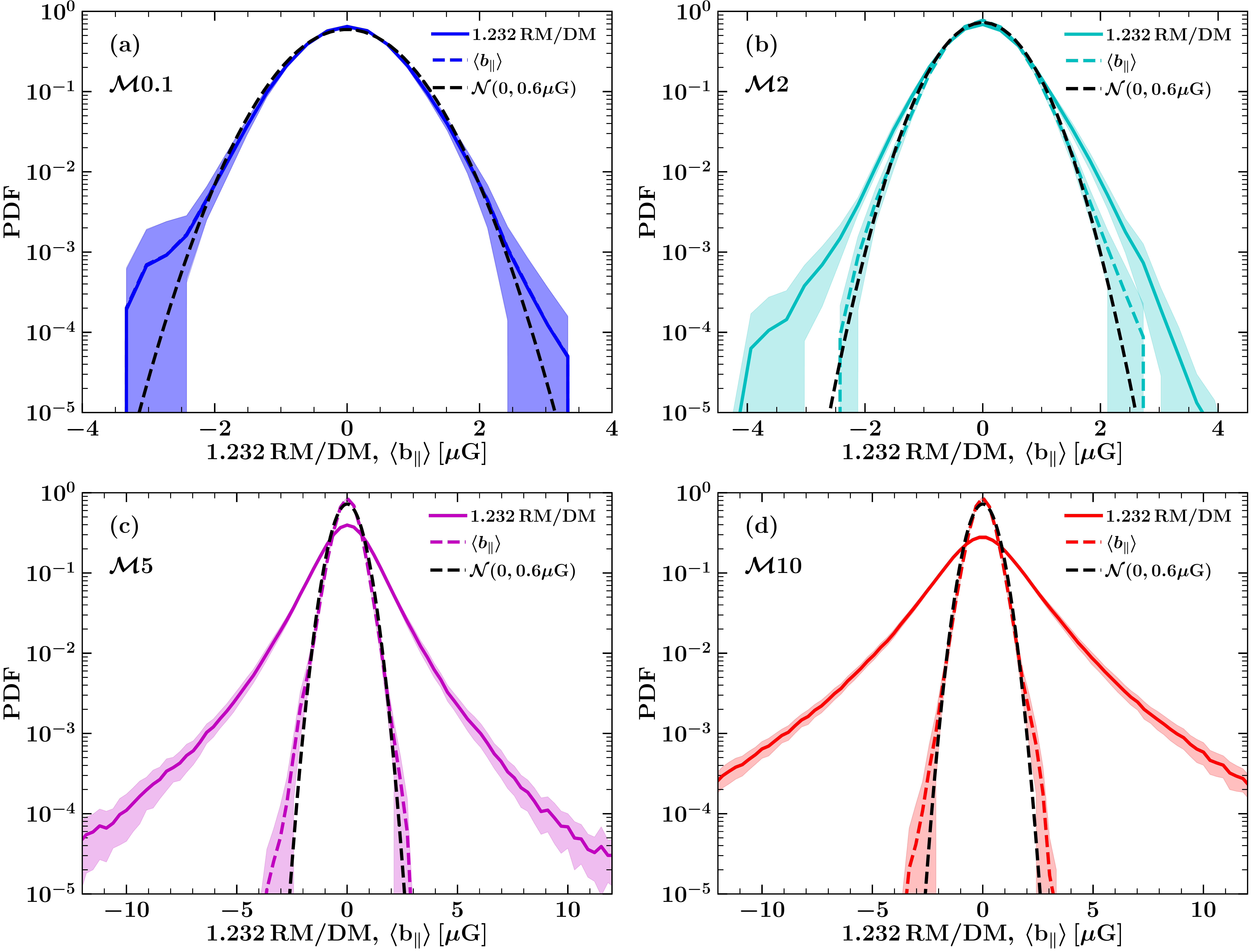}
\caption{The probability density function of the magnetic field estimated using \Eq{eq:obsb}, $\obsb$ (solid line) and the true average magnetic field $\bpar$ computed directly from the simulation (dashed line) for runs $\Mach0.1$ ((a), blue), $\Mach2$ ((b), cyan), $\Mach5$ ((c), magenta), and $\Mach10$ ((d), red). Note the difference in the $x$-axis for low and high Mach numbers. For each case,  the lines show the mean and the shaded regions show the one-sigma variations when averaged over ten eddy turnover times in the range $70 \le t/t_0 \le 80$ (\Fig{fig:ts}~(a)). All curves have a mean value very close to zero confirming the absence of any large-scale field. The distribution changes significantly for $\obsb$ (Gaussian at $\Mach=0.1$ and becomes non-Gaussian with the tail at larger magnetic field values becoming heavier with Mach number; see columns~7 and~9 in \Tab{tab:stats}), but the $\bpar$ distribution remains roughly Gaussian for all Mach numbers (because even if the random magnetic fields are non-Gaussian, the average over the large path length give rise to a Gaussian distribution due to the central limit theorem). For $\Mach=0.1$ (a), the subsonic case, both distributions are close to a Gaussian distribution (dashed black line) and the standard deviation is roughly equal to $1.232 \, (\sigma($\RM$)/\mu($\DM$)) \approx 1.232 (6 \rad \m^{-2} / 12 \pc \cm^{-3}) \simeq 0.6 \muG$. Thus, $\Mach\lesssim1$, where the density does not vary much, \Eq{eq:obsb} gives the correct estimate for the magnetic field. For $\Mach=2$ ((b), the transsonic case), the distribution of $\obsb$ differs from a Gaussian distribution, but not significantly (see kurtosis in column~6 of \Tab{tab:stats}). As the Mach number increases to $5$ and $10$ ((c) and (d), supersonic cases), the correlation between $\ne$ and $b$ becomes significant and the estimated magnetic field differs significantly from the true magnetic field.
}
\label{fig:obsbbpar}
\end{figure*}

\begin{table*}
	\caption{Properties of the $\RM$, $\DM$, $\obsb$, and $\bpar$ distributions for various Mach numbers computed from simulations assuming $\brms=5\muG$ and $\langle \ne \rangle = 0.02 \cm^{-3}$ over a numerical domain of size $L=600\pc$. The columns are as follows: 1. simulation name, 2. the standard deviation of the rotation measure in $\rad \m^{-2}$, $\sigma \left( \RM\right)$, 3. the kurtosis of the rotation measure, $\ku \left( \RM\right)$, 4. the mean of the dispersion measure in $\pc \cm^{-3}$, $\mu \left( \DM\right)$, 5. the standard deviation of the dispersion measure in $\pc \cm^{-3}$, $\mu \left( \DM\right)$, 6. the standard deviation of estimated magnetic field in $\muG$, $\sigma \left( \obsb \right)$, 7. the kurtosis of the estimated magnetic field, $\ku \left( \obsb \right)$, 8. the standard deviation of the true average parallel component of the magnetic field in $\muG$, $\sigma \left( \bpar \right)$, and 9. the kurtosis of the true average parallel component of magnetic field, $\ku \left( \bpar \right)$. For comparison, the kurtosis of a Gaussian distribution is $3$. All the reported errors are the maximum of the statistical and systematic errors.}
	\label{tab:stats}
	\begin{tabular}{lcccccccccc} 
		\hline 
		\hline
		 Simulation Name  & $\sigma \left( \RM\right)$ & $\ku \left( \RM\right)$ & $\mu \left( \DM\right)$ & $\sigma \left( \DM\right)$ & $\sigma \left( \obsb \right)$ & $\ku \left( \obsb \right)$ & $\sigma \left( \bpar \right)$ & $\ku \left( \bpar \right)$ \\
		--  & $[\rad \m^{-2}]$ & -- & $[\pc \cm^{-3}]$ &  $[\pc \cm^{-3}]$ & $[\muG]$ & -- & $[\muG]$ & -- \\
		\hline 
		$\Mach 0.1$  & $6.06 \pm 0.35$ &   $3.06 \pm 0.07$ &  $12.00 \pm 0.01$ & $0.015 \pm 0.001$  &  $0.64 \pm 0.06$ &  $3.46 \pm 0.91$ & $0.64 \pm 0.06$ &  $3.47 \pm 0.90$ \\
		
		$\Mach 2$  & $6.34 \pm 0.56$ &   $4.96 \pm 0.82$ &  $12.00 \pm 0.01$ & $2.18 \pm 0.20$  &   $0.62 \pm 0.04$ &  $3.76 \pm 0.58$ & $0.53 \pm 0.03$ &  $3.32 \pm 0.36$ \\
		
		$\Mach 5$  & $14.30 \pm 1.12$ &   $7.50 \pm 0.02$ &  $12.00 \pm 0.04$ & $3.49 \pm 0.27$  &   $1.36 \pm 0.14$ &  $8.87 \pm 2.40$ & $0.50 \pm 0.03$ &  $3.81 \pm 0.61$ \\
				
		$\Mach 10$  & $22.02 \pm 1.56$ &   $7.96 \pm 0.13$ &  $11.96 \pm 0.06$ & $4.15 \pm 0.21$  &  $2.07 \pm 0.18$ &  $9.56 \pm 1.22$ & $0.50 \pm 0.04$ &  $3.90 \pm 0.70$ \\	
			
		\hline
		\hline
		
   \end{tabular}
\end{table*}

\begin{table}
	\caption{\reva{Comparison of the standard deviation of rotation measure fluctuations obtained from numerical results with that estimated using the random walk model (\Eq{eq:rmrw}). The columns are as follows: 1. simulation name, 2. the standard deviation of rotation measure in $\rad \m^{-2}$ computed from the simulation results (\Fig{fig:rmdm}~(a)), $\sigma \left( \RM\right)$, 3. the standard deviation of rotation measure estimated using the random walk model (\Eq{eq:rmrw}) with $\langle \ne \rangle = 0.02 \cm^{-3}$, $\brms = 5\muG$, $L=600 \pc$, and $\lb/L$ from column~6 of \Tab{tab:corr}, $\sigma \left( \RM\right)_{\rm RW}$, and 4. the ratio of numbers in column 2. and 3., $\sigma \left( \RM\right)$/$\sigma \left( \RM\right)_{\rm RW}$. Errors in columns 3 and 4 are obtained by propagating errors in $\lb/L$ and $\sigma \left( \RM\right)$.}}
	\label{tab:rw}
	\begin{tabular}{lccc}
		\hline 
		\hline
		Simulation Name  & $\sigma \left( \RM\right)$ & $\sigma \left( \RM\right)_{\rm RW}$ &  $\sigma \left( \RM\right) / \sigma \left( \RM\right)_{\rm RW}$ \\
		--  & $[\rad \m^{-2}]$ &   $[\rad \m^{-2}]$ & -- \\
		\hline
		$\Mach 0.1$ & $\phantom{0}6.06 \pm 0.35$ & $12.10 \pm 0.32$ & $0.50 \pm 0.04$ \\
		$\Mach 2$  & $\phantom{0}6.34 \pm 0.56$ & $10.08 \pm 0.23$ & $0.63 \pm 0.07$ \\
		$\Mach 5$  & $14.30 \pm 1.12$ & $\phantom{0}9.72 \pm 0.37$ & $1.47 \pm 0.17$ \\
		$\Mach 10$  & $22.02 \pm 1.56$ & $\phantom{0}9.80 \pm 0.36$ & $2.25 \pm 0.24$ \\ 	
		\hline 
		\hline
   \end{tabular}
\end{table}

\Fig{fig:rmdm} shows the probability density function (PDF) of the rotation measure $\RM$ (a) and dispersion measure $\DM$ (b) for various Mach numbers and \Fig{fig:obsbbpar} shows the PDF of the estimated magnetic field $\obsb$ and the true average parallel magnetic field computed directly from simulations $\bpar$ for various Mach numbers ((a)--(d)). \Tab{tab:stats} lists key statistical measures of these distributions.

The PDF of $\RM$ (\Fig{fig:rmdm}~(a)) for $\Mach0.1$ (subsonic) follows a Gaussian distribution and $\Mach2$ (transsonic) is also close to a Gaussian distribution around the peak. However, the distribution for supersonic runs ($\Mach5$ and $\Mach10$) is non-Gaussian with long tails at higher values of $\RM$ and the tails become heavier as the Mach number increases. For the magnetic field amplified by subsonic and transsonic turbulence, it has been shown that weak magnetic field regions contribute to $\RM$ more than high-field regions \citep{BhatS2013, SurBS2018}. \reva{Assuming a simple random walk model for $\RM$ with $L/\lb$ magnetic correlation cells along the path length, the standard deviation of $\RM$ can be estimated as}
\begin{align} \label{eq:rmrw}
	\sigma \left( \RM\right)_{\rm RW} = 0.81 \langle \ne \rangle \frac{\brms}{\sqrt{3}} L (\lb/L)^{1/2}.
\end{align}
\reva{In \Tab{tab:rw}, we compute $\sigma \left( \RM\right)_{\rm RW}$ for all Mach numbers and compare it with $\sigma \left( \RM\right)$ obtained using numerical results (\Fig{fig:rmdm}~(a)). \citet{SurBS2018} show that the ratio $\sigma \left( \RM\right) / \sigma \left( \RM\right)_{\rm RW}$ remains close to $0.5$ for $\Mach \le 2.4$, which agrees with our $\Mach 0.1$ and $\Mach 2$ cases.} However, as the Mach number increases the correlated thermal electron density and magnetic field structures contribute significantly to $\sigma (\RM)$, causing it to increase \reva{and thus the ratio also increases (column 4 in \Tab{tab:rw})}. \revb{$\sigma \left( \RM\right)_{\rm RW}$ scales with $L^{1/2}$, so for a path-length of $5 \kpc$ (see \Sec{sec:obs:pulb}) instead of $600 \pc$, the numbers will be roughly three times higher than the values reported here.} The kurtosis $\ku(\RM)$ (column~3 in \Tab{tab:stats}) increases, confirming the non-Gaussianity of the distribution. \rev{In \App{app:m51}, we discuss the distribution of observed $\RM$s in the external galaxy M51.}

\Fig{fig:rmdm}~(b) shows that the standard deviation of $\DM$ increases as the Mach number increases, in line with previous expectations of the column density PDF of the interstellar medium \citep{PadoanEA1997, FederrathEA2008, KonstandinEA2012, KainulainenF2017, Federrath2018}.

\Fig{fig:obsbbpar} shows the PDF of the estimated magnetic field using $\RM$ and $\DM$, i.e., $\obsb$ \Eq{eq:obsb} and the true average parallel component of the magnetic field for all runs. For $\Mach=0.1$ (a), the distribution of $\obsb$ and $\bpar$ roughly follows the same Gaussian distribution of mean approximately equal to zero and standard deviation of $0.6 \mkG$ ($\approx 1.232 \, \sigma(\RM) / \mu(\DM) = 1.232 \, (6 \rad \m^{-2}/ 12 \pc \cm^{-3})  \simeq 0.6 \mkG$). This conclusion roughly remains the same for $\Mach2$ (b) case too. The standard deviation (column 6 and 8 in \Tab{tab:stats}) and kurtosis (column 7 and 9 in \Tab{tab:stats}) of the $\obsb$ and $\bpar$ distributions for $\Mach0.1$ and $\Mach2$ cases are also approximately equal. The kurtosis of the estimated and computed magnetic fields for $\Mach0.1$ and $\Mach2$ cases is close to the value $3$, the kurtosis of a Gaussian distribution. The equality of $\obsb$ and $\bpar$ distributions shows that for sub-sonic and transsonic cases the correlation between the thermal electron density and magnetic fields do not affect the magnetic field estimate and \Eq{eq:obsb} is applicable.

However, as the Mach number increases, the distribution of the estimated magnetic field $\obsb$ becomes non-Gaussian with heavy tails but the true average magnetic field $\bpar$ distribution remains Gaussian (\Fig{fig:obsbbpar}~(c, d)). Both the standard deviation (column 7 and 9 in \Tab{tab:stats}) and kurtosis (column 7 and 9 in \Tab{tab:stats}) of the $\obsb$ distribution for $\Mach5$ and $\Mach10$ is significantly higher than that of the $\bpar$ distribution. The kurtosis for $\obsb$ is much higher than the value $3$ confirming the non-Gaussian nature of the distribution. The difference in the $\obsb$ and $\bpar$ distribution at higher Mach numbers (supersonic runs) is due to the contribution of the correlation between the thermal electron density and magnetic fields to $\RM$ which is not accounted for in the $\DM$. Thus, the equation \Eq{eq:obsb} is not applicable in such cases, and using it might lead to a significant overestimation of magnetic field strengths. 

\begin{figure*}
\includegraphics[width=\columnwidth]{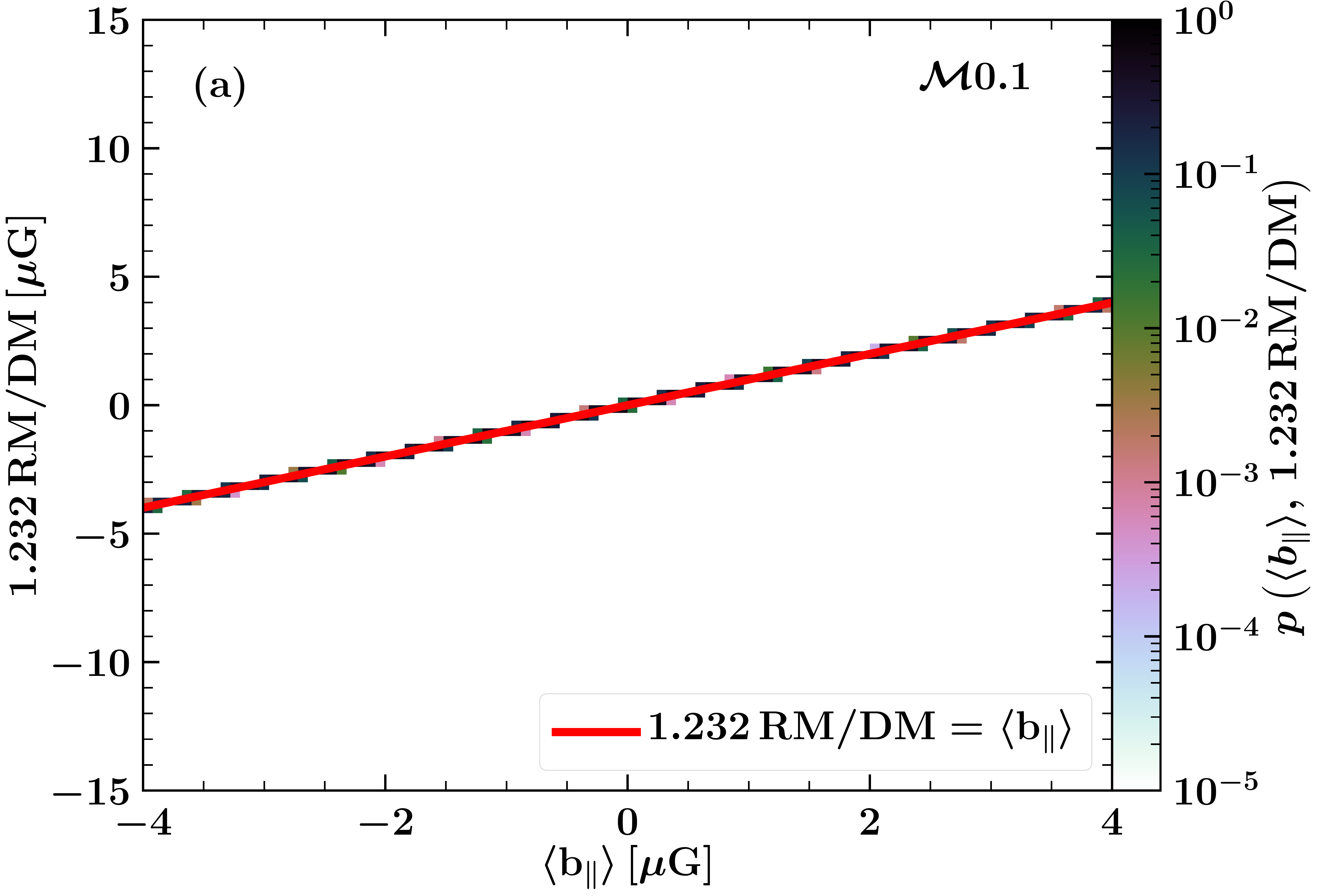} \hspace{0.5cm}
\includegraphics[width=\columnwidth]{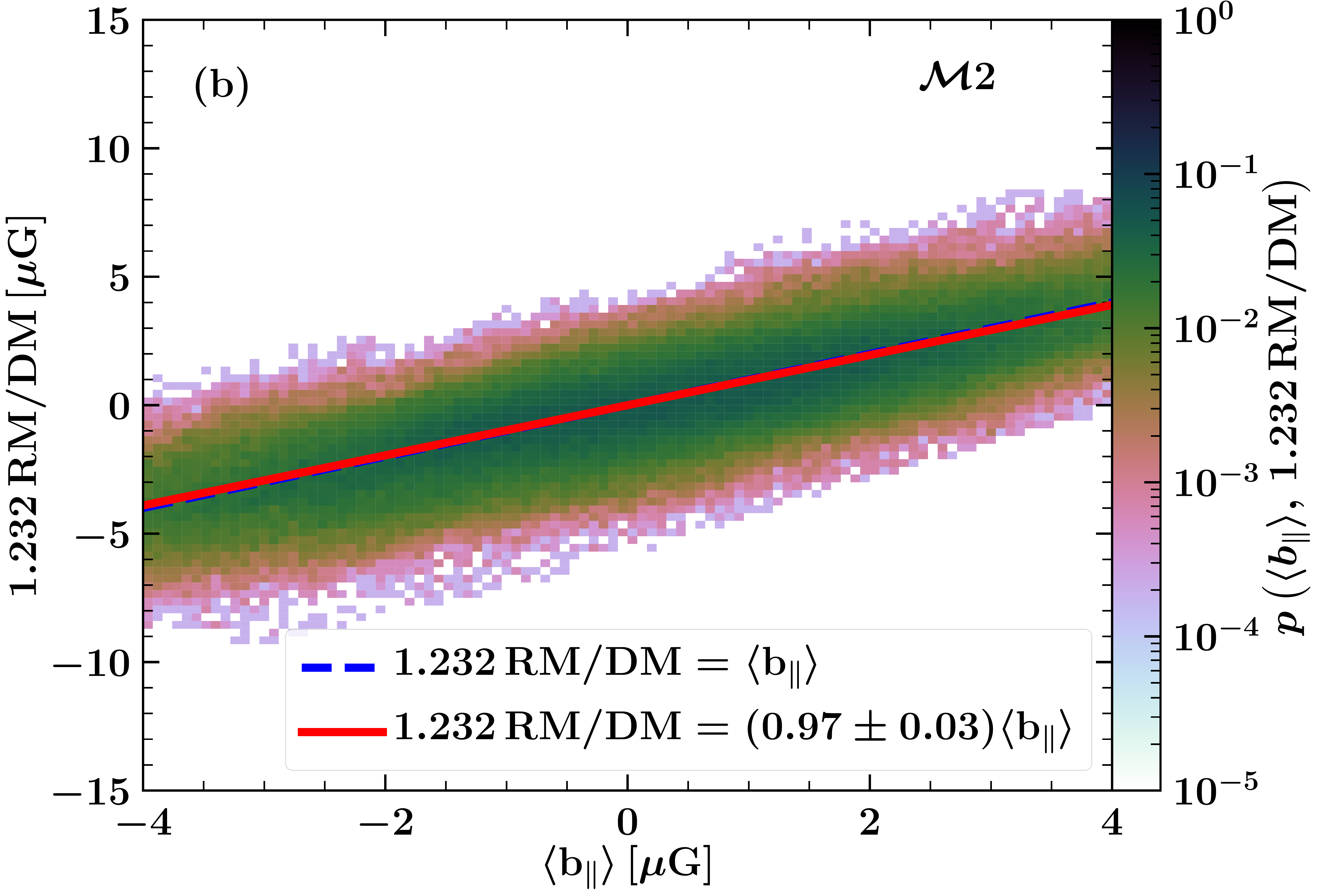}
\includegraphics[width=\columnwidth]{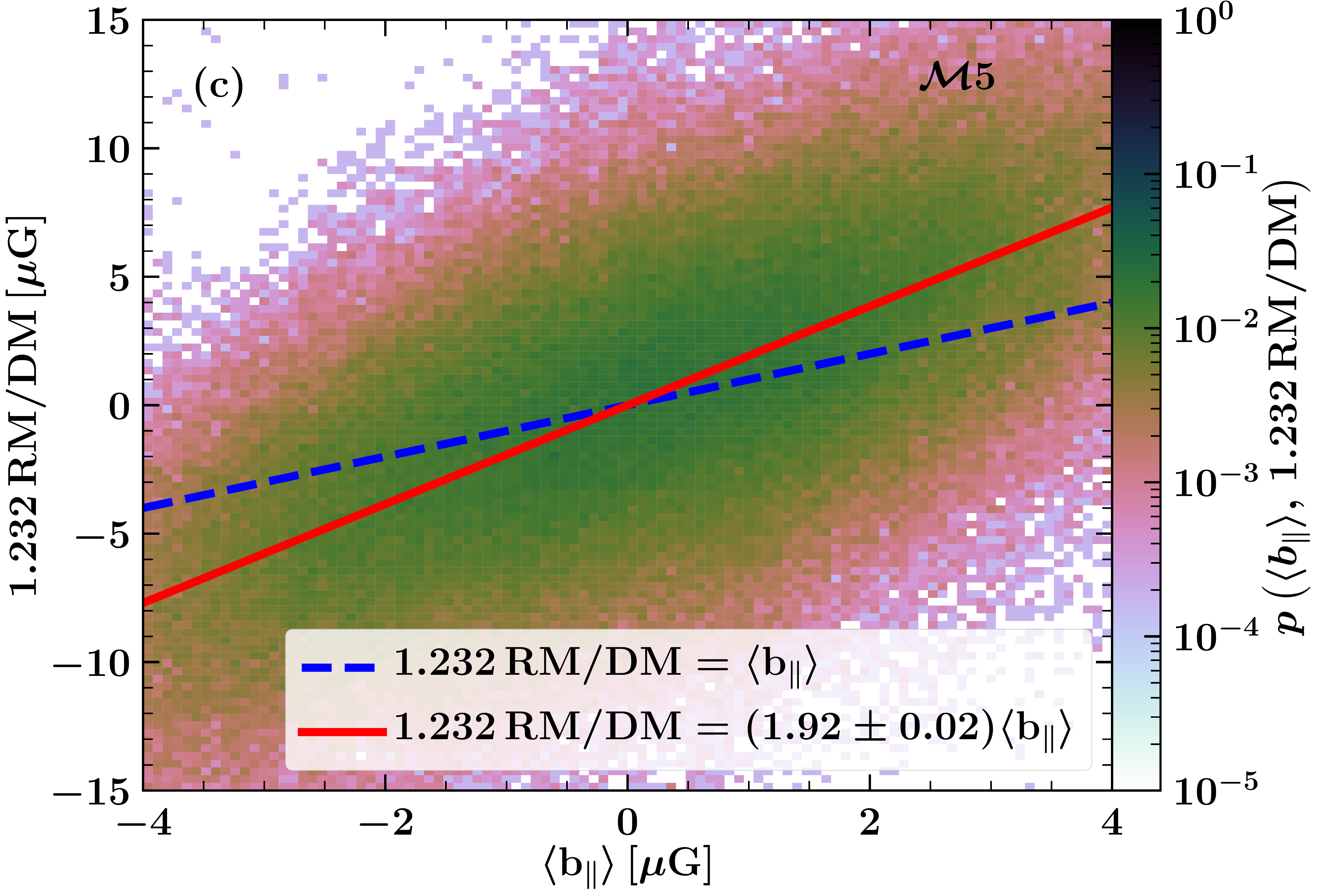} \hspace{0.5cm}
\includegraphics[width=\columnwidth]{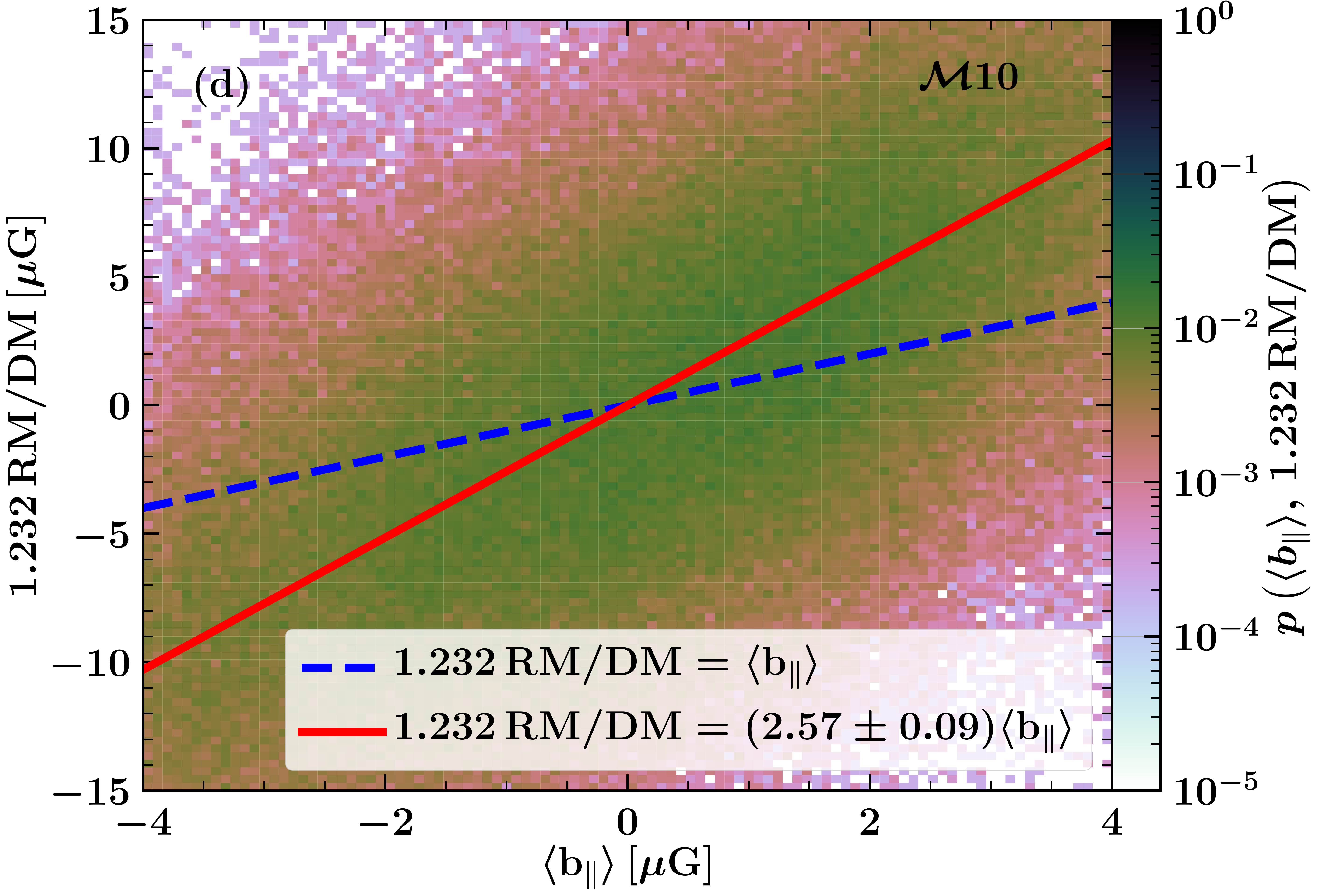}
\caption{The joint probability density function $p(\bpar, \obsb)$ of the true average parallel component of the magnetic field $\bpar$ and the estimated magnetic field $\obsb$ for different Mach number runs, $\Mach0.1$ (a), $\Mach2$ (b), $\Mach5$ (c), and $\Mach10$ (d). In each case, the solid red line shows that equality between two fields and the dashed blue line shows the best fit obtained from the joint probability density using a non-linear least squares optimisation. For $\Mach=0.1$ (subsonic case, (a)), both the red and dashed blue line overlap and thus the estimated magnetic field is equal to the true average magnetic field (\Eq{eq:obsb} provides an excellent estimate of $b$ in this case). This conclusion remains statistically the same even for $\Mach=2$ (transsonic case, (b)), although there is significant dispersion around the mean correlation. For supersonic cases ($\Mach=5$, (c) and $\Mach=10$, (d)), the best-fit line deviates significantly from 1:1 relation. In these cases the magnetic field can be over-estimated by a factor of $2$ and $3$, respectively. This is due to the enhanced correlation between $\ne$ and $b$ in supersonic magnetised turbulence. 
}
\label{fig:obsbbpar2d}
\end{figure*}

After comparing the $\obsb$ and $\bpar$ distributions, we explore the point-to-point spatial correlation between those two quantities in \Fig{fig:obsbbpar2d}, which shows their joint probability density $p(\bpar, \obsb)$ for various Mach numbers. In \Fig{fig:obsbbpar2d}, for each Mach number, the dashed blue line shows the equality between the two quantities (\Eq{eq:obsb}) and the red solid line shows the best fit constructed from the joint probability density. For $\Mach0.1$ (subsonic case), both lines are nearly identical, and thus the \Eq{eq:obsb} gives the correct estimate of the magnetic field. This conclusion remains statistically the same even for $\Mach2$ (transsonic) case; however, the spread around the $\obsb = \bpar$ line increases. As expected by now, this does not hold for $\Mach=5$ and $\Mach=10$ (supersonic) runs (panels c and d). The difference is significant, the slope differs by a factor of $2$ and $3$ for the $\Mach5$ and $\Mach10$ case, respectively. Thus, the correlation between $\ne$ and $b$ leads to an overestimation of the magnetic field by factors of a few.

The overall conclusion from the results of the simulations is as follows: for subsonic and transsonic cases, \Eq{eq:obsb} provides a very good estimate of the magnetic field. However, for supersonic runs, \Eq{eq:obsb} significantly overestimates the magnetic field, with the difference being proportional to the turbulent Mach number. This is because the correlation between $\ne$ and $b$, which increases with $\Mach$, is not accounted for in \Eq{eq:obsb}. In the next section, we explore the thermal electron density -- magnetic field correlation and its effect using observations.

\section{Observational study} \label{sec:obs}
After confirming the presence and the effect of the correlation between thermal electron density and magnetic fields on the magnetic field estimate obtained using \Eq{eq:obsb} in numerical simulations, we now explore the same question using observations. We primarily use the ATNF pulsar catalog for $\RM$ and $\DM$ measurements and correlate $\RM$, $\DM$, and $\obsb$ obtained from these observations (\Sec{sec:obs:pulb}) with various observational probes such as $\CO$ data of Milky Way molecular clouds (\Sec{sec:obs:pulbCO}), magnetic fields in the cold neutral medium obtained using Zeeman splitting of the 21~cm line (\Sec{sec:obs:pulbZeeman}), HI column density $\NHI$ (\Sec{sec:obs:nhicorr}), and $\IHa$ emission (\Sec{sec:obs:ihacorr}) in the Milky Way. \rev{Throughout the analysis, for the observed sample of pulsars, we compute and discuss only the statistical properties of the $\RM$, $\DM$, and $\obsb$ distributions, and their statistical correlations with various tracers of the thermal electron density and magnetic fields.}

\subsection{$\RM, \DM,$ and $\obsb$ from pulsar observations} \label{sec:obs:pulb}
\begin{figure*}
\includegraphics[width=\columnwidth]{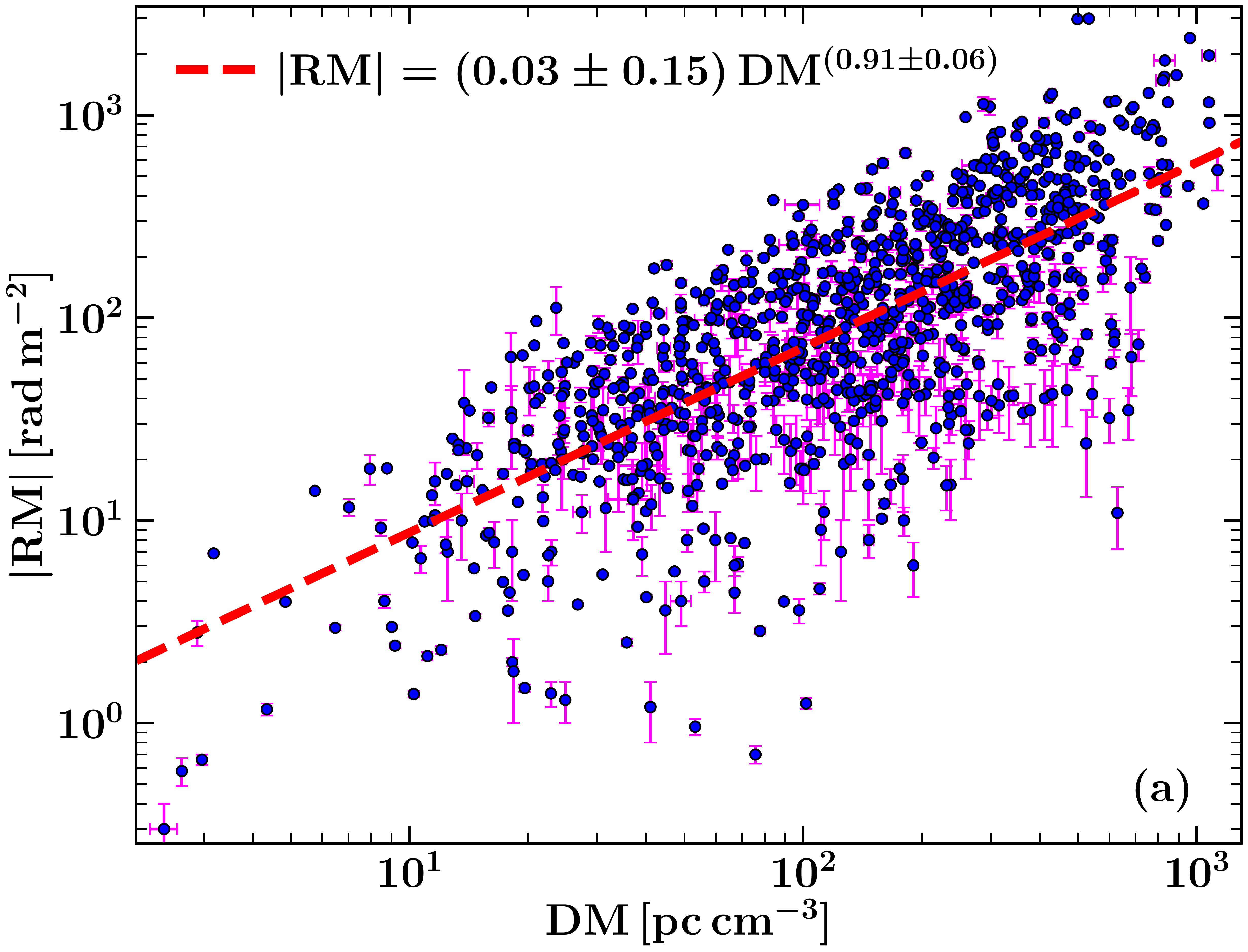} \hspace{0.5cm}
\includegraphics[width=\columnwidth]{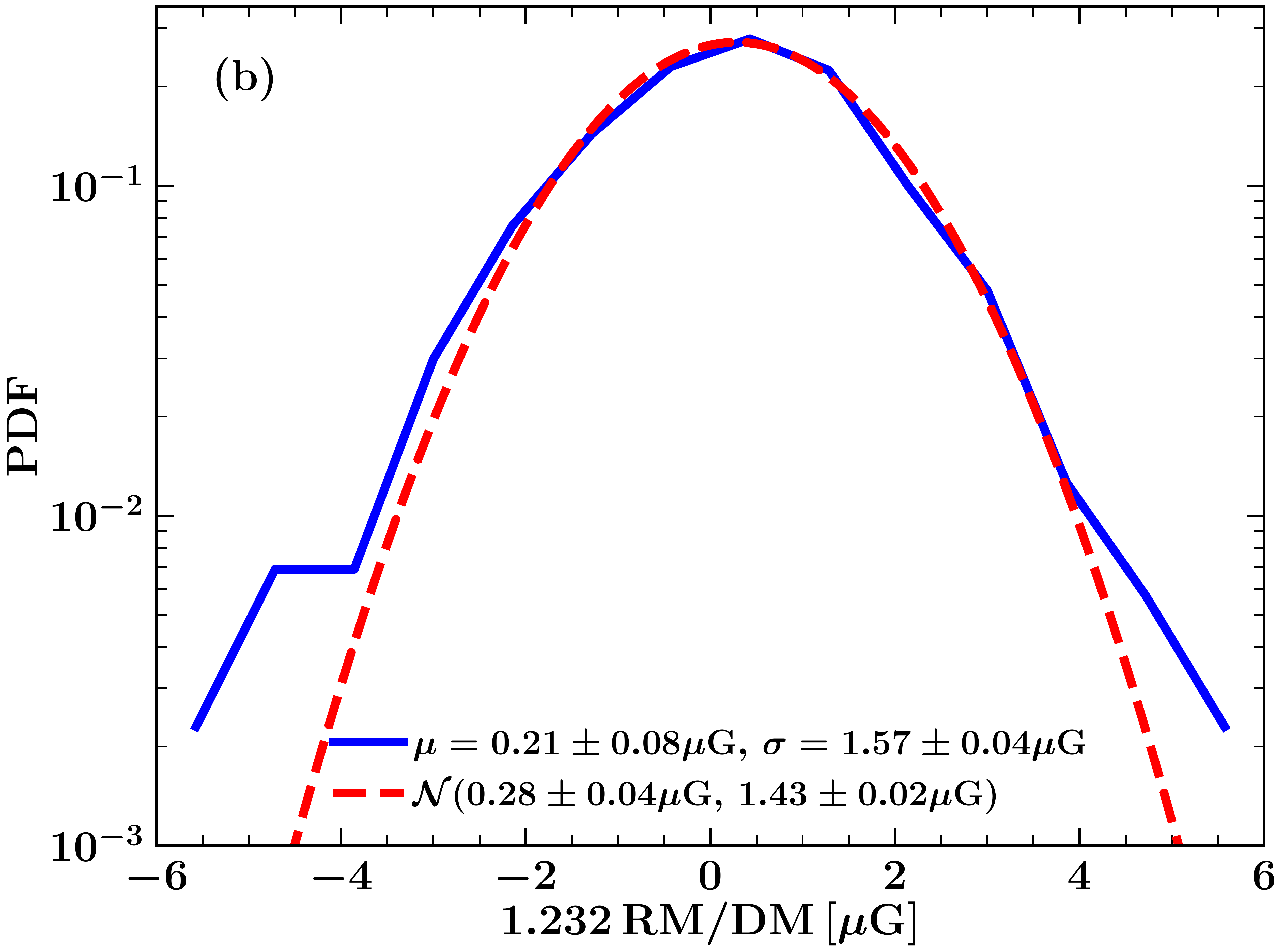}
\caption{(a) Scatter plot of $|\RM|$ and $\DM$ (blue points) with their respective errors (in magenta) for 1018~pulsars from the ATNF pulsar catalog \citep{ManchesterEA2005}. The red dashed line shows the best-fit relation between the two variables, $|\RM| = (0.03 \pm 0.15) \, \DM^{(0.91 \pm 0.06)}$. The relationship is roughly linear with some spread. The magnitude of $\RM$ is proportional to $\DM$, confirming significant contributions of $\langle \ne \rangle$ to the observed $\RM$. (b) PDF of the estimated value of the average (over various distances in the range $0.09$ -- $25 \kpc$) parallel component of the Galactic magnetic field $\obsb$ using the ATNF pulsar catalog. The distribution is close to a Gaussian distribution (dashed red line) with mean of $0.21 \pm 0.08 \muG$ and standard deviation of $1.57 \pm 0.04 \muG$. Also, the kurtosis of the distribution is $3.66 \pm 0.31$ (\Tab{tab:pulbCO}), very close to the Gaussian distribution. This suggests that over $\kpc$ scales, the ISM of the Milky Way is not supersonic and $\ne$ and $B$ are not correlated over such large distances.
}
\label{fig:pulb}
\end{figure*}

\begin{figure*}
\includegraphics[width=\columnwidth]{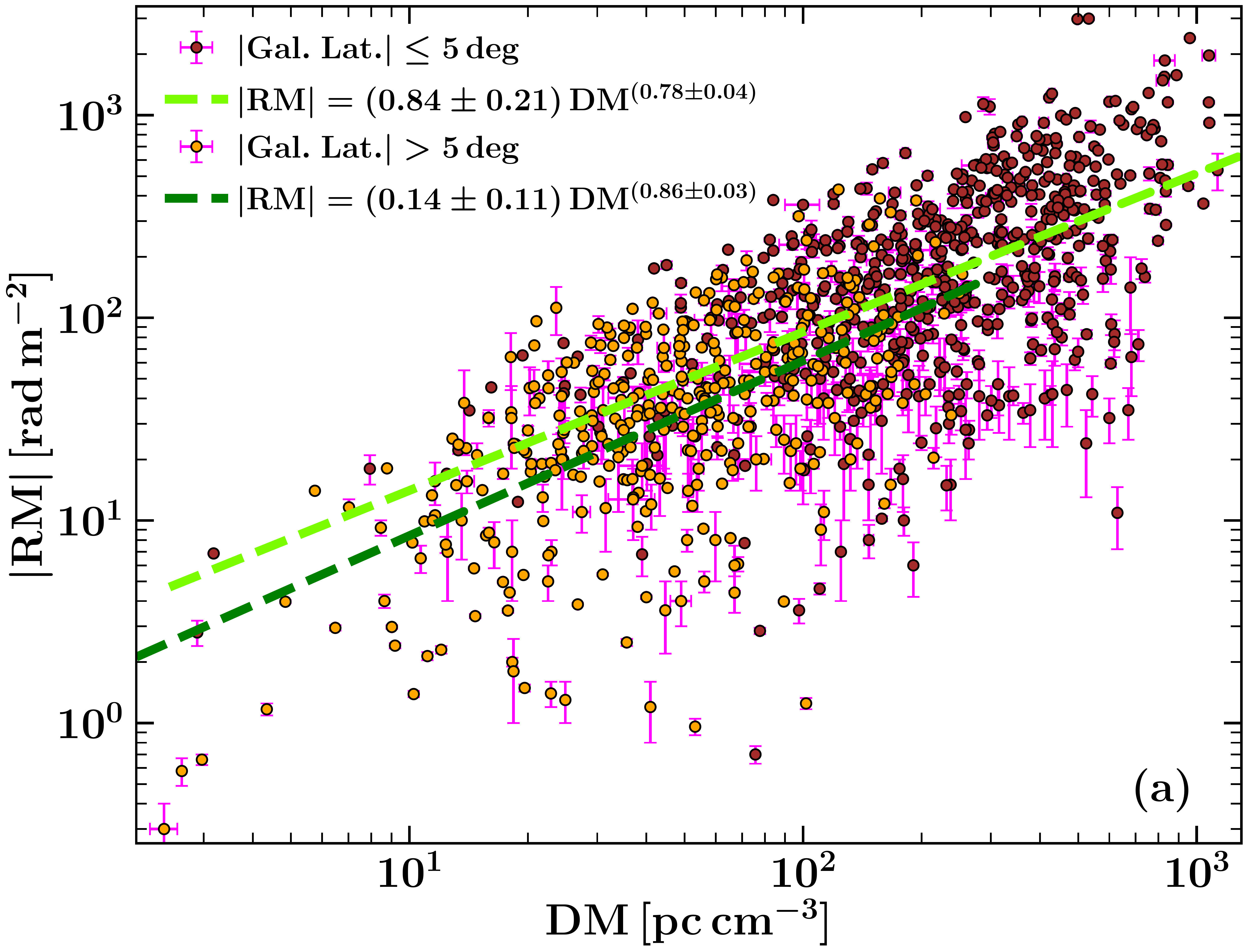} \hspace{0.5cm}
\includegraphics[width=\columnwidth]{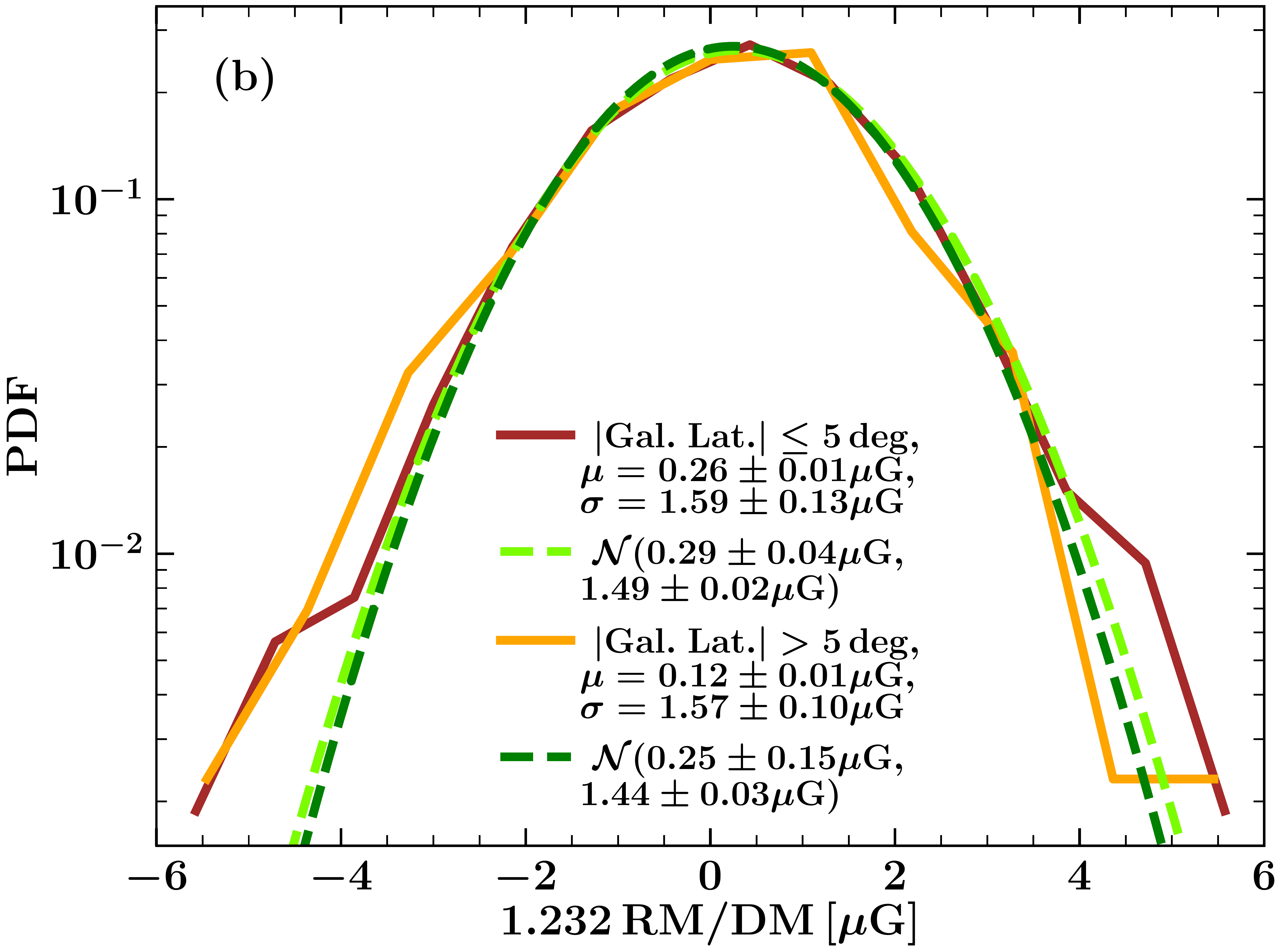}
\caption{\rev{Same as \Fig{fig:pulb}, but with the ATNF pulsar sample divided into two sub-samples: 1) pulsars in or close to the Galactic place, i.e., pulsars located with $\pm 5 \deg$ Galactic latitude (brown points in (a) and brown line in (b)), and 2) pulsars away from the Galactic plane, i.e., pulsar located outside $\pm 5 \deg$ Galactic latitude (orange points in (a) and orange line in (b)). The pulsars in or close to the Galactic plane have a statistically higher value of $|\RM|$ and $\DM$ (a), which is probably due to higher thermal electron density and magnetic fields close to the plane. However, the relationship between $\RM$ and $\DM$ (dashed light and dark green lines in (a)) and the statistical properties of the $\obsb$ distribution (dashed light and dark green lines in (b)) remain the same for both samples (see also \Tab{tab:pulbCO}). Thus, our conclusions remain the same as in \Fig{fig:pulb}.}}
\label{fig:pulbgp}
\end{figure*}

The ATNF pulsar catalog, updated version~1.63 (available at \href{http://www.atnf.csiro.au/research/pulsar/psrcat}{http://www.atnf.csiro.au/research/pulsar/psrcat}), contains 2812~pulsars, with 1173 of those having $\RM$ measurements. For our study, to ensure statistically significant results, we select all those pulsars for which the magnitude of $\RM$ is less than $3000 \rad \m^{-2}$ and the $\RM$ and $\DM$ are at least twice their respective non-zero error. This reduces our sample size to 1018 pulsars.  For these selected pulsars, the $\RM$ varies in the range $-2993$ to $+2400 \rad \m^{-2}$ with a mean error of $8.8 \%$, and $\DM$ varies in the range $2.38$ to $1130 \pc \cm^{-3}$ with a mean error of $0.6 \%$. They are at a distance ranging from $0.09 \kpc$ to $25 \kpc$ with a mean distance of $4.72 \kpc$. \reva{We do not intend to make one-to-one comparisons of simulation results with the observational data, because the path length in our simulations is $600 \pc$, whereas that for most pulsars is of the order of $>\kpc$. Increasing the path length in the numerical simulations to $\kpc$ scales would require removing the periodic boundary conditions and including large-scale galaxy properties such as differential rotation, gravity, and density stratification. We aim to do this in the future. However, we can still learn about the effects of the thermal electron density -- magnetic field correlation on the $\obsb$ distribution from the present simulations (e.g., the kurtosis of $\obsb$ is greater than that of a Gaussian when the magnetic field and the thermal electron density are correlated) and look for similar trends in the observations.}

\Fig{fig:pulb}~(a) shows the scatter plot (blue data points) of the magnitude of the $\RM$ and $\DM$ of these pulsars (with errors in magenta) and the dashed red line shows the best fit, which is of the form $|\RM| = (0.03 \pm 0.15) \, \DM^{(0.91 \pm 0.06)}$. \rev{The scatter in the data may be due to the following two reasons: 1) for pulsars with large path lengths, the magnetic field along the path length can have one or more reversals, which might lower the $\RM$, and 2) some of the older $\RM$s measurements might be affected by the $n \pi$ ambiguity (see Fig.~1 in \citet{RuzmaikinS1979} for a description of the problem and \citet{BrentjensB2005} for the resolution).} However, the approximate linear (best-fit) relationship between $\RM$ and $\DM$ confirms that the $\RM$ has a significant contribution from \reva{the average thermal electron density} and is not just dominated by the parallel component of the Galactic magnetic field. This motivates us further to study the role of $\ne$ and its correlation with $B$ in estimating Galactic magnetic fields using \Eq{eq:obsb}.  

The estimated average of the parallel component of the Galactic magnetic field $\obsb$ obtained using the $\RM$ and $\DM$ values from the ATNF pulsar catalog lies in the range $-7.37$ -- $+5.85 \muG$ with a mean error of $9\%$. \Fig{fig:pulb}(b) shows the PDF of $\obsb$ (in the range $-6$ to $+6 \mkG$, where the majority of points lie) and the distribution is very close to a Gaussian distribution (dashed red line). The mean and standard deviation of the distribution are $0.21 \pm 0.08 \muG$ and $1.57 \pm 0.04 \muG$, respectively. \reva{Since the small- and large-magnetic fields are possibly correlated over $100 \pc$ and a few $\kpc$ scales, respectively, and the path lengths to pulsars is of the order of $\kpc$, the mean of the $\obsb$ distribution is primarily controlled by the large-scale field (the small-scale field contribution to the mean $\obsb$ averages out to near zero), but the standard deviation of the $\obsb$ distribution has contributions from changes in both the small- and large-scale fields along the path length. The mean value does not directly correspond to the magnitude of the large-scale field because of large-scale magnetic field reversals along the path length.}

\reva{Using the random walk model (\Eq{eq:rmrw}), we can roughly estimate the contribution of small- and large-scale magnetic fields to the standard deviation of $\RM$ for a path length of $L=5 \kpc$ ($\approx$ mean of distances to pulsars), which in turn would provide a rough idea of the contribution of small- and large-scale field variations to the standard deviation of $\obsb$. We assume that over the path length, the ISM is primarily sub-sonic (the $\RM$ standard deviation is half of the value obtained using the random walk model, see column~4 in \Tab{tab:rw}) and the mean thermal electron density $\langle \ne \rangle$ is $0.02 \cm^{-3}$. For small-scale magnetic fields with $\lb=100 \pc$ and $\brms=5 \muG$, the standard deviation of $\RM$ is approximately $16.5 \rad \m^{-2}$, whereas for the large-scale magnetic field with $\lb = 2 \kpc$ and field strength of $1 \muG$ (for $\brms/\sqrt{3}$ in \Eq{eq:rmrw}), we obtain the standard deviation of $\RM$ to be around $25.5 \rad \m^{-2}$. Thus, both the small- and large-scale variations contribute to the standard deviation of $\obsb$ and the contribution of the large-scale field can be slightly higher.} 

Row~1 in \Tab{tab:pulbCO} shows the first four moments (mean $\mu$, standard deviation $\sigma$, skewness $\sk$, and kurtosis $\ku$) of the $\obsb$ distribution. The distribution of $\obsb$ is close to a Gaussian distribution ($\ku = 3.66 \pm 0.31$) and on comparing with the distributions in \Fig{fig:rmdm}~(a), we confirm that the \rev{part of the} ISM \rev{probed by pulsars} on large $\kpc$ scales is \reva{probably} not supersonic. \rev{Since the thermal electron density and magnetic fields are expected to be stronger close to the Galactic plane, we also divide the ATNF pulsar sample into two sub-samples: one close to the plane (within $\pm 5 \deg$ in Galactic latitude), and one away from the plane (outside $\pm 5\deg$ in Galactic latitude). \reva{} Even though $|\RM|$ and $\DM$ of pulsars close to the Galactic plane are statistically higher than those away from the plane, the relationship of $\RM$ and $\DM$ for these two sub-samples is roughly the same, as shown in \Fig{fig:pulbgp}(a). \reva{The mean of $\obsb$ for pulsars closer to the plane is higher than that of pulsars away from the plane, whereas the standard deviation of both the sub-samples is very similar (\Fig{fig:pulbgp}(b) and \Tab{tab:pulbCO}). This might be due to reasons such as a stronger large-scale field in the disc compared to outside the disc, similar small-scale magnetic fields in both regions, or a combination of both these reasons.} The PDFs of the estimated magnetic fields in the two sub-samples also have very similar \reva{kurtosis}; they are both roughly Gaussian\footnote{\rev{Further analysis along these lines requires dividing the pulsar sample into selected regions of the Milky Way, such as spiral arms, where the small-scale magnetic field is expected to be strong. However, for the current data, this procedure would result in a very low number of pulsars in each sub-region, and would consequently lead to low statistical significance. Similar analyses could be done with $\RM$s from a high-resolution polarisation survey of the Milky Way, such as the Galactic Magneto-Ionic Medium Survey, GMIMS \citep{DickeyEA2019}.}}.}  This also implies that the thermal electron density and the magnetic fields over such length scales are uncorrelated. This in turn implies that \Eq{eq:obsb} is applicable to estimate the average parallel Galactic magnetic field. However, to test this further, we correlate $\obsb$ from pulsars with various probes of locally strong magnetic fields and thermal electron density along the path length in Sec.~\ref{sec:obs:pulbCO}--\ref{sec:obs:ihacorr}.

\begin{table*}
	\caption{Statistical measures (mean $\mu$, standard deviation $\sigma$, skewness $\sk$, and kurtosis $\ku$) of the estimated magnetic field $\obsb$ distributions (\Fig{fig:pulb}~(b) and \Fig{fig:pulbCO}) obtained using the ATNF pulsar catalog \citep{ManchesterEA2005} and from $\CO$ data of molecular clouds \citep{MivilleME2017}. The rows are: 1.~pulsars in our sample from the ATNF catalog, \rev{2.~pulsars located within $\pm 5 \deg$ in Galactic latitude, i.e., in or close to the Galactic plane, 3.~pulsars located outside $\pm 5 \deg$ in Galactic latitude, i.e., away from the Galactic plane}, 4.~pulsars that have one or more molecular clouds along their line of sight based in $\CO$ data, and 5.~pulsars that do not have molecular clouds along their line of sight as per $\CO$ data. The columns are: 1.~the observational data set, 2.~the sample size, 3.~the mean of $\obsb$, $\mu(\obsb)$, 4.~the standard deviation of $\obsb$, $\sigma(\obsb)$, 5.~the skewness of $\obsb$, $\sk(\obsb)$, and 6.~the kurtosis of $\obsb$, $\ku(\obsb)$. The errors in the statistical measures are obtained by propagating reported observational uncertainties in $\RM$ and $\DM$.}
	\label{tab:pulbCO}
	\begin{tabular}{lccccc} 
		\hline
		\hline
		 Observational dataset & Sample size & $\mu \, (\obsb)$ & $\sigma \, (\obsb)$ &  $\sk \, (\obsb)$ & $\ku \, (\obsb)$ \\
		 -- & -- & $[\muG]$ & $[\muG]$ & -- & -- \\ 
		\hline
		ATNF & 1018 & $0.21 \pm 0.08$  & $1.57 \pm 0.04$ & $-0.16 \pm 0.09$ & $3.66 \pm 0.31$ \\
		ATNF (${\rm |Gal.~Lat.|} \le 5 \deg$) & 622 & $0.26 \pm 0.01$ & $1.59 \pm 0.13$ & $-0.08 \pm 0.04$ & $3.52 \pm 0.51$ \\
		ATNF (${\rm |Gal.~Lat.|} > 5 \deg$) & 396 & $0.12 \pm 0.01$ & $1.57 \pm 0.10$ & $-0.23 \pm 0.05$ & $3.47 \pm 0.04$ \\
		ATNF (with $\CO$) & 557 & $0.26 \pm 0.09$  & $1.62 \pm 0.11$ & $-0.08 \pm 0.09$ & $3.37 \pm 0.39$ \\
		ATNF (without $\CO$) & 461 & $0.12 \pm 0.02$  & $1.55 \pm 0.08$ & $-0.20 \pm 0.07$ & $3.43 \pm 0.16$ \\
		\hline
		\hline
   \end{tabular}
\end{table*}

\subsection{$\obsb$ from pulsar observations combined with $\CO$ data of molecular clouds} \label{sec:obs:pulbCO}

\begin{figure}
\includegraphics[width=\columnwidth]{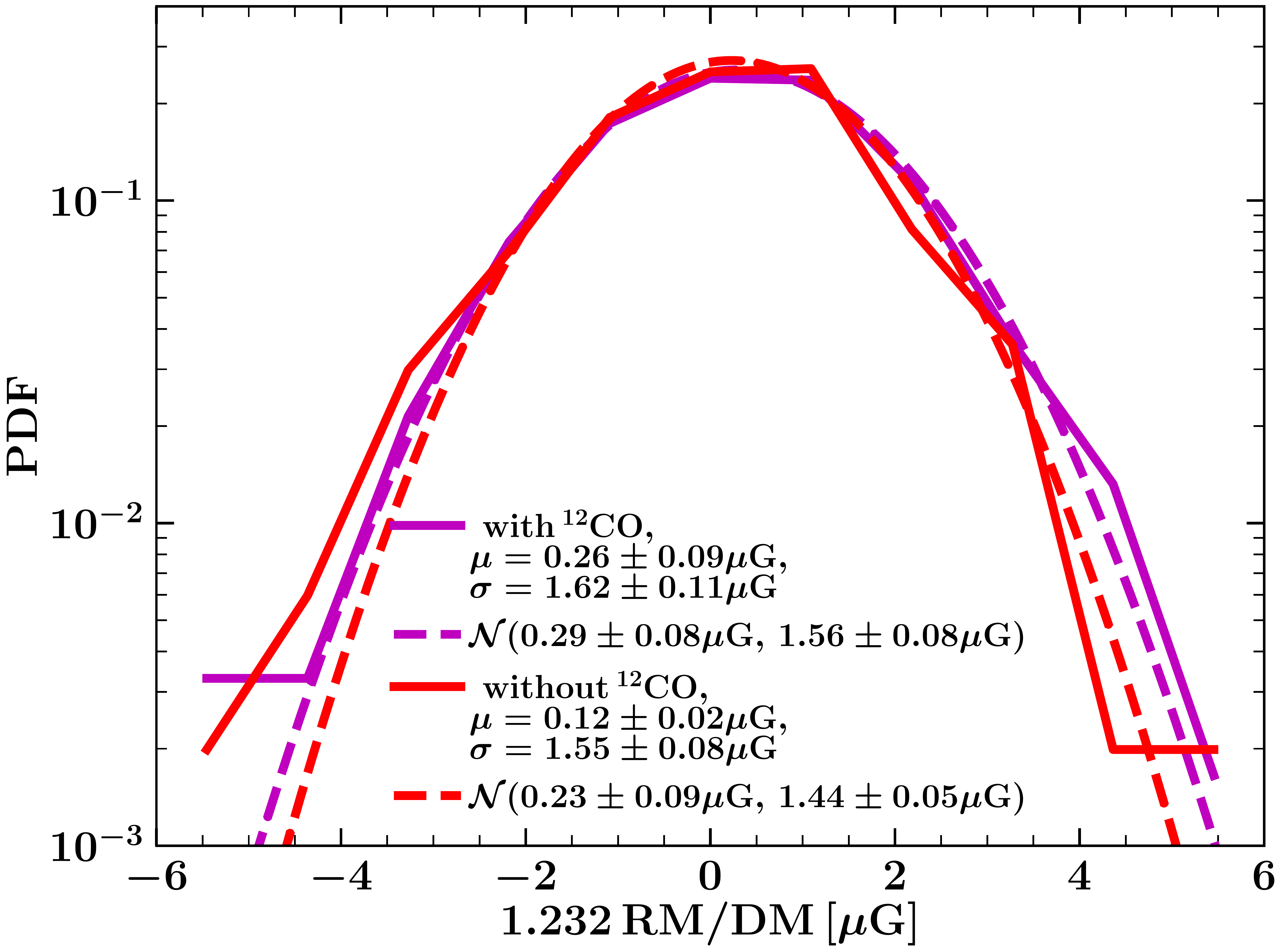} 
\caption{PDF of $\obsb$ from pulsars with and without CO along their line of sight \citep[data from][]{MivilleME2017}. The dashed lines show the corresponding fitted Gaussian distributions. The mean and standard deviation of the sample with $\CO$ along the line sight is slightly larger than that without $\CO$, but the difference is not statistically significant and lies within the error bars. Both distributions are very close to Gaussian (see \Tab{tab:pulbCO}). Thus, either the molecular clouds do not have significant $\ne$ to contribute to $\RM$ and $\DM$ and/or they occupy a very small region over the entire path length.}
\label{fig:pulbCO}
\end{figure}

We would expect the $\ne$ and $B$ to be correlated in the denser regions of the Milky Way. If the light from the pulsar passes through one or more of these clouds with correlated $\ne$ and $B$, the contribution of the correlation to $\RM$ would not cancel out on dividing by $\DM$. Thus, the pulsars with molecular clouds along their line of sight would (statistically) give a higher estimate of the magnetic field strength when calculated using \Eq{eq:obsb}. This higher estimate is not necessarily a result of the higher thermal electron density or stronger magnetic fields, but also because of the correlation between the two quantities.

One of the tracers of dense star-forming regions is the $\CO$ molecule. \cite{MivilleME2017} provides a catalog of 8107~molecular clouds in the Galactic plane (within $\pm \, 5^{\circ}$ in the Galactic latitude) based on $\CO$. We use the ATNF pulsars as before, but now combine them with the CO data, to select those pulsars that have emission passing through a molecular cloud based on the location and area of CO on the sky.

\Fig{fig:pulbCO} and \Tab{tab:pulbCO} (last two rows) shows the PDFs of $\obsb$ and the properties of the distribution with and without $\CO$ along the line of sight, respectively. We find that out of that 1018~pulsars in our sample, the lines of sight for 557 of those pass through molecular gas and for the remaining 461 it does not. \Fig{fig:pulbCO} and \Tab{tab:pulbCO} show that there is not much difference in both samples. \reva{The mean of the $\obsb$ distribution for pulsars with $\CO$ along their path length is greater than that without $\CO$, but their standard deviation is statistically the same. Since most of the molecular clouds lie close to the galactic plane, the reason for this result is as discussed in the previous section.} Moreover, the kurtosis of both distributions (column~6 in \Tab{tab:pulbCO}) is very similar to that of a Gaussian distribution. Thus, we conclude that over large length scales the $\ne$ and $B$ are uncorrelated. Since, statistically, there is no difference between the distribution with $\CO$ and without $\CO$, we can further conclude that either these molecular clouds occupy a very small path over the entire path length to the pulsar or that $\ne$ is very low inside molecular clouds.

\subsection{Correlation of $\obsb$ with magnetic fields measured in the CNM via Zeeman splitting of HI} \label{sec:obs:pulbZeeman}
\begin{figure}
\includegraphics[width=\columnwidth]{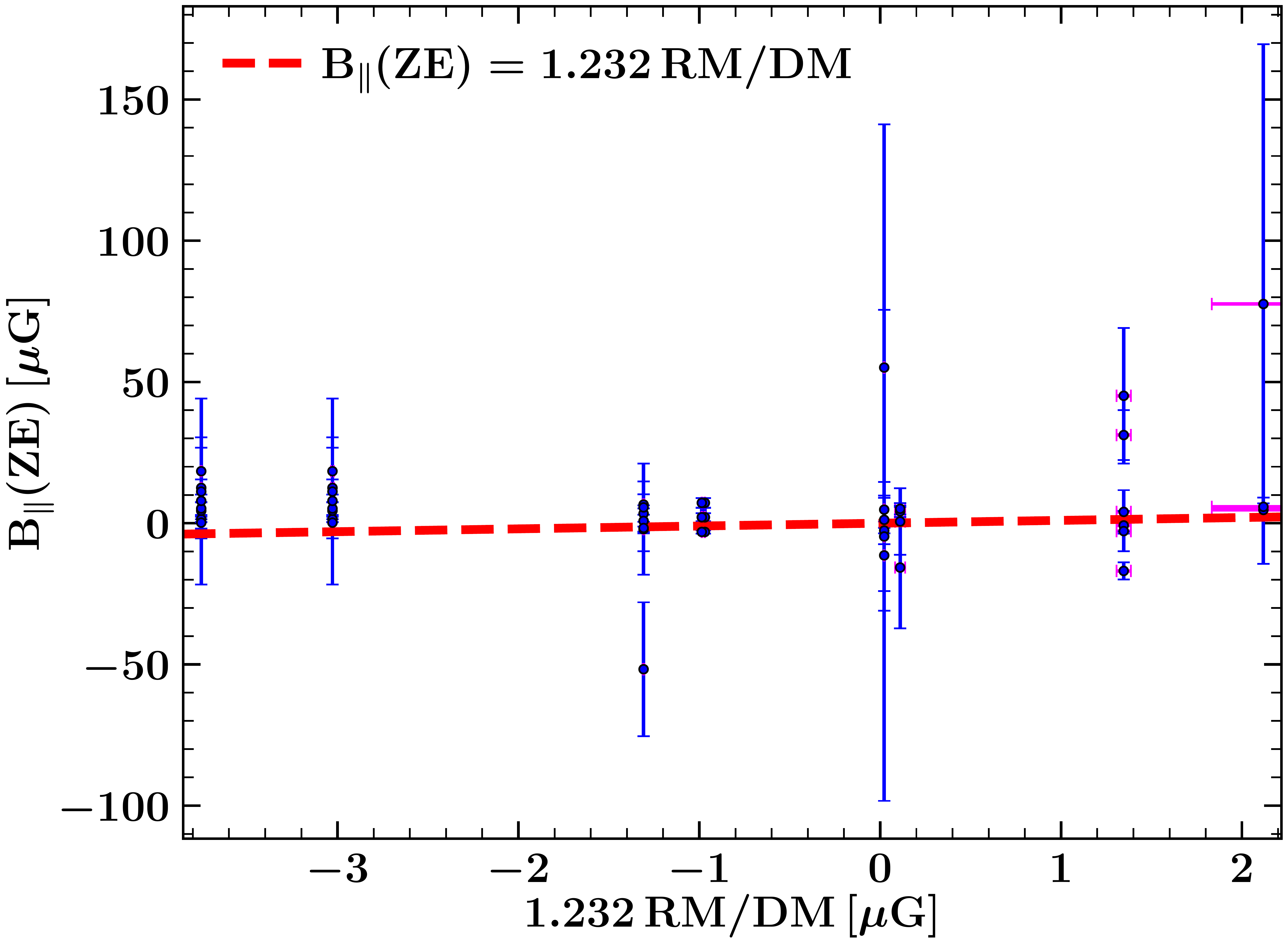} 
\caption{Scatter plot of the magnetic field strength in the CNM obtained from HI Zeeman splitting, $\rm B_{\parallel} (ZE)$ \citep[data from][]{HeilesT2004} as a function of $\obsb$ from pulsars (uncertainties shown in magenta). The red dashed line shows the one-to-one correlation. While both quantities seem to broadly agree, it is difficult to draw significant conclusions, because of the large uncertainties in $\rm B_{\parallel} (ZE)$.}
\label{fig:pulbZeeman}
\end{figure}

Another region in the ISM of the Milky Way where we would expect strong magnetic fields is the cold neutral medium (CNM). Such regions would probably have a lower $\ne$ but have a significantly strong $B$ in comparison to the average magnetic field of the ISM \citep[can be as high as $50 \muG$, see][]{HeilesT2004}. Such magnetic fields can be probed using Zeeman splitting of spectral lines. \citet{HeilesT2004} provides magnetic field measurements in the CNM using Zeeman splitting of the HI 21~cm absorption line for 41~sources. We examine how many of these sources are along the path length of the pulsars from the ATNF catalog and if there exists a correlation between $\obsb$ and the $B$ obtained from the Zeeman effect $\rm B_{\parallel} (ZE)$. Only 50~pulsars out of our full sample of 1018~pulsars have their lines of sight passing through the region probed by HI. \Fig{fig:pulbZeeman} shows the correlation between $\obsb$ from pulsars and $\rm B_{\parallel} (ZE)$ from HI, with the red dashed line showing the one-to-one correlation between the two quantities. Given the large errors in the magnetic field from HI and the small sample size of the overlap between the two datasets, it is difficult to draw significant conclusions.

\subsection{Correlation of $\RM, \DM,$ and $\obsb$ from pulsar observations with $\NHI$ data} \label{sec:obs:nhicorr}
\begin{figure*}
\includegraphics[width=\columnwidth]{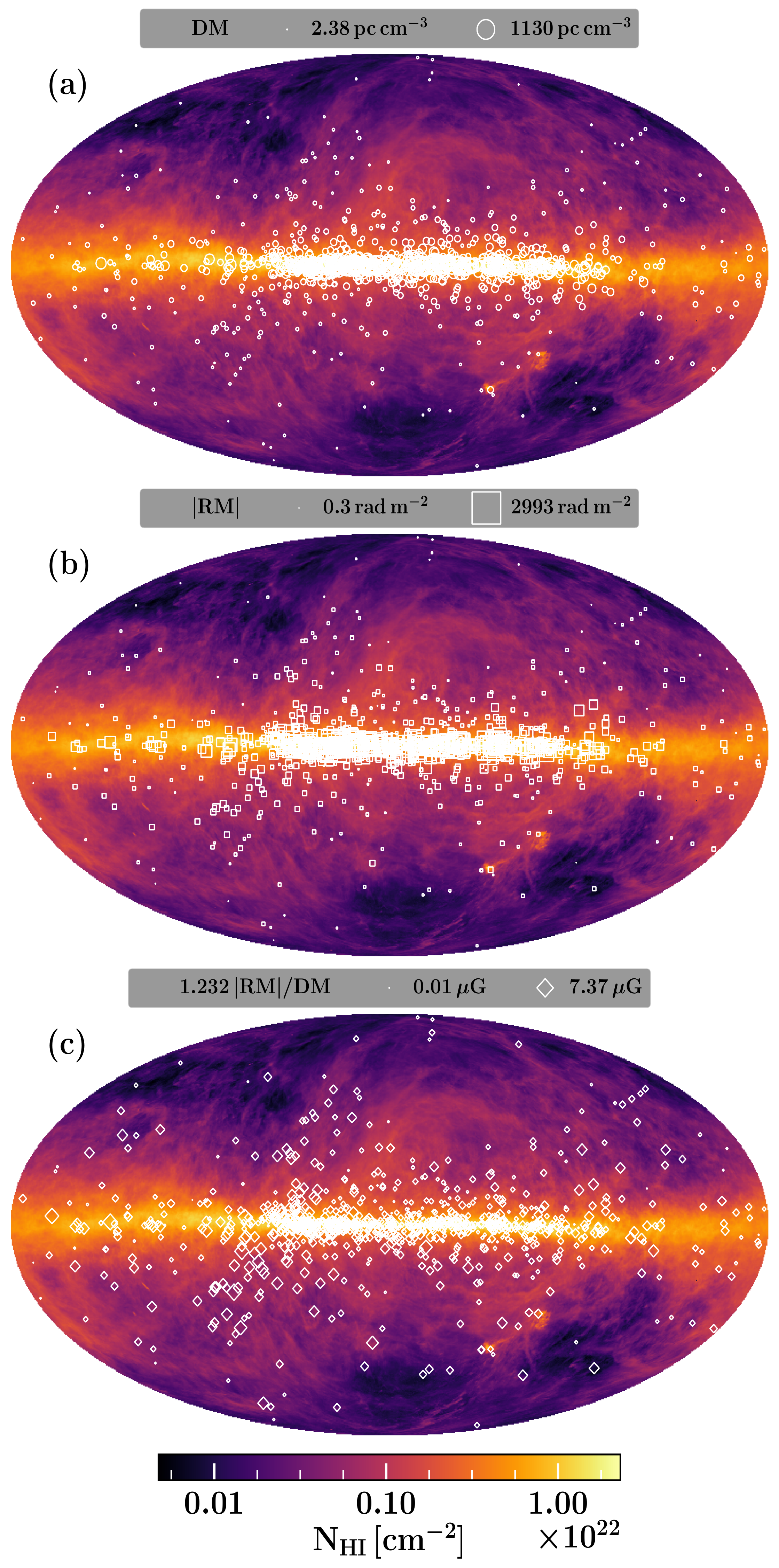}  \hspace{0.5cm}
\includegraphics[width=\columnwidth]{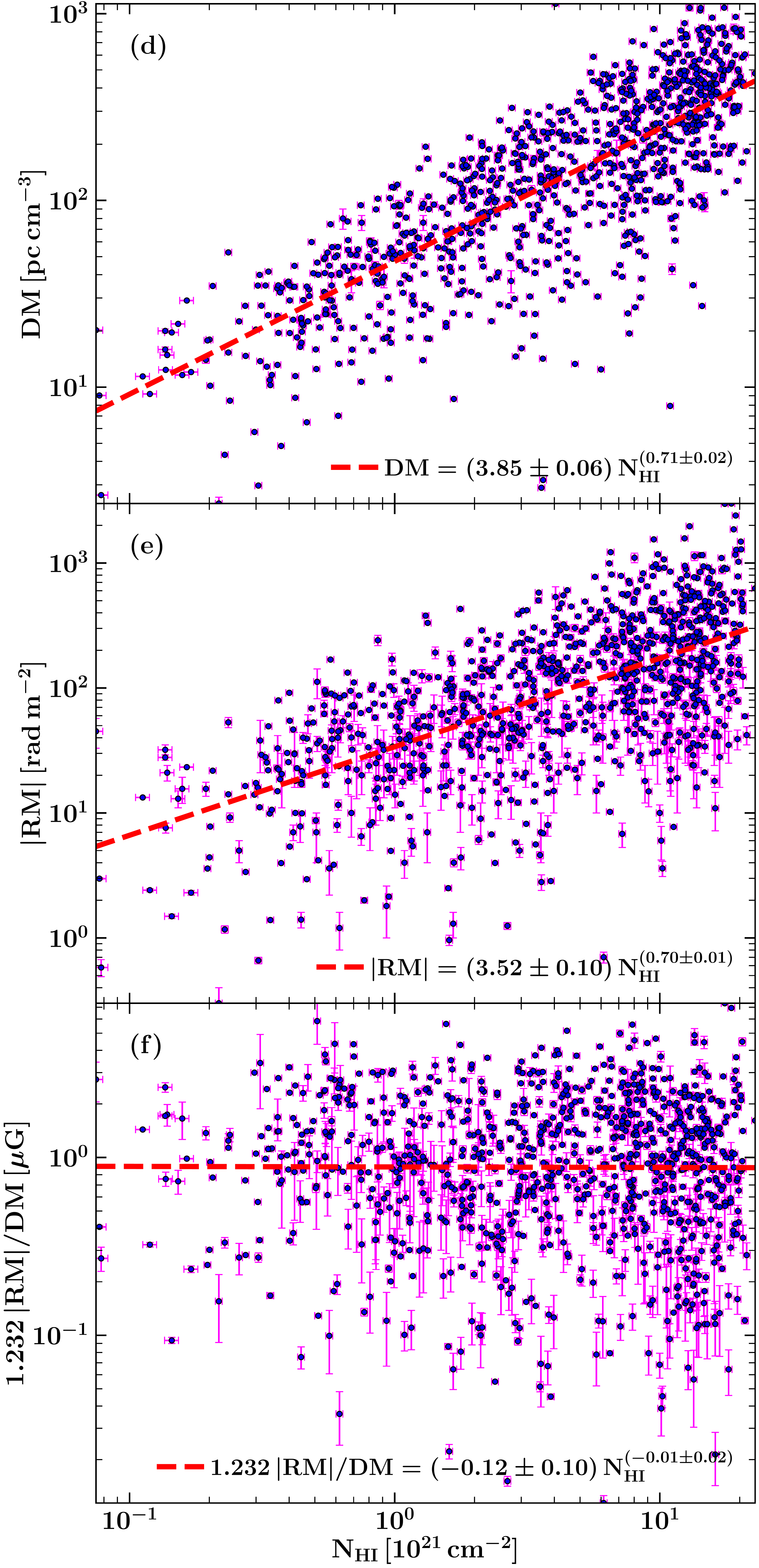}
\caption{Left panels: all-sky maps using Mollweide projection (HEALPix resolution N$_{\text{side}}$=1024, which corresponds to a spatial resolution of approximately $3.5 \arcmin$) of $\NHI$ \citep{HI4PI2016} with $\DM$ (a), $|\RM|$ (b), and $\obsbabs$ (c) from the ATNF pulsar catalog \citep{ManchesterEA2005}. Right panels: $\DM$ (d), $\RM$ (e), and $\obsbabs$ (f) as a function of $\NHI$. The uncertainties for $\NHI$ are computed based \citep[Fig.~15 of][]{WinkelEA2016}. The uncertainties in $\DM$ and $\RM$ are from the ATNF pulsar catalog. The red dashed line shows the best-fit relations. Both $|\RM|$ and $\DM$ have a positive correlation with $\NHI$, showing significant contributions of $\ne$ to $\RM$. Since they both have a similar correlation slope, the resulting estimated average parallel component of the magnetic field $\obsbabs$ (f) is not correlated with $\NHI$. This shows that the integrated $\ne$ along the path length, which is expected to be higher when $\NHI$ is higher, does not affect the magnetic field estimate, $\obsb$. Thus, the thermal electron density and magnetic fields are not \reva{locally} correlated for the most part of the path length.}
\label{fig:nhicorr}
\end{figure*}

\begin{table}
	\centering
	\caption{The exponent of relation between $\NHI$ \citep[row~1; right panel of \Fig{fig:nhicorr}, data from][]{HI4PI2016} and H$\alpha$ intensity \citep[row~2, right panel of \Fig{fig:ihacorr}, data from][]{Finkbeiner2003} with $\DM$ (column~1), $|\RM|$ (column~2), and $\obsbabs$ (column~3) from the ATNF pulsar catalog \citep{ManchesterEA2005}, respectively. The reported errors are from fitting, taking into account the reported uncertainties in all observational data.}
	\label{tab:pulbNHIIHa}
	\begin{tabular}{lccc} 
		\hline
		\hline
		Observational data & $\DM$ & $|\RM|$ & $\obsbabs$ \\
		\hline
		$\NHI$ & $0.71 \pm 0.02$ & $0.70 \pm 0.01$ & $-0.01 \pm 0.02$ \\
		$\IHa$ & $0.51 \pm 0.03$ & $0.43 \pm 0.02$ & $-0.08 \pm 0.03$ \\
		\hline
		\hline
   \end{tabular}
\end{table}

Galactic HI is a tracer of the atomic phase of the ISM and is primarily observed using radio observations of the 21~cm spectral line \citep{DickeyL1990}. In the limit of an optically thin medium, the HI column density $\NHI$ along any line of sight can be obtained. We would expect that the lines of sight for which $\NHI$ is higher, $\ne$ will also be high along the path length and thus would contribute significantly to $\DM$ and $\RM$. If along those path lengths, the $\ne$ and $B$ are correlated, the average parallel component of the Galactic magnetic field estimated via \Eq{eq:obsb} would be overestimated (\Sec{sec:simcorr:effect}, \Fig{fig:obsbbpar2d}), and thus, the estimated \reva{average} magnetic field value would be correlated with $\NHI$.

Here we use all-sky $\NHI$ data from the HI4PI survey \citep{HI4PI2016} and correlate it with the ATNF pulsar observations. \Fig{fig:nhicorr} (left panel) shows $\DM$ (a), $|\RM|$ (b), and $\obsbabs$ (c) from the pulsars overlayed on top of the all-sky $\NHI$ map.  Most pulsars lie in the Galactic plane, but overall they vary throughout the Milky Way. \Fig{fig:nhicorr} (right panel) shows the correlation of $\DM$ (d), $|\RM|$ (e), and $\obsbabs$ (f) with $\NHI$ obtained at the location of each pulsar using a bi-linear interpolation of the four nearest neighbouring points in the all-sky map. The errors in $\DM$ and $|\RM|$ are taken from the ATNF catalog and the error in $\obsbabs$ is computed from them. The error in the interpolated $\NHI$ is computed using the uncertainties given in Fig.~15 of \citet{WinkelEA2016}. The red dashed lines in \Fig{fig:nhicorr} (d), (e), and (f) are the best-fit relations between the resepective variables in each panel. The slopes of these lines are given in the first row of \Tab{tab:pulbNHIIHa}.  Both $\DM$ and $|\RM|$ are positively correlated with $\NHI$ and have similar correlation slopes. This shows that $\langle \ne \angle$ contributes significantly to $\RM$. However, the average of the parallel magnetic field strength estimated using \Eq{eq:obsb}, $\obsbabs$, is statistically uncorrelated with $\NHI$. This shows that the \revb{local} $\ne$ and $B$ are largely uncorrelated over $\kpc$ scales and the structures with \revb{locally} correlated $\ne$ and $B$ occupy a very small path length along the line of sight from the pulsar to us. Thus, \Eq{eq:obsb} is applicable on such large length scales. 

\subsection{Correlation of $\RM, \DM,$ and $\obsb$ from pulsar observations with $\IHa$ data} \label{sec:obs:ihacorr}

\begin{figure*}
\includegraphics[width=\columnwidth]{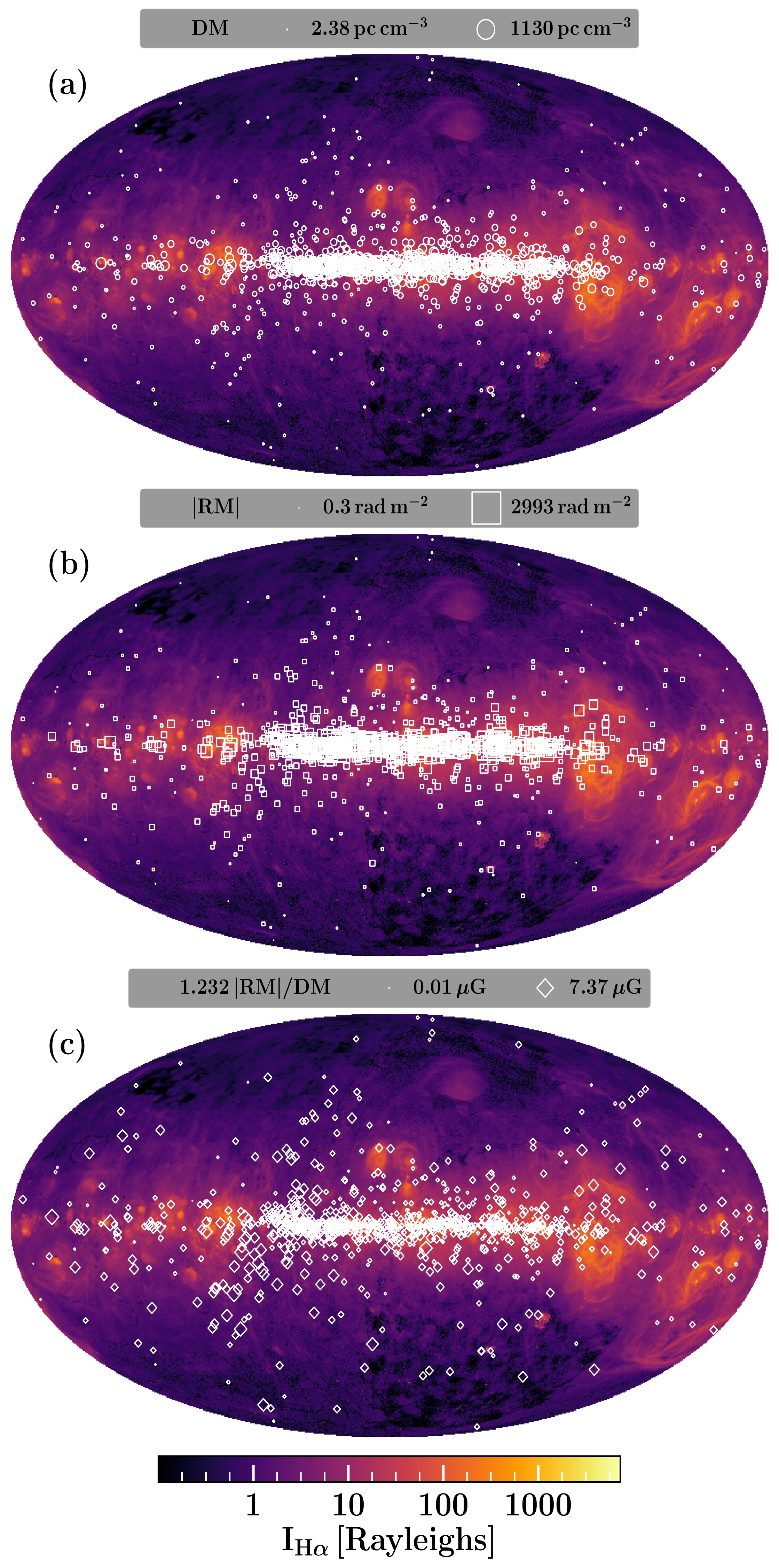} \hspace{0.5cm}
\includegraphics[width=\columnwidth]{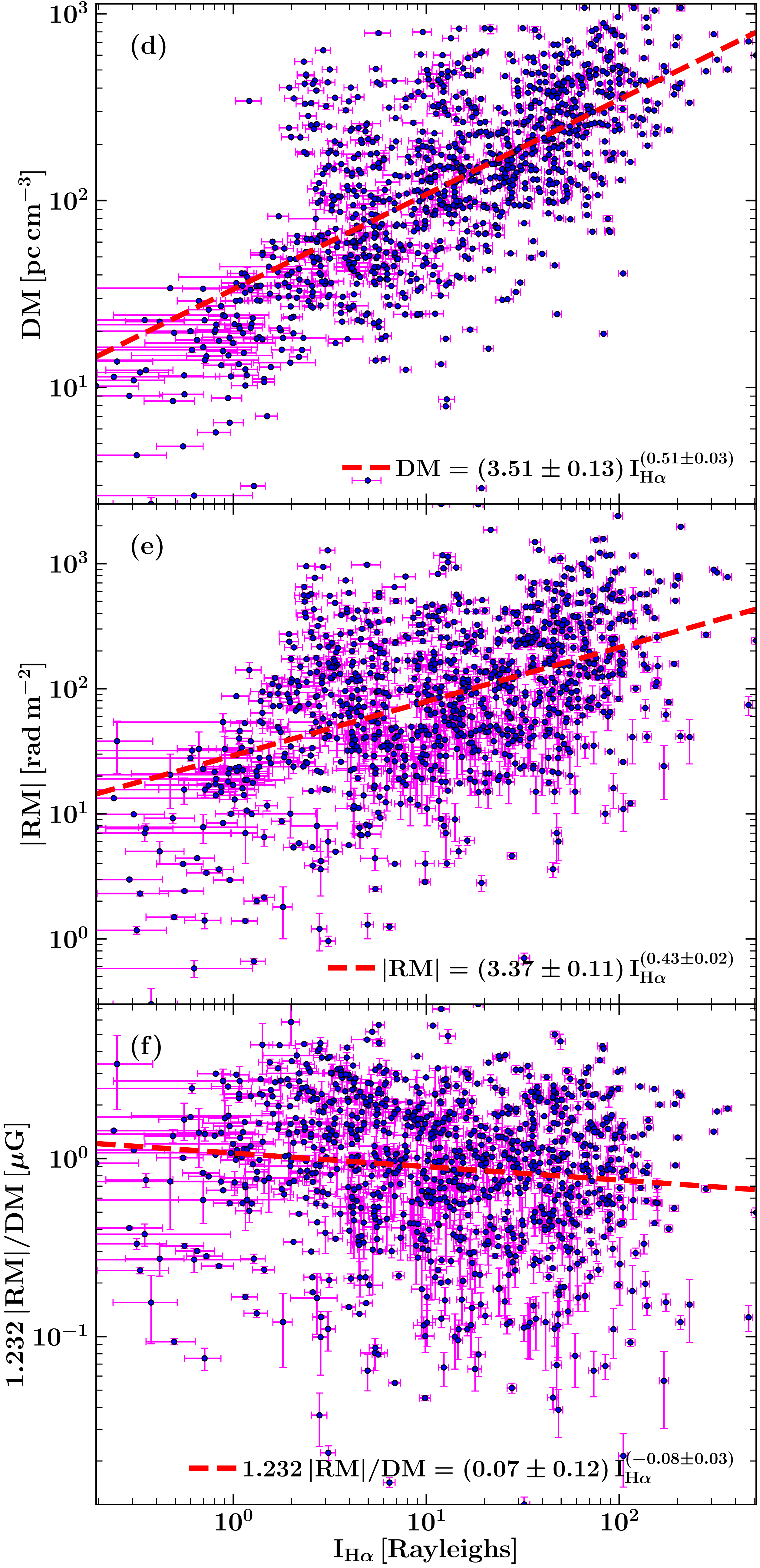}
\caption{Similar to \Fig{fig:nhicorr}, but for H$\alpha$ data taken from \citet{Finkbeiner2003}. The error in the H$\alpha$ intensity $\IHa$ is computed from the error map in \citet[][Sec.~3.5]{Finkbeiner2003}. Here again (similar to $\NHI$), both $\RM$ and $\DM$ show a positive correlation with $\IHa$, but $\obsb$ is largely un-correlated with $\IHa$ (or even shows a weak negative correlation). This confirms that over $\kpc$ scales, \revb{local} $\ne$ and $B$ are largely uncorrelated. Thus, \Eq{eq:obsb} is applicable on such scales.}
\label{fig:ihacorr}
\end{figure*}

Finally, we explore the correlation of the pulsar data with the ionised hydrogen intensity $\IHa$ in the Milky Way. We would expect that $\IHa$ is an even better probe of $\ne$ than $\NHI$ (\Sec{sec:obs:nhicorr}) because $\IHa$ directly probes ionised regions.

\Fig{fig:ihacorr} (left panel) shows the all-sky map of $\IHa$, taken from \citet{Finkbeiner2003} with $\DM$ (a), $|\RM|$ (b), and $\obsbabs$ (c) from the ATNF pulsars overlayed. $\IHa$ is strongest in the central regions of the Galaxy where many pulsars lie. \Fig{fig:ihacorr} (right panel) shows the correlation of $\DM$ (d), $|\RM|$ (e), and $\obsbabs$ (f) with $\IHa$ interpolated at the pulsar location (bi-linear interpolation with four nearest neighbouring points in the all-sky map). The error in $\IHa$ is computed from the uncertainty map provided in \citet[Sec.~3.5 in][]{Finkbeiner2003}. The red dashed line shows the correlation of $\DM$ (d), $\RM$ (e), and $\obsbabs$ (f) with $\IHa$, and the slopes of these lines are given in the second row of \Tab{tab:pulbNHIIHa}. Both $\DM$ and $|\RM|$ are positively correlated with $\IHa$ and their dependence is similar. \rev{The slope of the best-fit line for the dependence of $|\RM|$ and $\DM$ on $\IHa$ (red dashed line in \Fig{fig:ihacorr} (d, e)) is close to $0.5$ and this is probably because $\IHa$ is proportional to $\ne^2$.} However, the dependence of $\DM$ and $|\RM|$ on $\IHa$ is weaker than that on $\NHI$ (c.f., columns~2 and~3 in \Tab{tab:pulbNHIIHa}). Even though now both quantities ($\ne$ and $\IHa$) probe only ionised regions, the estimated average parallel component of the Galactic magnetic field $\obsb$ does not depend strongly on $\IHa$ (in fact the dependence is very mildly negative). This further confirms the fact that the \revb{local} $\ne$ and $B$ are uncorrelated over $\kpc$ scales and/\rev{or} the correlated ionised H$\alpha$ emitting gas in the ISM occupy only a small path length along the line of sight to the pulsars \rev{due to its low volume filling factor}.

\section{Discussion} \label{sec:dis}
In this paper, we studied the effect of $\ne$--$B$ correlations on magnetic-field estimates from pulsar observations using \Eq{eq:obsb}. From our results using numerical simulations (\Sec{sec:simres}), we find that using $\RM$ and $\DM$ as per \Eq{eq:obsb} in a system with strongly correlated $\ne$ and $B$ (expected in supersonic flows with $\Mach \gtrsim 2$), overestimates the average parallel component of the magnetic field. Then using various Milky Way observations in \Sec{sec:obs}, we show that over large $\kpc$ scales, \revb{locally} $\ne$ and $B$ are largely uncorrelated with correlated structures occupying relatively small path lengths. Here, we further discuss some more aspects of the aim of our study.

\subsection{Anti-correlation between $\ne$ and $B$} \label{sec:dis:ant}
In numerical simulations, we mainly studied the effect of the correlation between the thermal electron density and magnetic fields (specially the small-scale random magnetic fields). However, $\ne$ might also be anti-correlated with $B$. This, as expected, would underestimate the magnetic field when calculated using \Eq{eq:obsb}.

The anti-correlation between thermal electron density and magnetic fields can be due to a local pressure balance, but this further requires assumptions of isotropic random magnetic fields at smaller scales ($\sim 100 \pc$), roughly equal cosmic ray pressure and magnetic pressure, and energy equipartition between magnetic and turbulent kinetic energies at $\kpc$ scales \citep{BeckEA2003}. Also, the large-scale ordered field in external spiral galaxies is observed to occupy inter-arm regions and this can be physically due to a weaker mean-field dynamo in the gaseous arms (mostly because of a stronger stellar feedback) or a drift of magnetic fields with respect to the gaseous arms \citep[Sec.~4.9 in][]{Beck2016}. Assuming that the Milky Way is not a special spiral galaxy and is similar to one of the external spiral galaxies, the tendency of the large-scale magnetic fields to avoid gaseous arms might also give rise to an anti-correlation between the thermal electron density and large-scale magnetic fields. If there exists such a correlation, the magnetic fields obtained by using $\RM$ and $\DM$ from pulsar observations would be underestimated. On correlating the average parallel component of the Galactic magnetic field $\obsb$ with the Milky Way ionised hydrogen intensity $\IHa$ (\Fig{fig:ihacorr}~(f) and \Sec{sec:obs:ihacorr}), the slope of the best fit line is slightly negative (a negative slope suggests anti-correlation between thermal electron density and magnetic fields), but it is statistically not very significant to affect the magnetic field estimate $\obsb$.

Also, using various observational probes in \Sec{sec:obs}, we show that $\ne$ and $B$ are largely uncorrelated over $\kpc$ scales. The possibility of the correlation and anti-correlation cancelling out exactly for all the (1018) pulsars at different distances and with all the different observational probes is highly unlikely. Overall, we do not find any significant correlation or anti-correlation between $\ne$ and $B$ at such large scales. 

\subsection{Intrinsic $\RM$ of pulsars} \label{sec:dis:int}
In the entire study, we attribute the $\RM$ obtained from the pulsar observations to the thermal electron densities and magnetic fields of the Milky Way along the path from the pulsar to us. However, the pulsars themselves might also have some intrinsic $\RM$, which we discuss here.

\begin{table}
	\centering
	\caption{Parameters of the 12~pulsars for which \citet{IlieJW2019} associate the $\RM$ variations across the pulse profile to the magnetosphere of the pulsar. The columns are following: 1.~pulsar name, 2.~rotation measure from the ATNF pulsar catalog \citep{ManchesterEA2005}, $\RM$, 3.~mean of the rotation measure over the pulsar profile, $\langle \RM (\phi) \rangle$, where $\phi$ denotes the phase \citep[taken from Tab.~1 in][]{IlieJW2019}, and 4.~the computed difference between columns 2 and 3, $\delta(\RM)$. The errors in $\delta(\RM)$ are computed by propagating the uncertainties in $\RM$ and $\langle \RM (\phi) \rangle$. 
	}
	\label{tab:intrm}
	\begin{tabular}{lccc} 
		\hline
		\hline
		Pulsar name & $\RM$ & $\langle \RM (\phi) \rangle$ & $\delta(\RM)$ \\
		--  & $[\rad \m^{-2}]$ & $[\rad \m^{-2}]$ & $[\rad \m^{-2}]$ \\ 
		\hline
		J0738–4042& $+012.00 \pm 0.60$& $+011.97 \pm 0.04$& $0.03 \pm 0.60$ \\
		J0820–1350& $-001.20 \pm 0.40$& $-003.10 \pm 0.20$& $1.90 \pm 0.45$ \\
		J0907–5157& $-023.30 \pm 1.00$& $-025.90 \pm 0.10$& $2.60 \pm 1.00$ \\
		J0908–4913& $+010.00 \pm 1.60$& $+014.57 \pm 0.10$& $4.57 \pm 1.60$ \\
		J1243–6423& $+157.80 \pm 0.40$& $+161.93 \pm 0.10$& $4.13 \pm 0.41$ \\
		J1326–5859& $-579.60 \pm 0.90$& $-578.10 \pm 0.10$& $1.50 \pm 0.91$ \\
		J1359–6038& $+033.00 \pm 5.00$& $+038.50 \pm 0.10$& $5.50 \pm 5.00$ \\
		J1453–6413& $-018.60 \pm 0.20$& $-022.50 \pm 0.10$& $3.90 \pm 0.22$ \\
		J1456–6843& $-004.00 \pm 0.30$& $-000.90 \pm 0.10$& $3.10 \pm 0.32$ \\
		J1703–3241& $-021.70 \pm 0.50$& $-022.60 \pm 0.20$& $0.90 \pm 0.54$ \\
		J1745–3040& $+101.00 \pm 7.00$& $+096.00 \pm 0.10$& $5.00 \pm 7.00$ \\
		J1807–0847& $+166.00 \pm 9.00$& $+163.30 \pm 0.10$& $2.70 \pm 9.00$ \\
		\hline
		\hline
   \end{tabular}
\end{table}

$\RM$ from the pulsars is seen to vary across their pulse profile \citep{RamachandranEA2004, NoutsosEA2009, IlieJW2019}. These variations are only observed in a small sample of 42~pulsars \citep{IlieJW2019}. These variations can be due to the scattering of light by the ISM \citep{NoutsosEA2009} or intrinsic effects, such as the contribution from the pulsar magnetosphere \citep{IlieJW2019}. The contribution from each effect can vary between pulsars and it is usually difficult to disentangle various effects without a detailed study for each pulsar. However, if the ISM scattering is the dominant contribution, we would expect that the amplitude of $\RM$ variations will be correlated with the $\DM$ from the pulsar (since higher $\DM$ implies that the light from the pulsar either passes through a region with higher $\ne$ or moderate $\ne$ over large distances, both of which enhances scattering). Such a correlation is not seen \citep[Fig.~3 and Fig.~4 in][]{IlieJW2019}, and thus, the contribution from the pulsar magnetosphere can be significant. \citet{IlieJW2019} concluded that in 12 out of 42~pulsars with $\RM$ variations, the dominant contribution to the variation is from the magnetosphere. \Tab{tab:intrm} shows $\RM$ from the ATNF pulsar catalog \citep{ManchesterEA2005}, the mean of the $\RM$ variation across the pulse profile $\langle \RM (\phi) \rangle$ ($\phi$ is the phase) from the Table 1. in \citet{IlieJW2019}, and the computed difference between the two values $\delta(\RM)$ for those 12~pulsars. The mean and standard deviation of $\delta(\RM)$ are approximately $3 \rad \m^{-2}$ and $1.63 \rad \m^{-2}$, respectively. Thus, the intrinsic $\RM$ contribution is very small compared to that of the ISM. We also conclude that the intrinsic $\RM$ would have a negligible effect on the statistical properties of the Galactic magnetic fields estimated using $\RM$ and $\DM$ from pulsar observations.

\subsection{Similar concerns for Fast Radio Bursts (FRBs)} \label{sec:dis:frb}

\begin{figure}
	\includegraphics[width=\columnwidth]{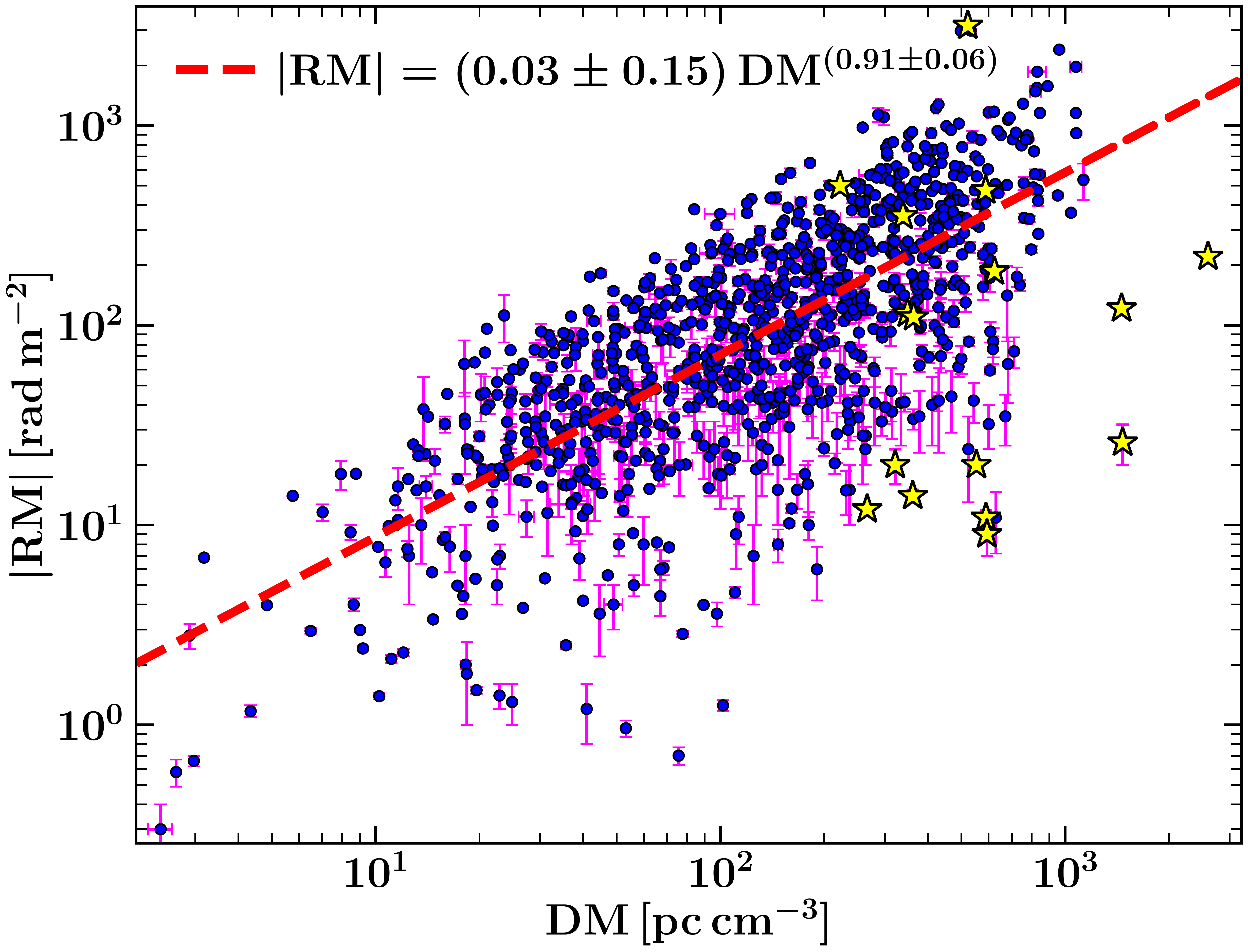} 
	\caption{Scatter plot of $|\RM|$ and $\DM$ (blue points) with their respective errors (in magenta) for 1018~pulsars from the ATNF pulsar catalog (same as in \Fig{fig:pulb}~(a)) and 16~Fast Radio Bursts (FRBs) (yellow stars with uncertainties in magenta) from the FRB catalog \citep{PetroffEA2016}. The red dashed line shows the best-fit relation between the two variables, $|\RM| = (0.03 \pm 0.15) \, \DM^{(0.91 \pm 0.06)}$, obtained with the pulsar data only. The FRB points are not completely inconsistent with the pulsar data trend. However, the FRB sample is too small to draw significant conclusions and it is also difficult to disentangle contributions to $\RM$ and $\DM$ of FRBs from various components along the path length (source, host, intergalactic medium, intervening galaxies, Milky Way, and observational uncertainties). 
	}
	\label{fig:frb}
\end{figure}

Fast Radio Bursts (FRBs) are bright ($50 \mJy$ -- $100 \Jy$) millisecond (or even shorter) pulses of radio emission and though their origin is unclear, it is known that they are from extragalactic sources as their observed dispersion measure exceeds the Milky Way contribution \citep{PetroffHL2019, CordesC2019}. Very much like pulsars, the dispersion measure $\DM$ and the rotation measure $\RM$ can be obtained from the observations of FRBs and can be used to probe extragalactic magnetic fields \citep{RaviEA2016, AkahoriRG2016, VazzaEA2018b, HacksteinEA2019}.

We use the FRB catalog given in \citet{PetroffEA2016} (updated version available at \href{http://www.frbcat.org}{http://www.frbcat.org}) to explore the correlation between the observed $|\RM|$ and $\DM$ from FRBs (similar to \Fig{fig:pulb}~(a)). Out of the 129~FRBs in the catalog, $\RM$ measurements are available for 19 of those. Similar to the pulsar data selection criteria, we chose those FRBs for which $\RM$ and $\DM$ are at least twice the reported uncertainty, which is the case for 16 of the 19~FRBs. \Fig{fig:frb} shows a scatter plot of $|\RM|$ and $\DM$ for these 16~FRBs (yellow stars) overlaid on the pulsar data (blue points). The red dashed line shows the best-fit line obtained from the pulsar data. The FRB points are not inconsistent with the trend obtained using the pulsar data set, but the scatter is significant. \rev{Even though the path length of the light from FRBs is much larger than that of the pulsars in the Milky Way, the average parallel component of the magnetic fields estimated via \Eq{eq:obsb} for some of the FRBs is seen to be smaller than that of the pulsars (yellow stars below the blue points and also below the red dashed line in \Fig{fig:frb}). This is probably due to magnetic field reversals along the path length.} 

It is difficult to conclude much from this, due to the small sample size of the existing FRBs with $\RM$ measurements in comparison to the pulsars, and also because the $\RM$ and $\DM$ of FRBs contain contributions from various regions along the path length. The $\RM$ and $\DM$ of a FRB would probably have contributions from the following components: source, host, intergalactic medium, intervening galaxies, Milky Way, and the uncertainties of the observing telescope. \rev{In fact, a significant fraction of $\RM$ and $\DM$ from FRBs is possibly due to the host galaxy \citep{CordesEA2016}.} Thus, it becomes difficult to isolate the contribution of each component. However, if there exists a correlation (or equivalently anti-correlation) between thermal electron density and magnetic fields in any of those components (for example, possibly a positive correlation is expected in the host and intervening galaxies), it must be taken into account when using FRBs as probes of extragalactic magnetic fields. We would expect that with the existing, upcoming, and future FRB surveys using telescopes such as the Australian Square Kilometre Array Pathfinder \citep[ASKAP;][]{BannisterEA2017}, Canadian Hydrogen Intensity Mapping Experiment \citep[CHIME;][]{CHIME2018}, upgraded Molonglo telescope \citep[UTMOST;][]{CalebEA2017}, and eventually Square Kilometre Array \citep[SKA;][]{FialkovL2017}, a large number of FRBs will be detected and $\RM$s for a large population of those will be determined. Then the question of the importance of $\ne$--$B$ correlations in extracting extragalactic magnetic field information from FRBs can be studied in much greater detail.

\section{Conclusions} \label{sec:con}
Using numerical simulations and observations, we comprehensively studied the correlation between the thermal election density ($\ne$) and the magnetic field ($B$), and the effect of such a correlation on the dispersion and rotation measures, $\DM$ and $\RM$ of pulsars and the estimated average parallel component of the Galactic magnetic field from them, $\obsb$. This study tests the validity of \Eq{eq:obsb}. We used non-ideal (prescribed viscosity and resistivity) numerical simulations of driven MHD turbulence initialised with a weak random magnetic field having mean zero, established that the Mach number $\Mach$ controls the correlation of the thermal electron density and small-scale magnetic fields ($b$), and then compared the computed $\obsb$ and true $\bpar$ in the simulations. We also used $\DM$, $\RM$, and $\obsb$ of pulsars taken from the ATNF pulsar catalog \citep{ManchesterEA2005} and correlated them with the local (star-forming regions traced by $\CO$ with data provided by \citet{MivilleME2017} and magnetic fields in the dense CNM probed by Zeeman splitting $\rm B_{\parallel} (ZE)$ using data from \citet{HeilesT2004}) and all sky (neutral hydrogen column density $\NHI$ data from \citet{HI4PI2016}, and ionised hydrogen intensity $\IHa$ data from \citet{Finkbeiner2003}) probes of the enhanced thermal electron density and/or magnetic fields along the path of light from the pulsars to us. To try and study the correlation between thermal electron density and magnetic fields for Fast Radio Bursts (FRBs), as done for pulsars, we also use the FRB catalog \citep{PetroffEA2016}. Based on our results, we arrive at the following conclusions:

\begin{itemize}
	\item In simulations, the seeded random magnetic field evolves, magnetic energy grows exponentially, it becomes strong enough to back react on the turbulent flow, and then it eventually reaches a statistically steady state (\Fig{fig:ts}~(a) and \Tab{tab:corr}). The growth rate in the exponential phase and the saturated level (the ratio of the magnetic to kinetic energy) decreases as the Mach number $\Mach$ of the turbulent flow increases. The correlation between $\ne$ and the small-scale random magnetic fields $b$ is largely controlled by $\Mach$. The correlation decreases when the field strength becomes significant to back react on the flow due to locally enhanced magnetic pressure (\Fig{fig:ts}~(b)). But still, the overall correlation remains negative for subsonic turbulence and positive for transsonic and supersonic turbulence. The correlation coefficient increases as the Mach number increases (column~5 in \Tab{tab:corr}).

	\item The probability density function (PDF) of the simulated $\RM$ is Gaussian for the subsonic case and is non-Gaussian for higher Mach numbers ($\Mach=2, 5, 10$), with a long tail of the distribution at higher $\RM$ becoming heavier (more probable) with increasing Mach number (\Fig{fig:rmdm}~(a) and column~3 in \Tab{tab:stats}). Thus, at higher values of $\Mach$, $\RM$ cannot be described completely by the mean and standard deviation of the distribution and higher-order moments are required. $\DM$ is approximately constant for the $\Mach=0.1$ case since the fluctuations in $\ne$ are negligible (\Fig{fig:struc3d}~(a)), but $\DM$ varies significantly as the Mach number increases (\Fig{fig:rmdm}~(b) and the columns 4 and 5 in \Tab{tab:stats}).

	\item The PDF of $\obsb$ from the simulations is similar to the true $\bpar$ PDF only for subsonic and transsonic cases (\Fig{fig:obsbbpar}). For higher Mach numbers ($\Mach=5, 10$), the PDF of $\obsb$ is non-Gaussian, whereas the $\bpar$ PDF is Gaussian for all Mach numbers. Both the standard deviation and kurtosis of the $\obsb$ distribution for supersonic runs ($\Mach=5, 10$) are significantly higher than that of the $\bpar$ distribution (columns~6--9 of \Tab{tab:stats}). Thus, the \reva{local} correlation between \reva{ the thermal electron density and the magnetic field} leads to an overestimation of the average parallel component of the Galactic magnetic field when calculated using $\RM$ and $\DM$ via \Eq{eq:obsb}. The average field strengths, obtained using \Eq{eq:obsb} can be overestimated by a factor as high as three in the case of $\Mach=10$ (\Fig{fig:obsbbpar2d}).

	\item After quantifying the effect of the correlation between thermal electron density and small-scale magnetic fields using simulations, we use $\DM$ and $\RM$ of pulsars from the ATNF pulsar catalog and show that $|\RM|$ is proportional to $\DM$ (\Fig{fig:pulb}~(a) and \Fig{fig:pulbgp}~(a)). This confirms that $\langle \ne \rangle$ contributes significantly to the observed $\RM$, and the $\RM$ value is not completely dominated by the Galactic magnetic field. We then show that the PDF of $\obsb$ from pulsar observations is very close to a Gaussian distribution (\Fig{fig:pulb}~(b) and \Fig{fig:pulbgp}~(b)). This confirms that the ISM of the Milky Way, \rev{as probed by pulsars}, is \reva{probably} not supersonic over the scales comparable to the distance to the pulsar, which are roughly $\kpc$ scales for the sample.

	\item We study the effect of the locally enhanced $\ne$--$B$ correlation along the path length of the pulsars by dividing the sample into two groups: one for which the line of sight passes through one or more star-forming molecular clouds as per $\CO$ data and another for which it does not. The PDF of $\obsb$ for pulsars with $\CO$ along their lines of sight has a slightly higher mean and standard deviation than that without $\CO$ data (\Fig{fig:pulbCO}), but the difference is not statistically significant. Moreover, the distribution for both samples is very close to a Gaussian distribution and the kurtosis of all three distributions (pulsar sample from the ATNF catalog, pulsar sample with $\CO$, and pulsar sample without $\CO$) are very similar (column~6 in \Tab{tab:pulbCO}). This shows that localised correlated $\ne$--$B$ structures contribute very little to $\obsb$, and that \revb{local} $\ne$ and $B$ are largely uncorrelated over such $\kpc$ scales. 
	
	We also perform a similar analysis, but with pulsar lines of sight passing through localised regions of the CNM. However, due to the small sample size and large uncertainties, it is difficult to conclude much from this analysis (\Fig{fig:pulbZeeman}).

	\item Finally, we correlate $\obsb$ from pulsars with indirect probes of $\ne$, i.e., with $\NHI$ and H$\alpha$. We find that $\DM$ and $|\RM|$ from pulsars are positively correlated with both the probes from $\NHI$ and $\IHa$ but the average parallel component of the Galactic magnetic field $\obsb$ is statistically uncorrelated with any of them. Thus, we again conclude that \revb{local} $\ne$ and $B$ are primarily uncorrelated on $\kpc$ scales. However, the $\ne$--$B$ correlation and its effect on the magnetic field estimate might be important on smaller, sub-kpc scales.
	
	\item Based on our results and conclusions, we also discuss the following additional aspects of the study: lack of significant anti-correlation between $\ne$ and $B$ from our observational results (indirectly expected from a weaker mean-field dynamo in the gaseous spiral arms, magnetic fields drifting away from the gaseous arms, and observations of magnetic fields in external spiral galaxies), negligible intrinsic $\RM$ of pulsars, and possible effects of $\ne$--$B$ correlations in extracting extragalactic magnetic fields from FRBs.
\end{itemize}

\section*{Acknowledgements}
\rev{We thank our referee, Rainer Beck, for very useful comments and suggestions.} A.~S.~is grateful to Naomi McClure-Griffiths for discussions regarding the HI data, to Aris Tritsis for discussions regarding the Zeeman splitting, and to Aritra Basu, Aristeidis Noutsos, Nataliya Porayko, and Patrick Weltevrede for discussions regarding the intrinsic contribution of the pulsar to the observed $\RM$. C.~F.~acknowledges funding provided by the Australian Research Council (Discovery Project DP170100603 and Future Fellowship FT180100495), and the Australia-Germany Joint Research Cooperation Scheme (UA-DAAD). We further acknowledge high-performance computing resources provided by the Leibniz Rechenzentrum and the Gauss Centre for Supercomputing (grants~pr32lo, pr48pi and GCS Large-scale project~10391), and the Australian National Computational Infrastructure (grant~ek9) in the framework of the National Computational Merit Allocation Scheme and the ANU Merit Allocation Scheme. 

\section*{Data availability}
The data from the simulations used in the \Sec{sec:simres} is available upon a resonable request to the corresponding author, Amit Seta (\href{mailto:Amit.seta@anu.adu.au}{amit.seta@anu.adu.au}).

The observational data used in \Sec{sec:obs}, \Sec{sec:dis}, and \App{app:m51} are publicly available. The data used in \Sec{sec:obs} are taken from the ATNF pulsar catalog in \cite{ManchesterEA2005} (version~1.63: \href{http://www.atnf.csiro.au/research/pulsar/psrcat}{http://www.atnf.csiro.au/research/pulsar/psrcat}) for the pulsar $\RM$ and $\DM$ data, \cite{MivilleME2017} for the $\CO$ data, \cite{HeilesT2004} for the magnetic field in the CNM obtained from Zeeman splitting of the 21~cm line, \cite{HI4PI2016} for the $\NHI$ data, and \cite{Finkbeiner2003} for the $\IHa$ data. The data used in \Sec{sec:dis} are taken from \cite{IlieJW2019} for the $\RM$ variations across the pulsar profiles and the FRB catalog in \cite{PetroffEA2016} (\href{http://www.frbcat.org}{http://www.frbcat.org}) for $\RM$ and $\DM$ from FRBs. \rev{The $\RM$ data for M51 used in \App{app:m51} are taken from \cite{KierdorfEA2020}.}



\bibliographystyle{mnras}
\bibliography{pul} 

\begin{thebibliography}{}
\makeatletter
\relax
\def\mn@urlcharsother{\let\do\@makeother \do\$\do\&\do\#\do\^\do\_\do\%\do\~}
\def\mn@doi{\begingroup\mn@urlcharsother \@ifnextchar [ {\mn@doi@}
  {\mn@doi@[]}}
\def\mn@doi@[#1]#2{\def\@tempa{#1}\ifx\@tempa\@empty \href
  {http://dx.doi.org/#2} {doi:#2}\else \href {http://dx.doi.org/#2} {#1}\fi
  \endgroup}
\def\mn@eprint#1#2{\mn@eprint@#1:#2::\@nil}
\def\mn@eprint@arXiv#1{\href {http://arxiv.org/abs/#1} {{\tt arXiv:#1}}}
\def\mn@eprint@dblp#1{\href {http://dblp.uni-trier.de/rec/bibtex/#1.xml}
  {dblp:#1}}
\def\mn@eprint@#1:#2:#3:#4\@nil{\def\@tempa {#1}\def\@tempb {#2}\def\@tempc
  {#3}\ifx \@tempc \@empty \let \@tempc \@tempb \let \@tempb \@tempa \fi \ifx
  \@tempb \@empty \def\@tempb {arXiv}\fi \@ifundefined
  {mn@eprint@\@tempb}{\@tempb:\@tempc}{\expandafter \expandafter \csname
  mn@eprint@\@tempb\endcsname \expandafter{\@tempc}}}

\bibitem[\protect\citeauthoryear{{Akahori}, {Ryu}  \& {Gaensler}}{{Akahori}
  et~al.}{2016}]{AkahoriRG2016}
{Akahori} T.,  {Ryu} D.,   {Gaensler} B.~M.,  2016, \mn@doi [\apj]
  {10.3847/0004-637X/824/2/105}, \href
  {https://ui.adsabs.harvard.edu/abs/2016ApJ...824..105A} {824, 105}

\bibitem[\protect\citeauthoryear{{Bannister} et~al.,}{{Bannister}
  et~al.}{2017}]{BannisterEA2017}
{Bannister} K.~W.,  et~al., 2017, \mn@doi [\apjl] {10.3847/2041-8213/aa71ff},
  \href {https://ui.adsabs.harvard.edu/abs/2017ApJ...841L..12B} {841, L12}

\bibitem[\protect\citeauthoryear{{Beattie}, {Federrath}  \& {Seta}}{{Beattie}
  et~al.}{2020}]{BeattieFS2020}
{Beattie} J.~R.,  {Federrath} C.,   {Seta} A.,  2020, \mn@doi [\mnras]
  {10.1093/mnras/staa2257}, \href
  {https://ui.adsabs.harvard.edu/abs/2020MNRAS.498.1593B} {498, 1593}

\bibitem[\protect\citeauthoryear{{Beck}}{{Beck}}{2015}]{Beck2015}
{Beck} R.,  2015, \mn@doi [\aap] {10.1051/0004-6361/201425572}, \href
  {https://ui.adsabs.harvard.edu/abs/2015A&A...578A..93B} {578, A93}

\bibitem[\protect\citeauthoryear{{Beck}}{{Beck}}{2016}]{Beck2016}
{Beck} R.,  2016, \araa, \href
  {http://adsabs.harvard.edu/abs/2016A%26ARv..24....4B} {24, 4}

\bibitem[\protect\citeauthoryear{{Beck}, {Shukurov}, {Sokoloff}  \&
  {Wielebinski}}{{Beck} et~al.}{2003}]{BeckEA2003}
{Beck} R.,  {Shukurov} A.,  {Sokoloff} D.,   {Wielebinski} R.,  2003, \mn@doi
  [\aap] {10.1051/0004-6361:20031101}, \href
  {http://adsabs.harvard.edu/abs/2003A%26A...411...99B} {411, 99}

\bibitem[\protect\citeauthoryear{{Berkhuijsen} \& {M{\"u}ller}}{{Berkhuijsen}
  \& {M{\"u}ller}}{2008}]{BerkhuijsenM2008}
{Berkhuijsen} E.~M.,  {M{\"u}ller} P.,  2008, \mn@doi [\aap]
  {10.1051/0004-6361:200809675}, \href
  {https://ui.adsabs.harvard.edu/abs/2008A&A...490..179B} {490, 179}

\bibitem[\protect\citeauthoryear{{Bhat} \& {Subramanian}}{{Bhat} \&
  {Subramanian}}{2013}]{BhatS2013}
{Bhat} P.,  {Subramanian} K.,  2013, \mn@doi [\mnras] {10.1093/mnras/sts516},
  \href {https://ui.adsabs.harvard.edu/#abs/2013MNRAS.429.2469B} {429, 2469}

\bibitem[\protect\citeauthoryear{Bouchut, Klingenberg  \& Waagan}{Bouchut
  et~al.}{2007}]{BouchutKW2007}
Bouchut F.,  Klingenberg C.,   Waagan K.,  2007, Numerische Mathematik, 108, 7

\bibitem[\protect\citeauthoryear{Bouchut, Klingenberg  \& Waagan}{Bouchut
  et~al.}{2010}]{BouchutKW2010}
Bouchut F.,  Klingenberg C.,   Waagan K.,  2010, Numerische Mathematik, 115,
  647

\bibitem[\protect\citeauthoryear{{Boulares} \& {Cox}}{{Boulares} \&
  {Cox}}{1990}]{BoularesC1990}
{Boulares} A.,  {Cox} D.~P.,  1990, \mn@doi [\apj] {10.1086/169509}, \href
  {https://ui.adsabs.harvard.edu/abs/1990ApJ...365..544B} {365, 544}

\bibitem[\protect\citeauthoryear{{Brentjens} \& {de Bruyn}}{{Brentjens} \& {de
  Bruyn}}{2005}]{BrentjensB2005}
{Brentjens} M.~A.,  {de Bruyn} A.~G.,  2005, \mn@doi [\aap]
  {10.1051/0004-6361:20052990}, \href
  {https://ui.adsabs.harvard.edu/#abs/2005A&A...441.1217B} {441, 1217}

\bibitem[\protect\citeauthoryear{{Burkhart}, {Lazarian}  \&
  {Gaensler}}{{Burkhart} et~al.}{2012}]{BurkhartLG2012}
{Burkhart} B.,  {Lazarian} A.,   {Gaensler} B.~M.,  2012, \mn@doi [\apj]
  {10.1088/0004-637X/749/2/145}, \href
  {http://adsabs.harvard.edu/abs/2012ApJ...749..145B} {749, 145}

\bibitem[\protect\citeauthoryear{{CHIME/FRB Collaboration} et~al.,}{{CHIME/FRB
  Collaboration} et~al.}{2018}]{CHIME2018}
{CHIME/FRB Collaboration} et~al., 2018, \mn@doi [\apj]
  {10.3847/1538-4357/aad188}, \href
  {https://ui.adsabs.harvard.edu/abs/2018ApJ...863...48C} {863, 48}

\bibitem[\protect\citeauthoryear{{Caleb} et~al.,}{{Caleb}
  et~al.}{2017}]{CalebEA2017}
{Caleb} M.,  et~al., 2017, \mn@doi [\mnras] {10.1093/mnras/stx638}, \href
  {https://ui.adsabs.harvard.edu/abs/2017MNRAS.468.3746C} {468, 3746}

\bibitem[\protect\citeauthoryear{{Cesarsky}}{{Cesarsky}}{1980}]{Cesarsky1980}
{Cesarsky} C.~J.,  1980, \mn@doi [\araa] {10.1146/annurev.aa.18.090180.001445},
  \href {http://adsabs.harvard.edu/abs/1980ARA%26A..18..289C} {18, 289}

\bibitem[\protect\citeauthoryear{{Cordes} \& {Chatterjee}}{{Cordes} \&
  {Chatterjee}}{2019}]{CordesC2019}
{Cordes} J.~M.,  {Chatterjee} S.,  2019, \mn@doi [\araa]
  {10.1146/annurev-astro-091918-104501}, \href
  {https://ui.adsabs.harvard.edu/abs/2019ARA&A..57..417C} {57, 417}

\bibitem[\protect\citeauthoryear{{Cordes}, {Wharton}, {Spitler}, {Chatterjee}
  \& {Wasserman}}{{Cordes} et~al.}{2016}]{CordesEA2016}
{Cordes} J.~M.,  {Wharton} R.~S.,  {Spitler} L.~G.,  {Chatterjee} S.,
  {Wasserman} I.,  2016, arXiv e-prints, \href
  {https://ui.adsabs.harvard.edu/abs/2016arXiv160505890C} {p. arXiv:1605.05890}

\bibitem[\protect\citeauthoryear{{Crocker}, {Jones}, {Melia}, {Ott}  \&
  {Protheroe}}{{Crocker} et~al.}{2010}]{CrockerEA2010}
{Crocker} R.~M.,  {Jones} D.~I.,  {Melia} F.,  {Ott} J.,   {Protheroe} R.~J.,
  2010, \mn@doi [\nat] {10.1038/nature08635}, \href
  {https://ui.adsabs.harvard.edu/abs/2010Natur.463...65C} {463, 65}

\bibitem[\protect\citeauthoryear{{Crutcher} \& {Kemball}}{{Crutcher} \&
  {Kemball}}{2019}]{CrutcherK2019}
{Crutcher} R.~M.,  {Kemball} A.~J.,  2019, \mn@doi [Frontiers in Astronomy and
  Space Sciences] {10.3389/fspas.2019.00066}, \href
  {https://ui.adsabs.harvard.edu/abs/2019FrASS...6...66C} {6, 66}

\bibitem[\protect\citeauthoryear{{Crutcher}, {Wandelt}, {Heiles}, {Falgarone}
  \& {Troland}}{{Crutcher} et~al.}{2010}]{CrutcherEA2010}
{Crutcher} R.~M.,  {Wandelt} B.,  {Heiles} C.,  {Falgarone} E.,   {Troland}
  T.~H.,  2010, \mn@doi [\apj] {10.1088/0004-637X/725/1/466}, \href
  {https://ui.adsabs.harvard.edu/abs/2010ApJ...725..466C} {725, 466}

\bibitem[\protect\citeauthoryear{{Dickey} \& {Lockman}}{{Dickey} \&
  {Lockman}}{1990}]{DickeyL1990}
{Dickey} J.~M.,  {Lockman} F.~J.,  1990, \mn@doi [\araa]
  {10.1146/annurev.aa.28.090190.001243}, \href
  {https://ui.adsabs.harvard.edu/abs/1990ARA&A..28..215D} {28, 215}

\bibitem[\protect\citeauthoryear{{Dickey} et~al.,}{{Dickey}
  et~al.}{2019}]{DickeyEA2019}
{Dickey} J.~M.,  et~al., 2019, \mn@doi [\apj] {10.3847/1538-4357/aaf85f}, \href
  {https://ui.adsabs.harvard.edu/abs/2019ApJ...871..106D} {871, 106}

\bibitem[\protect\citeauthoryear{{Dubey} et~al.,}{{Dubey}
  et~al.}{2008}]{DubeyEA2008}
{Dubey} A.,  et~al., 2008, {Challenges of Extreme Computing using the FLASH
  code}.
p.~145

\bibitem[\protect\citeauthoryear{{Evirgen}, {Gent}, {Shukurov}, {Fletcher}  \&
  {Bushby}}{{Evirgen} et~al.}{2017}]{EvirgenEA17}
{Evirgen} C.~C.,  {Gent} F.~A.,  {Shukurov} A.,  {Fletcher} A.,   {Bushby} P.,
  2017, \mn@doi [\mnras] {10.1093/mnrasl/slw196}, \href
  {http://adsabs.harvard.edu/abs/2017MNRAS.464L.105E} {464, L105}

\bibitem[\protect\citeauthoryear{{Federrath}}{{Federrath}}{2016}]{Federrath2016}
{Federrath} C.,  2016, \mn@doi [Journal of Plasma Physics]
  {10.1017/S0022377816001069}, \href
  {https://ui.adsabs.harvard.edu/#abs/2016JPlPh..82f5301F} {82, 535820601}

\bibitem[\protect\citeauthoryear{{Federrath}}{{Federrath}}{2018}]{Federrath2018}
{Federrath} C.,  2018, \mn@doi [Physics Today] {10.1063/PT.3.3947}, \href
  {https://ui.adsabs.harvard.edu/abs/2018PhT....71f..38F} {71, 38}

\bibitem[\protect\citeauthoryear{{Federrath}, {Klessen}  \&
  {Schmidt}}{{Federrath} et~al.}{2008}]{FederrathEA2008}
{Federrath} C.,  {Klessen} R.~S.,   {Schmidt} W.,  2008, \mn@doi [\apjl]
  {10.1086/595280}, \href
  {https://ui.adsabs.harvard.edu/abs/2008ApJ...688L..79F} {688, L79}

\bibitem[\protect\citeauthoryear{{Federrath}, {Roman-Duval}, {Klessen},
  {Schmidt}  \& {Mac Low}}{{Federrath} et~al.}{2010}]{FederrathEA2010}
{Federrath} C.,  {Roman-Duval} J.,  {Klessen} R.~S.,  {Schmidt} W.,   {Mac Low}
  M.-M.,  2010, \mn@doi [\aap] {10.1051/0004-6361/200912437}, \href
  {http://adsabs.harvard.edu/abs/2010A%26A...512A..81F} {512, A81}

\bibitem[\protect\citeauthoryear{{Federrath}, {Chabrier}, {Schober},
  {Banerjee}, {Klessen}  \& {Schleicher}}{{Federrath}
  et~al.}{2011}]{FederrathEA2011}
{Federrath} C.,  {Chabrier} G.,  {Schober} J.,  {Banerjee} R.,  {Klessen}
  R.~S.,   {Schleicher} D.~R.~G.,  2011, \mn@doi [\prl]
  {10.1103/PhysRevLett.107.114504}, \href
  {https://ui.adsabs.harvard.edu/#abs/2011PhRvL.107k4504F} {107, 114504}

\bibitem[\protect\citeauthoryear{{Federrath}, {Schober}, {Bovino}  \&
  {Schleicher}}{{Federrath} et~al.}{2014}]{FederrathEA2014}
{Federrath} C.,  {Schober} J.,  {Bovino} S.,   {Schleicher} D. R.~G.,  2014,
  \mn@doi [\apj] {10.1088/2041-8205/797/2/L19}, \href
  {https://ui.adsabs.harvard.edu/#abs/2014ApJ...797L..19F} {797, L19}

\bibitem[\protect\citeauthoryear{{Federrath} et~al.,}{{Federrath}
  et~al.}{2016}]{FederrathEA2016}
{Federrath} C.,  et~al., 2016, \mn@doi [\apj] {10.3847/0004-637X/832/2/143},
  \href {https://ui.adsabs.harvard.edu/abs/2016ApJ...832..143F} {832, 143}

\bibitem[\protect\citeauthoryear{{Federrath}, {Klessen}, {Iapichino}  \&
  {Beattie}}{{Federrath} et~al.}{2020}]{FederrathEA2020}
{Federrath} C.,  {Klessen} R.~S.,  {Iapichino} L.,   {Beattie} J.~R.,  2020,
  arXiv e-prints, \href {https://ui.adsabs.harvard.edu/abs/2020arXiv201106238F}
  {p. arXiv:2011.06238}

\bibitem[\protect\citeauthoryear{{Fialkov} \& {Loeb}}{{Fialkov} \&
  {Loeb}}{2017}]{FialkovL2017}
{Fialkov} A.,  {Loeb} A.,  2017, \mn@doi [\apjl] {10.3847/2041-8213/aa8905},
  \href {https://ui.adsabs.harvard.edu/abs/2017ApJ...846L..27F} {846, L27}

\bibitem[\protect\citeauthoryear{{Finkbeiner}}{{Finkbeiner}}{2003}]{Finkbeiner2003}
{Finkbeiner} D.~P.,  2003, \mn@doi [\apjs] {10.1086/374411}, \href
  {https://ui.adsabs.harvard.edu/abs/2003ApJS..146..407F} {146, 407}

\bibitem[\protect\citeauthoryear{{Fosalba}, {Lazarian}, {Prunet}  \&
  {Tauber}}{{Fosalba} et~al.}{2002}]{FosalbaEA2002}
{Fosalba} P.,  {Lazarian} A.,  {Prunet} S.,   {Tauber} J.~A.,  2002, \mn@doi
  [\apj] {10.1086/324297}, \href
  {https://ui.adsabs.harvard.edu/abs/2002ApJ...564..762F} {564, 762}

\bibitem[\protect\citeauthoryear{{Fryxell} et~al.,}{{Fryxell}
  et~al.}{2000}]{FryxellEA2000}
{Fryxell} B.,  et~al., 2000, \mn@doi [\apjs] {10.1086/317361}, \href
  {https://ui.adsabs.harvard.edu/abs/2000ApJS..131..273F} {131, 273}

\bibitem[\protect\citeauthoryear{{Gaensler}, {Haverkorn}, {Staveley-Smith},
  {Dickey}, {McClure-Griffiths}, {Dickel}  \& {Wolleben}}{{Gaensler}
  et~al.}{2005}]{Gaenslar2005}
{Gaensler} B.~M.,  {Haverkorn} M.,  {Staveley-Smith} L.,  {Dickey} J.~M.,
  {McClure-Griffiths} N.~M.,  {Dickel} J.~R.,   {Wolleben} M.,  2005, \mn@doi
  [Science] {10.1126/science.1108832}, \href
  {http://adsabs.harvard.edu/abs/2005Sci...307.1610G} {307, 1610}

\bibitem[\protect\citeauthoryear{{Gaensler}, {Madsen}, {Chatterjee}  \&
  {Mao}}{{Gaensler} et~al.}{2008}]{GaenslerEA2008}
{Gaensler} B.~M.,  {Madsen} G.~J.,  {Chatterjee} S.,   {Mao} S.~A.,  2008,
  \mn@doi [\pasa] {10.1071/AS08004}, \href
  {https://ui.adsabs.harvard.edu/abs/2008PASA...25..184G} {25, 184}

\bibitem[\protect\citeauthoryear{{Gaensler} et~al.,}{{Gaensler}
  et~al.}{2011}]{Gaenslar2011}
{Gaensler} B.~M.,  et~al., 2011, \mn@doi [\nat] {10.1038/nature10446}, \href
  {http://adsabs.harvard.edu/abs/2011Natur.478..214G} {478, 214}

\bibitem[\protect\citeauthoryear{{Grete}, {O'Shea}  \& {Beckwith}}{{Grete}
  et~al.}{2020}]{GreteOB2020}
{Grete} P.,  {O'Shea} B.~W.,   {Beckwith} K.,  2020, \mn@doi [\apj]
  {10.3847/1538-4357/ab5aec}, \href
  {https://ui.adsabs.harvard.edu/abs/2020ApJ...889...19G} {889, 19}

\bibitem[\protect\citeauthoryear{{HI4PI Collaboration} et~al.,}{{HI4PI
  Collaboration} et~al.}{2016}]{HI4PI2016}
{HI4PI Collaboration} et~al., 2016, \mn@doi [\aap]
  {10.1051/0004-6361/201629178}, \href
  {https://ui.adsabs.harvard.edu/abs/2016A&A...594A.116H} {594, A116}

\bibitem[\protect\citeauthoryear{{Hackstein}, {Br{\"u}ggen}, {Vazza},
  {Gaensler}  \& {Heesen}}{{Hackstein} et~al.}{2019}]{HacksteinEA2019}
{Hackstein} S.,  {Br{\"u}ggen} M.,  {Vazza} F.,  {Gaensler} B.~M.,   {Heesen}
  V.,  2019, \mn@doi [\mnras] {10.1093/mnras/stz2033}, \href
  {https://ui.adsabs.harvard.edu/abs/2019MNRAS.488.4220H} {488, 4220}

\bibitem[\protect\citeauthoryear{{Han}, {Manchester}  \& {Qiao}}{{Han}
  et~al.}{1999}]{HanMQ1999}
{Han} J.~L.,  {Manchester} R.~N.,   {Qiao} G.~J.,  1999, \mn@doi [\mnras]
  {10.1046/j.1365-8711.1999.02544.x}, \href
  {https://ui.adsabs.harvard.edu/abs/1999MNRAS.306..371H} {306, 371}

\bibitem[\protect\citeauthoryear{{Han}, {Manchester}, {Lyne}, {Qiao}  \& {van
  Straten}}{{Han} et~al.}{2006}]{HanEA2006}
{Han} J.~L.,  {Manchester} R.~N.,  {Lyne} A.~G.,  {Qiao} G.~J.,   {van Straten}
  W.,  2006, \mn@doi [\apj] {10.1086/501444}, \href
  {https://ui.adsabs.harvard.edu/abs/2006ApJ...642..868H} {642, 868}

\bibitem[\protect\citeauthoryear{{Han}, {Manchester}, {van Straten}  \&
  {Demorest}}{{Han} et~al.}{2018}]{HanEA2018}
{Han} J.~L.,  {Manchester} R.~N.,  {van Straten} W.,   {Demorest} P.,  2018,
  \mn@doi [\apjs] {10.3847/1538-4365/aa9c45}, \href
  {https://ui.adsabs.harvard.edu/abs/2018ApJS..234...11H} {234, 11}

\bibitem[\protect\citeauthoryear{{Haugen}, {Brandenburg}  \& {Mee}}{{Haugen}
  et~al.}{2004}]{HaugenBM2004}
{Haugen} N. E.~L.,  {Brandenburg} A.,   {Mee} A.~J.,  2004, \mn@doi [\mnras]
  {10.1111/j.1365-2966.2004.08127.x}, \href
  {https://ui.adsabs.harvard.edu/#abs/2004MNRAS.353..947H} {353, 947}

\bibitem[\protect\citeauthoryear{{Haverkorn}}{{Haverkorn}}{2015}]{Haverkorn2015}
{Haverkorn} M.,  2015, in {Lazarian} A.,  {de Gouveia Dal Pino} E.~M.,
  {Melioli} C.,  eds,  Astrophysics and Space Science Library Vol. 407,
  Magnetic Fields in Diffuse Media. p.~483 (\mn@eprint {arXiv} {1406.0283}),
  \mn@doi{10.1007/978-3-662-44625-6_17}

\bibitem[\protect\citeauthoryear{{Haverkorn}, {Katgert}  \& {de
  Bruyn}}{{Haverkorn} et~al.}{2004}]{HaverkornKdB2004}
{Haverkorn} M.,  {Katgert} P.,   {de Bruyn} A.~G.,  2004, \mn@doi [\aap]
  {10.1051/0004-6361:200400051}, \href
  {http://adsabs.harvard.edu/abs/2004A%26A...427..549H} {427, 549}

\bibitem[\protect\citeauthoryear{{Heiles} \& {Troland}}{{Heiles} \&
  {Troland}}{2004}]{HeilesT2004}
{Heiles} C.,  {Troland} T.~H.,  2004, \mn@doi [\apjs] {10.1086/381753}, \href
  {https://ui.adsabs.harvard.edu/abs/2004ApJS..151..271H} {151, 271}

\bibitem[\protect\citeauthoryear{{Hewish}, {Bell}, {Pilkington}, {Scott}  \&
  {Collins}}{{Hewish} et~al.}{1968}]{HewishEA1968}
{Hewish} A.,  {Bell} S.~J.,  {Pilkington} J.~D.~H.,  {Scott} P.~F.,   {Collins}
  R.~A.,  1968, \mn@doi [\nat] {10.1038/217709a0}, \href
  {https://ui.adsabs.harvard.edu/abs/1968Natur.217..709H} {217, 709}

\bibitem[\protect\citeauthoryear{{Ilie}, {Johnston}  \& {Weltevrede}}{{Ilie}
  et~al.}{2019}]{IlieJW2019}
{Ilie} C.~D.,  {Johnston} S.,   {Weltevrede} P.,  2019, \mn@doi [\mnras]
  {10.1093/mnras/sty3315}, \href
  {https://ui.adsabs.harvard.edu/abs/2019MNRAS.483.2778I} {483, 2778}

\bibitem[\protect\citeauthoryear{{Indrani} \& {Deshpande}}{{Indrani} \&
  {Deshpande}}{1999}]{IndraniD1999}
{Indrani} C.,  {Deshpande} A.~A.,  1999, \mn@doi [\na]
  {10.1016/S1384-1076(98)00038-4}, \href
  {https://ui.adsabs.harvard.edu/abs/1999NewA....4...33I} {4, 33}

\bibitem[\protect\citeauthoryear{{Kainulainen} \& {Federrath}}{{Kainulainen} \&
  {Federrath}}{2017}]{KainulainenF2017}
{Kainulainen} J.,  {Federrath} C.,  2017, \mn@doi [\aap]
  {10.1051/0004-6361/201731028}, \href
  {https://ui.adsabs.harvard.edu/abs/2017A&A...608L...3K} {608, L3}

\bibitem[\protect\citeauthoryear{{Kalberla}, {Kerp}, {Haud}  \&
  {Haverkorn}}{{Kalberla} et~al.}{2017}]{KalberlaEA2017}
{Kalberla} P.~M.~W.,  {Kerp} J.,  {Haud} U.,   {Haverkorn} M.,  2017, \mn@doi
  [\aap] {10.1051/0004-6361/201629627}, \href
  {https://ui.adsabs.harvard.edu/#abs/2017A&A...607A..15K} {607, A15}

\bibitem[\protect\citeauthoryear{{Kierdorf} et~al.,}{{Kierdorf}
  et~al.}{2020}]{KierdorfEA2020}
{Kierdorf} M.,  et~al., 2020, \mn@doi [\aap] {10.1051/0004-6361/202037847},
  \href {https://ui.adsabs.harvard.edu/abs/2020A&A...642A.118K} {642, A118}

\bibitem[\protect\citeauthoryear{{Klein} \& {Fletcher}}{{Klein} \&
  {Fletcher}}{2015}]{KleinFletcher2015}
{Klein} U.,  {Fletcher} A.,  2015, {Galactic and Intergalactic Magnetic
  Fields}.
{Springer Praxis Books, Springer International Publishing, Heidelberg, Germany}

\bibitem[\protect\citeauthoryear{{Konstandin}, {Girichidis}, {Federrath}  \&
  {Klessen}}{{Konstandin} et~al.}{2012}]{KonstandinEA2012}
{Konstandin} L.,  {Girichidis} P.,  {Federrath} C.,   {Klessen} R.~S.,  2012,
  \mn@doi [\apj] {10.1088/0004-637X/761/2/149}, \href
  {https://ui.adsabs.harvard.edu/abs/2012ApJ...761..149K} {761, 149}

\bibitem[\protect\citeauthoryear{{Krumholz} \& {Federrath}}{{Krumholz} \&
  {Federrath}}{2019}]{KrumholzF2019}
{Krumholz} M.~R.,  {Federrath} C.,  2019, \mn@doi [Frontiers in Astronomy and
  Space Sciences] {10.3389/fspas.2019.00007}, \href
  {https://ui.adsabs.harvard.edu/abs/2019FrASS...6....7K} {6, 7}

\bibitem[\protect\citeauthoryear{{Lyne} \& {Smith}}{{Lyne} \&
  {Smith}}{1989}]{LyneS1989}
{Lyne} A.~G.,  {Smith} F.~G.,  1989, \mn@doi [\mnras]
  {10.1093/mnras/237.3.533}, \href
  {https://ui.adsabs.harvard.edu/abs/1989MNRAS.237..533L} {237, 533}

\bibitem[\protect\citeauthoryear{{Manchester}}{{Manchester}}{1972}]{Manchester1972}
{Manchester} R.~N.,  1972, \mn@doi [\apj] {10.1086/151326}, \href
  {https://ui.adsabs.harvard.edu/abs/1972ApJ...172...43M} {172, 43}

\bibitem[\protect\citeauthoryear{{Manchester}}{{Manchester}}{1974}]{Manchester1974}
{Manchester} R.~N.,  1974, \mn@doi [\apj] {10.1086/152757}, \href
  {https://ui.adsabs.harvard.edu/abs/1974ApJ...188..637M} {188, 637}

\bibitem[\protect\citeauthoryear{{Manchester}, {Hobbs}, {Teoh}  \&
  {Hobbs}}{{Manchester} et~al.}{2005}]{ManchesterEA2005}
{Manchester} R.~N.,  {Hobbs} G.~B.,  {Teoh} A.,   {Hobbs} M.,  2005, \mn@doi
  [\aj] {10.1086/428488}, \href
  {https://ui.adsabs.harvard.edu/abs/2005AJ....129.1993M} {129, 1993}

\bibitem[\protect\citeauthoryear{{Mao}, {Gaensler}, {Haverkorn}, {Zweibel},
  {Madsen}, {McClure-Griffiths}, {Shukurov}  \& {Kronberg}}{{Mao}
  et~al.}{2010}]{MaoEA2010}
{Mao} S.~A.,  {Gaensler} B.~M.,  {Haverkorn} M.,  {Zweibel} E.~G.,  {Madsen}
  G.~J.,  {McClure-Griffiths} N.~M.,  {Shukurov} A.,   {Kronberg} P.~P.,  2010,
  \mn@doi [\apj] {10.1088/0004-637X/714/2/1170}, \href
  {https://ui.adsabs.harvard.edu/abs/2010ApJ...714.1170M} {714, 1170}

\bibitem[\protect\citeauthoryear{Martins~Afonso, Mitra  \&
  Vincenzi}{Martins~Afonso et~al.}{2019}]{AfonsoMV2019}
Martins~Afonso M.,  Mitra D.,   Vincenzi D.,  2019, \mn@doi [Proceedings of the
  Royal Society A: Mathematical, Physical and Engineering Sciences]
  {10.1098/rspa.2018.0591}, 475, 20180591

\bibitem[\protect\citeauthoryear{{Mestel} \& {Spitzer}}{{Mestel} \&
  {Spitzer}}{1956}]{MestelS1956}
{Mestel} L.,  {Spitzer} L. J.,  1956, \mn@doi [\mnras]
  {10.1093/mnras/116.5.503}, \href
  {https://ui.adsabs.harvard.edu/abs/1956MNRAS.116..503M} {116, 503}

\bibitem[\protect\citeauthoryear{{Mitra}, {Wielebinski}, {Kramer}  \&
  {Jessner}}{{Mitra} et~al.}{2003}]{MitraEA2003}
{Mitra} D.,  {Wielebinski} R.,  {Kramer} M.,   {Jessner} A.,  2003, \mn@doi
  [\aap] {10.1051/0004-6361:20021702}, \href
  {https://ui.adsabs.harvard.edu/abs/2003A&A...398..993M} {398, 993}

\bibitem[\protect\citeauthoryear{{Miville-Desch{\^e}nes}, {Murray}  \&
  {Lee}}{{Miville-Desch{\^e}nes} et~al.}{2017}]{MivilleME2017}
{Miville-Desch{\^e}nes} M.-A.,  {Murray} N.,   {Lee} E.~J.,  2017, \mn@doi
  [\apj] {10.3847/1538-4357/834/1/57}, \href
  {https://ui.adsabs.harvard.edu/abs/2017ApJ...834...57M} {834, 57}

\bibitem[\protect\citeauthoryear{{Noutsos}, {Karastergiou}, {Kramer},
  {Johnston}  \& {Stappers}}{{Noutsos} et~al.}{2009}]{NoutsosEA2009}
{Noutsos} A.,  {Karastergiou} A.,  {Kramer} M.,  {Johnston} S.,   {Stappers}
  B.~W.,  2009, \mn@doi [\mnras] {10.1111/j.1365-2966.2009.14806.x}, \href
  {https://ui.adsabs.harvard.edu/abs/2009MNRAS.396.1559N} {396, 1559}

\bibitem[\protect\citeauthoryear{{On}, {Chan}, {Wu}, {Saxton}  \& {van
  Driel-Gesztelyi}}{{On} et~al.}{2019}]{OnEA2019}
{On} A. Y.~L.,  {Chan} J. Y.~H.,  {Wu} K.,  {Saxton} C.~J.,   {van
  Driel-Gesztelyi} L.,  2019, \mn@doi [\mnras] {10.1093/mnras/stz2683}, \href
  {https://ui.adsabs.harvard.edu/abs/2019MNRAS.490.1697O} {490, 1697}

\bibitem[\protect\citeauthoryear{{Padoan}, {Nordlund}  \& {Jones}}{{Padoan}
  et~al.}{1997}]{PadoanEA1997}
{Padoan} P.,  {Nordlund} A.,   {Jones} B. J.~T.,  1997, \mn@doi [\mnras]
  {10.1093/mnras/288.1.145}, \href
  {https://ui.adsabs.harvard.edu/abs/1997MNRAS.288..145P} {288, 145}

\bibitem[\protect\citeauthoryear{{Petroff} et~al.,}{{Petroff}
  et~al.}{2016}]{PetroffEA2016}
{Petroff} E.,  et~al., 2016, \mn@doi [\pasa] {10.1017/pasa.2016.35}, \href
  {https://ui.adsabs.harvard.edu/abs/2016PASA...33...45P} {33, e045}

\bibitem[\protect\citeauthoryear{{Petroff}, {Hessels}  \& {Lorimer}}{{Petroff}
  et~al.}{2019}]{PetroffHL2019}
{Petroff} E.,  {Hessels} J.~W.~T.,   {Lorimer} D.~R.,  2019, \mn@doi [\aapr]
  {10.1007/s00159-019-0116-6}, \href
  {https://ui.adsabs.harvard.edu/abs/2019A&ARv..27....4P} {27, 4}

\bibitem[\protect\citeauthoryear{{Pillai}, {Kauffmann}, {Tan}, {Goldsmith},
  {Carey}  \& {Menten}}{{Pillai} et~al.}{2015}]{PillaiEA2015}
{Pillai} T.,  {Kauffmann} J.,  {Tan} J.~C.,  {Goldsmith} P.~F.,  {Carey} S.~J.,
    {Menten} K.~M.,  2015, \mn@doi [\apj] {10.1088/0004-637X/799/1/74}, \href
  {https://ui.adsabs.harvard.edu/abs/2015ApJ...799...74P} {799, 74}

\bibitem[\protect\citeauthoryear{{Planck Collaboration} et~al.,}{{Planck
  Collaboration} et~al.}{2016a}]{Planck2016filg}
{Planck Collaboration} et~al., 2016a, \mn@doi [\aap]
  {10.1051/0004-6361/201526506}, \href
  {http://adsabs.harvard.edu/abs/2016A%26A...586A.141P} {586, A141}

\bibitem[\protect\citeauthoryear{{Planck Collaboration} et~al.,}{{Planck
  Collaboration} et~al.}{2016b}]{PlanckLSF2016}
{Planck Collaboration} et~al., 2016b, \mn@doi [\aap]
  {10.1051/0004-6361/201528033}, \href
  {https://ui.adsabs.harvard.edu/abs/2016A&A...596A.103P} {596, A103}

\bibitem[\protect\citeauthoryear{{Planck Collaboration\vspace{0cm}}
  et~al.,}{{Planck Collaboration\vspace{0cm}} et~al.}{2016}]{Planck2016fil}
{Planck Collaboration\vspace{0cm}} et~al., 2016, \mn@doi [\aap]
  {10.1051/0004-6361/201425305}, \href
  {http://adsabs.harvard.edu/abs/2016A%26A...586A.136P} {586, A136}

\bibitem[\protect\citeauthoryear{{Ramachandran}, {Backer}, {Rankin}, {Weisberg}
   \& {Devine}}{{Ramachandran} et~al.}{2004}]{RamachandranEA2004}
{Ramachandran} R.,  {Backer} D.~C.,  {Rankin} J.~M.,  {Weisberg} J.~M.,
  {Devine} K.~E.,  2004, \mn@doi [\apj] {10.1086/383179}, \href
  {https://ui.adsabs.harvard.edu/abs/2004ApJ...606.1167R} {606, 1167}

\bibitem[\protect\citeauthoryear{{Rand} \& {Kulkarni}}{{Rand} \&
  {Kulkarni}}{1989}]{RandK1989}
{Rand} R.~J.,  {Kulkarni} S.~R.,  1989, \mn@doi [\apj] {10.1086/167747}, \href
  {https://ui.adsabs.harvard.edu/abs/1989ApJ...343..760R} {343, 760}

\bibitem[\protect\citeauthoryear{{Ravi} et~al.,}{{Ravi}
  et~al.}{2016}]{RaviEA2016}
{Ravi} V.,  et~al., 2016, \mn@doi [Science] {10.1126/science.aaf6807}, \href
  {https://ui.adsabs.harvard.edu/abs/2016Sci...354.1249R} {354, 1249}

\bibitem[\protect\citeauthoryear{{Raymond}}{{Raymond}}{1992}]{Raymond1992}
{Raymond} J.~C.,  1992, \mn@doi [\apj] {10.1086/170892}, \href
  {https://ui.adsabs.harvard.edu/abs/1992ApJ...384..502R} {384, 502}

\bibitem[\protect\citeauthoryear{{Ruzmaikin} \& {Sokoloff}}{{Ruzmaikin} \&
  {Sokoloff}}{1979}]{RuzmaikinS1979}
{Ruzmaikin} A.~A.,  {Sokoloff} D.~D.,  1979, \aap, \href
  {https://ui.adsabs.harvard.edu/#abs/1979A&A....78....1R} {78, 1}

\bibitem[\protect\citeauthoryear{{Seta} \& {Beck}}{{Seta} \&
  {Beck}}{2019}]{SetaB2019}
{Seta} A.,  {Beck} R.,  2019, \mn@doi [Galaxies] {10.3390/galaxies7020045},
  \href {https://ui.adsabs.harvard.edu/abs/2019Galax...7...45S} {7, 45}

\bibitem[\protect\citeauthoryear{{Seta} \& {Federrath}}{{Seta} \&
  {Federrath}}{2020}]{SetaF2020}
{Seta} A.,  {Federrath} C.,  2020, \mn@doi [\mnras] {10.1093/mnras/staa2978},
  \href {https://ui.adsabs.harvard.edu/abs/2020MNRAS.499.2076S} {499, 2076}

\bibitem[\protect\citeauthoryear{{Seta}, {Shukurov}, {Wood}, {Bushby}  \&
  {Snodin}}{{Seta} et~al.}{2018}]{SetaEA2018}
{Seta} A.,  {Shukurov} A.,  {Wood} T.~S.,  {Bushby} P.~J.,   {Snodin} A.~P.,
  2018, \mn@doi [\mnras] {10.1093/mnras/stx2606}, \href
  {http://adsabs.harvard.edu/abs/2018MNRAS.473.4544S} {473, 4544}

\bibitem[\protect\citeauthoryear{Seta, Bushby, Shukurov  \& Wood}{Seta
  et~al.}{2020}]{SetaEA2020}
Seta A.,  Bushby P.~J.,  Shukurov A.,   Wood T.~S.,  2020, \mn@doi [Phys. Rev.
  Fluids] {10.1103/PhysRevFluids.5.043702}, 5, 043702

\bibitem[\protect\citeauthoryear{{Shetty} \& {Ostriker}}{{Shetty} \&
  {Ostriker}}{2006}]{ShettyO2016}
{Shetty} R.,  {Ostriker} E.~C.,  2006, \mn@doi [\apj] {10.1086/505594}, \href
  {https://ui.adsabs.harvard.edu/abs/2006ApJ...647..997S} {647, 997}

\bibitem[\protect\citeauthoryear{{Shukurov}, {Snodin}, {Seta}, {Bushby}  \&
  {Wood}}{{Shukurov} et~al.}{2017}]{ShukurovEA2017}
{Shukurov} A.,  {Snodin} A.~P.,  {Seta} A.,  {Bushby} P.~J.,   {Wood} T.~S.,
  2017, \mn@doi [\apjl] {10.3847/2041-8213/aa6aa6}, \href
  {http://adsabs.harvard.edu/abs/2017ApJ...839L..16S} {839, L16}

\bibitem[\protect\citeauthoryear{{Shukurov}, {Rodrigues}, {Bushby}, {Hollins}
  \& {Rachen}}{{Shukurov} et~al.}{2019}]{ShukurovEA2019}
{Shukurov} A.,  {Rodrigues} L. F.~S.,  {Bushby} P.~J.,  {Hollins} J.,
  {Rachen} J.~P.,  2019, \mn@doi [\aap] {10.1051/0004-6361/201834642}, \href
  {https://ui.adsabs.harvard.edu/abs/2019A&A...623A.113S} {623, A113}

\bibitem[\protect\citeauthoryear{{Smith}}{{Smith}}{1968}]{Smith1968}
{Smith} F.~G.,  1968, \mn@doi [\nat] {10.1038/218325a0}, \href
  {https://ui.adsabs.harvard.edu/abs/1968Natur.218..325S} {218, 325}

\bibitem[\protect\citeauthoryear{{Sobey} et~al.,}{{Sobey}
  et~al.}{2019}]{SobeyEA2019}
{Sobey} C.,  et~al., 2019, \mn@doi [\mnras] {10.1093/mnras/stz214}, \href
  {https://ui.adsabs.harvard.edu/abs/2019MNRAS.484.3646S} {484, 3646}

\bibitem[\protect\citeauthoryear{{Sun}, {Gaensler}, {Carretti}, {Purcell},
  {Staveley-Smith}, {Bernardi}  \& {Haverkorn}}{{Sun} et~al.}{2014}]{SunEA2014}
{Sun} X.~H.,  {Gaensler} B.~M.,  {Carretti} E.,  {Purcell} C.~R.,
  {Staveley-Smith} L.,  {Bernardi} G.,   {Haverkorn} M.,  2014, \mn@doi
  [\mnras] {10.1093/mnras/stt2110}, \href
  {https://ui.adsabs.harvard.edu/abs/2014MNRAS.437.2936S} {437, 2936}

\bibitem[\protect\citeauthoryear{{Sur}, {Bhat}  \& {Subramanian}}{{Sur}
  et~al.}{2018}]{SurBS2018}
{Sur} S.,  {Bhat} P.,   {Subramanian} K.,  2018, \mn@doi [\mnras]
  {10.1093/mnrasl/sly007}, \href
  {https://ui.adsabs.harvard.edu/#abs/2018MNRAS.475L..72S} {475, L72}

\bibitem[\protect\citeauthoryear{{Vazza}, {Br{\"u}ggen}, {Hinz}, {Wittor},
  {Locatelli}  \& {Gheller}}{{Vazza} et~al.}{2018}]{VazzaEA2018b}
{Vazza} F.,  {Br{\"u}ggen} M.,  {Hinz} P.~M.,  {Wittor} D.,  {Locatelli} N.,
  {Gheller} C.,  2018, \mn@doi [\mnras] {10.1093/mnras/sty1968}, \href
  {https://ui.adsabs.harvard.edu/abs/2018MNRAS.480.3907V} {480, 3907}

\bibitem[\protect\citeauthoryear{Waagan, Federrath  \& Klingenberg}{Waagan
  et~al.}{2011}]{WaaganFK2011}
Waagan K.,  Federrath C.,   Klingenberg C.,  2011, J. Comput. Phys., 230, 3331

\bibitem[\protect\citeauthoryear{{Winkel}, {Kerp}, {Fl{\"o}er}, {Kalberla},
  {Ben Bekhti}, {Keller}  \& {Lenz}}{{Winkel} et~al.}{2016}]{WinkelEA2016}
{Winkel} B.,  {Kerp} J.,  {Fl{\"o}er} L.,  {Kalberla} P.~M.~W.,  {Ben Bekhti}
  N.,  {Keller} R.,   {Lenz} D.,  2016, \mn@doi [\aap]
  {10.1051/0004-6361/201527007}, \href
  {https://ui.adsabs.harvard.edu/abs/2016A&A...585A..41W} {585, A41}

\bibitem[\protect\citeauthoryear{{Wu}, {Kim}, {Ryu}, {Cho}  \&
  {Alexander}}{{Wu} et~al.}{2009}]{WuEA2009}
{Wu} Q.,  {Kim} J.,  {Ryu} D.,  {Cho} J.,   {Alexander} P.,  2009, \mn@doi
  [\apjl] {10.1088/0004-637X/705/1/L86}, \href
  {https://ui.adsabs.harvard.edu/abs/2009ApJ...705L..86W} {705, L86}

\bibitem[\protect\citeauthoryear{{Wu}, {Kim}  \& {Ryu}}{{Wu}
  et~al.}{2015}]{WuKR2015}
{Wu} Q.,  {Kim} J.,   {Ryu} D.,  2015, \mn@doi [\na]
  {10.1016/j.newast.2014.02.006}, \href
  {https://ui.adsabs.harvard.edu/abs/2015NewA...34...21W} {34, 21}

\bibitem[\protect\citeauthoryear{{Yao}, {Manchester}  \& {Wang}}{{Yao}
  et~al.}{2017}]{YaoEA2017}
{Yao} J.~M.,  {Manchester} R.~N.,   {Wang} N.,  2017, \mn@doi [\apj]
  {10.3847/1538-4357/835/1/29}, \href
  {https://ui.adsabs.harvard.edu/abs/2017ApJ...835...29Y} {835, 29}

\bibitem[\protect\citeauthoryear{{Zaroubi} et~al.,}{{Zaroubi}
  et~al.}{2015}]{ZaroubiEA2015}
{Zaroubi} S.,  et~al., 2015, \mn@doi [\mnras] {10.1093/mnrasl/slv123}, \href
  {http://adsabs.harvard.edu/abs/2015MNRAS.454L..46Z} {454, L46}

\makeatother
\end{thebibliography}



\appendix
\section{$\RM$ distribution for the external spiral galaxy M51} \label{app:m51}

\begin{figure}
	\includegraphics[width=\columnwidth]{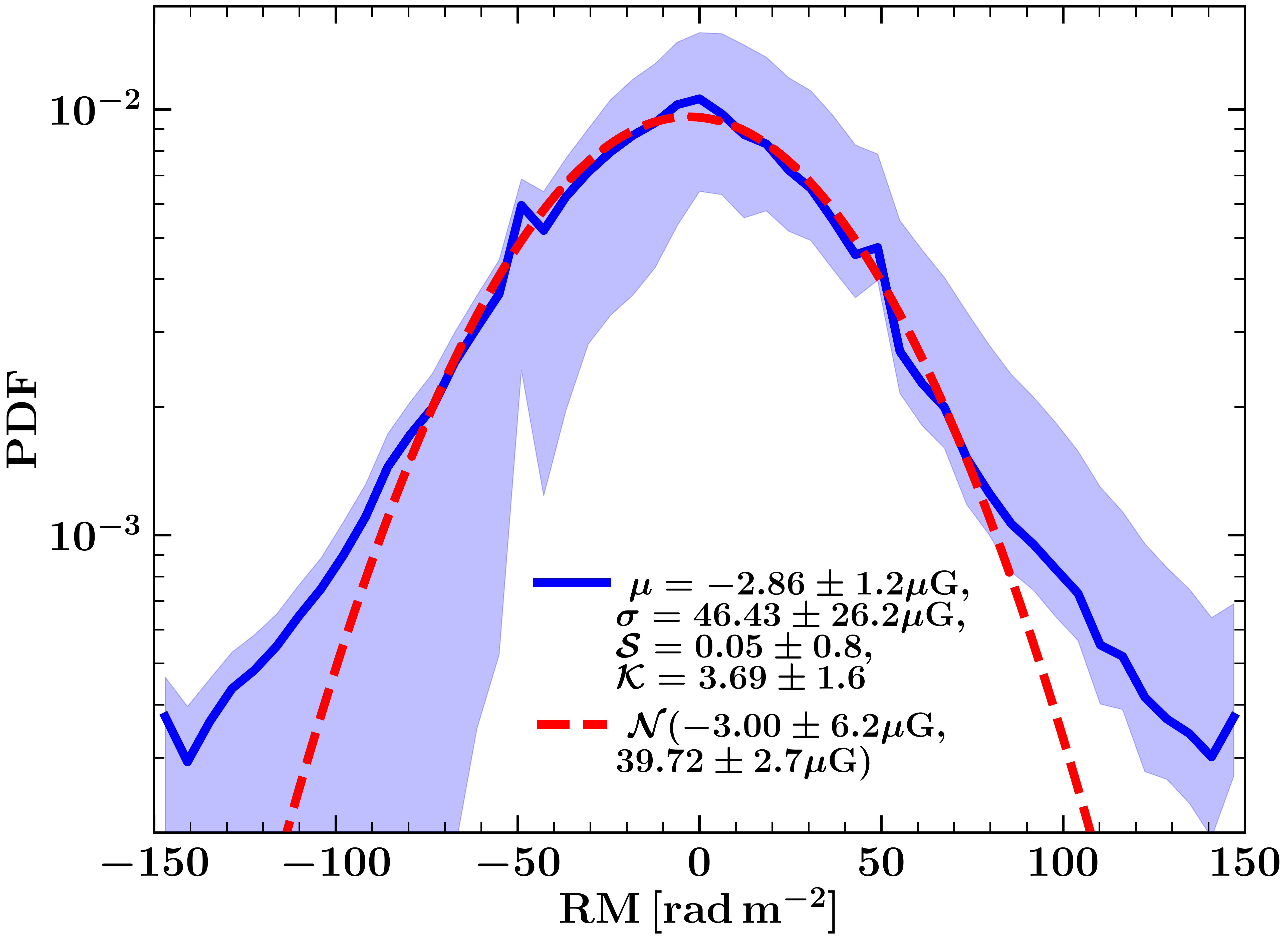} 
	\caption{\rev{Probability distribution function of $\RM$ (solid blue line with the shaded region showing the corresponding uncertainty computed from the error map of $\RM$) in the external galaxy M51 \citep[$\RM$ data and the associated error taken from][Fig.~6]{KierdorfEA2020}. The mean $\mu$, standard deviation $\sigma$, skewness $\sk$, and kurtosis $\ku$ of the $\RM$ distribution is shown in the legend. The red dashed lines show a fitted Gaussian distribution with its mean and standard deviation given in the legend. The distribution is slightly non-Gaussian (kurtosis greater than three) but more data with lower $\RM$ error is needed to reduce the uncertainties in the statistical properties of the $\RM$ distribution.}}
	\label{fig:m51}
\end{figure}

\rev{In \Fig{fig:m51}, we show the probability distribution function of $\RM$s observed in the grand design external spiral galaxy M51 using the data from \cite{KierdorfEA2020}. The primary source of $\RM$, in this case, is the polarised emission from the galaxy itself and not due to point sources such as pulsars. \cite{KierdorfEA2020} associate the high standard deviation of the $\RM$ distribution to the presence of tangled regular fields and/or vertical fields. In \Fig{fig:m51}, we show that the distribution can be non-Gaussian (kurtosis greater than three) and that might give rise to a higher $\RM$ standard deviation when computed directly from the data. The fitted Gaussian distribution has a smaller standard deviation ($\approx 39.72 \rad \m^{-2}$) than the directly computed value of $46.63 \rad \m^{-2}$. This might lower the strength of the small-scale random magnetic fields computed from the standard deviation of the $\RM$ distribution. However, since the uncertainties in the statistical properties of the $\RM$ distribution (calculated using the $\RM$ error map) is relatively high, it is difficult to isolate this effect. Polarised intensity with a high signal-to-noise ratio is required to reduce the uncertainties in the observed $\RM$s, which in turn would reduce the uncertainties in the statistical properties of the $\RM$ distribution.
}


\bsp	
\label{lastpage}
\end{document}